\documentclass{article}
\usepackage{natbib}
\usepackage{epubtk}    
\usepackage{booktabs}  
\usepackage{graphicx}  
\usepackage{amssymb}
\usepackage{footnote}

\newcommand\degree{\mbox{$^\circ$}}%

\newcommand{\matriz}[1]{\mbox{\bf #1}}

\renewcommand{\vector}[1]{\mbox{\boldmath $#1$}}
\newcommand{\kms}{\mbox{$\:$km$\,$s$^{-1}$}}
\newcommand{\ms}{\mbox{$\:$m$\,$s$^{-1}$}}

\newcommand{\trans}{^{\scriptscriptstyle {\rm T}}}

\newcommand{\bbbone}{{\mathchoice {\rm 1\mskip-4mu l} {\rm 1\mskip-4mu l}{\rm 1\mskip-4.5mu l} {\rm 1\mskip-5mu l}}}
\newcommand{\bbbr}{{\mathchoice {\rm I\mskip-4mu R} {\rm I\mskip-4mu R}{\rm I\mskip-4.5mu R} {\rm IRmskip-5mu R}}}
\newcommand{\bbbl}{{\mathchoice {\rm I\mskip-4mu L} {\rm I\mskip-4mu L}{\rm I\mskip-4.5mu L} {\rm IRmskip-5mu L}}}

\begin{document}

\title{Inversion of the radiative transfer equation for polarized light}

\author{\epubtkAuthorData{Jose Carlos del Toro Iniesta}{%
Instituto de Astrof\'{\i}sica de Andaluc\'{\i}a (CSIC) \\
Apdo. de Correos 3004, E-18080, Granada, Spain}{%
jti@iaa.es} { }
\and \epubtkAuthorData{Basilio Ruiz Cobo}{%
Instituto de Astrof\'{\i}sica de Canarias \\
V\'{\i}a L\'{a}ctea, s/n, E-38200, La Laguna (Tenerife), Spain}{%
brc@iac.es} { }
}

\date{}
\maketitle

\begin{abstract}
Since the early 1970s, inversion techniques have become the most useful tool for inferring the magnetic, dynamic, and thermodynamic properties of the solar atmosphere. Inversions have been proposed in the literature with a sequential increase in model complexity: astrophysical inferences depend not only on measurements but also on the physics assumed to prevail both on the formation of the spectral line Stokes profiles and on their detection with the instrument. Such an intrinsic model dependence makes it necessary to formulate specific means that include the physics in a properly quantitative way. The core of this physics lies in the radiative transfer equation (RTE), where the properties of the atmosphere are assumed to be known while the unknowns are the four Stokes profiles. The solution of the (differential) RTE is known as the direct or forward problem. From an observational point of view, the problem is rather the opposite: the data are made up of the observed Stokes profiles and the unknowns are the solar physical quantities. Inverting the RTE is therefore mandatory. Indeed, the formal solution of this equation can be considered an integral equation. The solution of such an integral equation is called the inverse problem. Inversion techniques are automated codes aimed at solving the inverse problem. 

The foundations of inversion techniques are critically revisited with an emphasis on making explicit the many assumptions underlying each of them. An incremental complexity procedure is advised for the implementation in practice. Coarse details of the profiles or coarsely sampled profiles should be  reproduced first with simple model atmospheres (with, for example, a few physical quantities that are constant with optical depth). If the Stokes profiles are well sampled and differences between synthetic and observed ones are greater than the noise, then the inversion should proceed by using more complex models (that is, models where physical quantities vary with depth or, eventually, with more than one component). Significant improvements are expected as well from the use of new inversion techniques that take the spatial degradation by the instruments into account.

\end{abstract}

\epubtkKeywords{Solar magnetic fields, polarimetry, spectropolarimetry, inversion techniques}

\newpage
\tableofcontents
\newpage

\section{Introduction}
\label{sec:introduction}

Unlike other branches of physics, astrophysics cannot apply the third pillar of the scientific method, experimentation. After observing nature and conjecturing laws that govern its behavior, astronomers cannot carry out experiments that confirm or falsify the theory. Experimentation is then substituted by new observations conducted to check the theoretical predictions. The intrinsic inability for directly measuring the celestial objects adds a special difficulty to the astrophysical tasks. We do not have thermometers, weighing scales, tachometers, magnetometers that can directly gauge the physical conditions in the object. Rather we have to be content with indirect evidence or inferences obtained from the only real astrophysical measurements, namely those related to light. The intensity and polarization properties for visible light, the associated electric field for radio frequencies, or the energy or momentum of high energy photons, as functions of space, wavelength, and time, can be fully quantified with \emph{errors} that are directly related to the accuracy of the instruments.\footnote{Actual experiments have also been devised to directly detect neutrinos, other particles, and gravitational waves, but such projects are way beyond the scope of this paper.} From these real measurements, the observational astronomer must deduce or infer the physical quantities that characterize the object with \emph{uncertainties} that depend on both the experimental errors and the assumptions that allow him/her to translate light-derived quantities into the object quantities. Observational astrophysics could hence probably be defined as the art of \emph{inferring} the physical quantities of heavenly bodies from real measurements of the light received from them.

Somehow, these astrophysical tasks can be mathematically seen as a mapping between two spaces, namely the space of observables and that of the object's physical quantities. The success of the astronomer then depends on his/her ability (\emph{the art}) to characterize not only the mapping but the two spaces. On the observable side, what really matters is the specific choice of measurable parameters  and how well they are measured; that is, how many light parameters are obtained (the signal) and which are the measurement errors (the noise). On the object's physical condition side, what is substantive is the selection of quantities to be inferred. Of course, the finer the ---affordable--- detail in describing any of the two spaces, the better. The keyword is \emph{affordable} because infinite resolution does not exist in the real world: a compromise is always in order between the number of available observables and the number of inferred physical quantities. The representation of both spaces therefore needs approximations that constrain the sub-spaces to be explored and how they are described: which Stokes parameters, with which wavelength and time sampling, and  with which instrument profile and resolution on the one hand, and, on the other, which quantities and how they are assumed to vary on the object with time and space. Concerning the mapping, this should represent the physics that generates the observables from the given physical conditions in the object and thus illustrates the dependence of the observables on given physical quantities. Understanding this physics is crucial if the researcher is to select observables that are as ``orthogonal'' as possible; that is, that depend mostly on one physical quantity and not on the others. Certainly, the physics mapping needs approximations as well. These approximations depend a great deal on the observables and on the object's physical quantities; for example, the assumptions cannot be the same if you have fully sampled Stokes profiles or just a few wavelength samples; different hypotheses apply for physical quantities that do or do not vary with depth in the atmosphere, or that are expected to present a given range of magnitudes. Therefore, mappings may include (often over-simplistic) one-dimensional calibration curves between a given observable parameter and a given physical quantity, or complicated multidimensional relationships between observables and quantities that require the definition of a metric or distance in at least one of the two spaces. 

Even in the simplest situations, the relationship between observables and quantities does not have to be linear and may depend on the specific sub-space of the physical parameters. For example, a calibration curve based on the weak-field approximation may apply for a given range of magnetic fields but saturate for stronger ones (see Sections~\ref{sec:weakfield} and \ref{sec:weakatmosphere}). But, when the problem can be assumed to be multidimensional, covariances appear because single observables rarely depend on just a single quantity (see Section~\ref{section:techniques}). For example, a given spectral line Stokes $V$ profile can seemingly grow or weaken by the same amount owing to changes in temperature or magnetic field strength \citep[e.g.,][]{1996SoPh..164..169D}. An example can be seen in Figure~\ref{fig:stokes}, where two apparently equal $V$ profiles come from two different atmospheres. With all these ingredients at hand, the astrophysical analysis of observations is a non-linear, fully involved, topological task where many decisions have to be made (\emph{the art}) and, hence, cannot be taken for granted. 

\epubtkImage{}{%
  \begin{figure}[htbp]
   \centerline{\includegraphics[width=\textwidth]{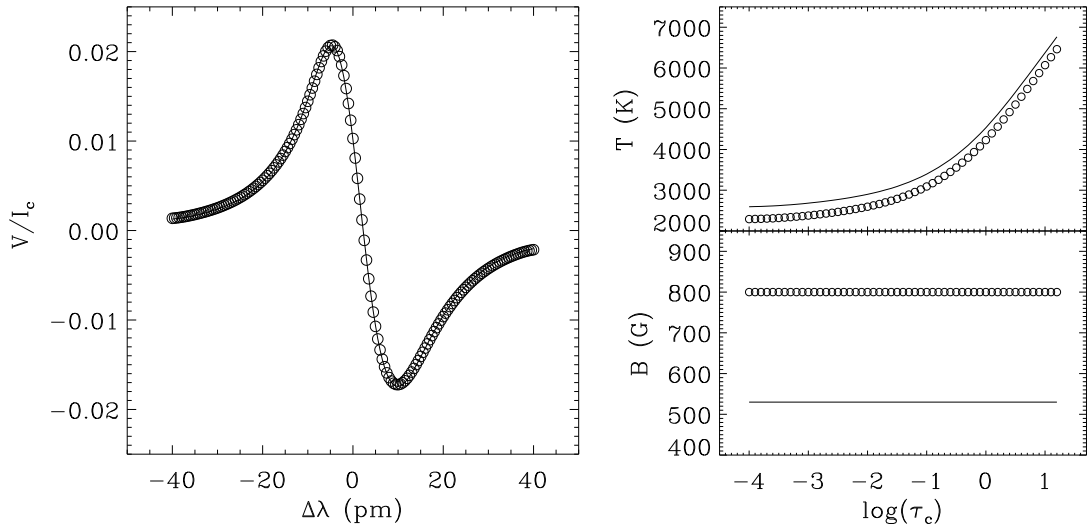}}
    \caption{Left panel: Open circles: Stokes $V$ profile in units of the continuum intensity of the Fe~{\sc i} line at 630.25 nm synthesized in a model atmosphere in hydrostatic equilibrium, 2000~K cooler than the \cite{1994ApJ...436..400D} model, with a constant longitudinal magnetic field of 800~G, a gradient in velocity from 2~\kms at the bottom and 0~\kms at the top of the photosphere,  and  a macroturbulence velocity of 1~\kms. Solid line: Stokes $V$ profile of the same line (normalized the same way), synthes\-ized in a model atmosphere 305~K hotter that the former, 270~G weaker, and with a higher macroturbulence velocity of 2.06~\kms. Right panel: $T$ and $B$ stratifications for the two models.}
    \label{fig:stokes}
\end{figure}}

The techniques by which astronomers have obtained information about the physical conditions in the object have evolved in parallel to technological advancements; that is, to the available means we have of gathering such information. 
The community has gradually enhanced its knowledge from medium-band measurements including one or several spectral lines to very fine wavelength sampling of the four Stokes profiles of single or multiple spectral lines; from old curves of growth for equivalent widths to highly sophisticated techniques that include the solution of the radiative transfer equation (RTE). The finer the information, the more complete the physical description.

Following \citet{2001ASPC..236..487S}, let us consider the simplest case of having a single observable parameter, the Doppler displacement with respect to the rest position of the spectral line, $\Delta \lambda$, and a single physical quantity to derive, the line-of-sight (LOS) velocity, $v_{\rm LOS}$. Imagine that we measure $\Delta \lambda$ by finding the minimum (or the maximum in the case of an emission line) of the intensity profile. The biunivocal mapping between the one-dimensional space of observables ---that containing all possible Doppler displacements--- and the one-dimensional space of physical quantities ---that of LOS velocities--- is given by the Doppler formula \begin{equation}
\label{eq:doppler}
v_{\rm LOS} = \frac{\Delta\lambda}{\lambda_0} c,
\end{equation}
where $\lambda_0$ stands for the vacuum rest wavelength position of the line and $c$ for the speed of light. This simple inference relationship requires at least three implicit physical assumptions for the Doppler displacement to be properly defined and measured; namely that a) the solar feature is spatially resolved, b)  the line is in pure absorption (or pure emission), and c) $v_{\rm LOS}$ is constant along the LOS. First, if we have unresolved structures we cannot ascribe the inferred velocity to any of them. Second, lines with core reversals, either in absorption or in emission, do not qualify for the extremum-finding method. And third, as soon as we have an asymmetric profile, $\Delta\lambda$ can no longer be properly defined for the line but for a given height through the profile, and then the mapping in Eq.\ (\ref{eq:doppler}) immediately loses its meaning. While in the case of a constant velocity, we properly infer that velocity, in the presence of gradients we infer a value corresponding only to the ---in principle unknown--- layers where the core of our line has been formed (typically the highest layers of the atmosphere). \emph{We measure a velocity but we do not know which one.} Strictly speaking, the same measurement corresponds to different physical quantities depending on the assumptions. Of course we could complicate our problem a little and try to determine the stratification of LOS velocities with height, or simply estimate a gradient, by measuring the so-called bisector, the geometric position of those points equidistant from both wings of the profile at a given depth. At that point, our spaces have increased their dimensions and Eq.\ (\ref{eq:doppler}) is no longer the sole ingredient of our mapping because we must add some more physical assumptions to interpret the different displacements of the bisector in terms of velocities at different heights in the atmosphere. Hence, depending on the assumed physics, the quantitative results may change. This easy example has been used to illustrate that even the simplest inference is dependent on  physical assumptions. This is an inherent property of astrophysical measurement and no one can escape from it: the same observable can mean different things depending on the assumed underlying physics. Most of the criticisms of the inversion techniques that are reviewed in this paper often come from this lack of uniqueness of the results. Many authors claim that the inversion of the RTE is an ill-posed problem. This being true, one should realize that astrophysics itself is indeed ill-conditioned, and this is a fact we have to deal with, either willingly or not.

The physics connecting the object quantities with the observable parameters is of paramount significance and deserves a little consideration at this point. Radiative transfer is the discipline encompassing the generation and transport of electromagnetic radiation through the solar (stellar) atmosphere. Hence, the mapping between the two spaces will be based upon it and depend on its degrees of approximation. The specification of the radiation field through a scattering atmosphere was first formulated as a physical problem by \cite{1871PhilMag..41...107,1871PhilMag..41...274,1881PhilMag..12...81,1899PhilMag..47...375}. In the astrophysical realm, the problem was posed in the works by \cite{1905ApJ....21....1S} and \cite{1906GottNach...41S} without taking polarization into account. After that, although not known to the astrophysical community, \citet{1929AnnPhys12..23S} presented a theory of anisotropic absorption that is nothing but a rigorous formulation of the radiative transfer equation. Very importantly, he used the formalism proposed by \citet{1852TransCamb..9...399S} to deal with partially polarized light. It was not, however, until the works by \citet{1946ApJ...103..351C,1946ApJ...104..110C,1947Apj...105..424C} that the transfer problem of polarized light was settled as an astrophysical problem on its own. The Stokes formalism has regularly been used since then in the astronomical literature. After Hale's  (\citeyear{hale1908}) discovery of sunspot magnetic fields, the interpretation of the solar (stellar) spectrum of polarized light became necessary and a full theory has been developed since the mid 1950s. The first modern formulation of an equation of radiative transfer for polarized light was presented by \cite{1956PASJ....8..108U}, who also provided a solution in the simplified case of a Milne--Eddington (ME) atmosphere. Only absorption processes were taken into account and a complete description had to wait until the works by \cite{1962IZKry..27...148R,1962IzKry..28...259R,1967IzKry..37...56R}, who also included dispersion effects (the so-called magneto-optical effects). These two derivations were phenomenological and somewhat heuristic. A rigorous derivation of the radiative transfer equation (RTE) based on quantum electrodynamics was obtained by \cite{landi+landi1972}. Later, four derivations of the RTE from basic principles of classical physics were published by \cite{jefferies+etal1989}, \citeauthor{1991sopo.work..416S} (\citeyear{1991sopo.work..416S}; see also \citeauthor{1994KAP...book...S} \citeyear{1994KAP...book...S}),  \citeauthor{1992soti.book...71L} (\citeyear{1992soti.book...71L}; see also \citeauthor{2004ASSL..307.....L} \citeyear{2004ASSL..307.....L}), and \cite{2003isp..book.....D}. A discussion of the RTE and the several assumptions used in various available inference techniques is deferred to Section~\ref{section:assumptions}.

Certainly, any inference has to be based on solutions of the RTE because it relates the observable Stokes spectrum with the unknowns of the problem; namely, the physical quantities characterizing the state of the atmosphere they come from. No matter how simplified such solutions can be, it is natural to compare the observations with theoretical calculations in prescribed sets of physical quantities. The comparison of observational and synthetic parameters results in values for the sought-for quantities that may be refined in further iterations by changing the theoretical prescriptions. This trial-and-error method can be practical when the problem is very simple (involving a few free parameters) but can become unsuitable for practical use if the number of free parameters is large. Even automated trial-and-error ---i.e., Monte Carlo--- methods may fail to converge to a reliable set of physical conditions in the medium. Some more educated techniques are needed to finally work out that convergence between observed and synthetic parameters. 

Generally speaking, any method in which information about the integrand of an integral equation is obtained from the resulting value of the integral is called an inversion method. In our particular case, it is straightforward to write the synthetic Stokes spectra as an integral involving a kernel that depends on the physical conditions of the atmosphere (see Eq.\ \ref{eq:rteformalsolution}). In fact, the emergent formal solution of the RTE is the most basic type of integral equation, namely a Fredholm equation of the first type, because both integration limits are fixed. Consequently, we will call inversion codes or inversion techniques those methods that (almost) automatically succeed in finding reliable physical quantities from a set of observed Stokes spectra because we shall understand that they indeed automatically solve that integral equation. There is a whole variety of flavors depending on the several hypotheses that can be assumed, but all of them share the characteristic feature of automatically minimizing a distance in the topological space of observables. The idea had already been  clearly explained in the seminal work by \cite{1972lfpm.conf..227H}: ``Solve for \vector{B} on the bases of best fit of the observed profiles to the theoretical profiles''. And the free parameters for such a best fit were found through least squares minimization of the profile differences. They obtained only an average longitudinal field component because their Stokes $Q$ and $U$ observations were not fully reliable and  magneto-optical effects were not taken into account, but the fundamental idea underlying many of the current techniques can already be found in that very paper, including a simple two-component model to describe the possible existence of spatially unresolved magnetic fields. 

In a thorough study using synthetic Stokes profiles, \cite{auer+etal1977} proposed a new inversion method based on Unno's theory and tested its behavior in the presence of several realistic circumstances, such as asymmetric profiles, magnetic field gradients, magneto-optical effects, and unresolved magnetic features. This technique was later generalized by \cite{1984SoPh...93..269L} to include magneto-optical and damping effects. The numerical check of the code was fairly successful but neither the original code by \cite{auer+etal1977} nor the new one by \cite{1984SoPh...93..269L} were applied to observations. Independently of the latter authors, the preliminary studies by \cite{1985NASCP2374..341S}, \cite{1985NASCP2374..342L}, and \cite{1985NASCP2374..306S} jelled in what has been one of the most successful ME inversion codes so far by \citet{skumanich+lites1987}, later extended by \citet{1988ApJ...330..493L} to mimic a chromospheric rise in the source function (see Section~\ref{sec:milne}). This code has been extensively used with observational data, most notably those obtained with the Advanced Stokes Polarimeter \citep{1992SPIE.1746...22E}. 

Based on the thin flux tube approximation, \citet{1990A&amp;A...233..583K} proposed an inversion code for extracting physical information not from the Stokes profiles themselves but from several parameters calculated from $I$ and $V$ observations of a plage and a network. Two years later, \citet{1992A&amp;A...263..339S} presented a new inversion code whereby from the whole Stokes $I$ and $V$ profiles they selected among a handful of prescribed temperature stratifications and inferred height-independent magnetic field strength and inclination, Doppler shift, filling factor (surface fraction in the resolution element covered by magnetic fields), macro- and microturbulent velocities, and some  atomic parameters of the spectral line. The very same year, \citet{1992ApJ...398..375R} introduced SIR, an acronym for Stokes Inversion based on Response functions. Like the former codes, SIR ran a non-linear, least-squares, iterative Levenberg--Marquardt algorithm but with a remarkable step-forward feature: physical quantities characterizing the atmosphere were allowed to vary with optical depth. The increase of free parameters can generate a singularity problem: the variation of some atmospheric parameters may not produce any change on the synthetic spectra or, in other cases, different combinations of the perturbation of several parameters may produce the same change in the spectra. The success of SIR lies in regularizing the problem through a \emph{tailored} Singular Value Decomposition method (SVD). This allows, in principle, to look for any arbitrarily complex atmospheric stratification. The three components of the magnetic field, the LOS velocity, the temperature stratification, and the microturbulence may have any height profile. The code also infers height-independent microturbulent velocity and filling factor. The possibility exists for also fitting some atomic parameters \citep[e.g.,][]{2001ApJ...558..830A}. but they are typically fixed in practice. The code can be applied to any number of spectral lines that are observed simultaneously. SIR has been successful in a large number of observing cases and its use is still spreading among the community. 

Following SIR's strategy (that is, using response functions, nodes, Levenberg--Marquardt, and SVD), an evolution of the \citet{1992A&amp;A...263..339S} code called SPINOR was presented by \citet{1998A&amp;A...336L..65F} that also allowed for height variations of the physical quantities and included the possibility of multi-ray calculations assuming the thin flux tube approximation. \citet{1997ApJ...491..993S} proposed an original inversion code under the MISMA (MIcro-Structured Magnetic Atmosphere) hypothesis (see Sections~\ref{sec:misma} and \ref{sec:mismaatmos}). In 2000, the codes by \citet[][NICOLE ---NLTE Inversion Code based on the Lorien Engine---]{2000ApJ...530..977S} and by \citeauthor{2000ApJ...535..475B} (\citeyear{2000ApJ...535..475B}; see also \citeyear{1997ApJ...478L..45B}) were presented. The first (based on an earlier code by \citeauthor{1998ApJ...507..470S} \citeyear{1998ApJ...507..470S} without taking either polarization or magnetic fields into account) included non-LTE radiative transfer (see Section~\ref{sec:NLTE}), and the second was specifically designed for analyzing Stokes $I$ and $V$ profiles in terms of the thin flux tube approximation by using an analytic shortcut for radiative transfer proposed by \citet[][see Section~\ref{sec:interlaced}]{1995A&A...294..855D}. On their hand, \citet{rees+etal2000} proposed a Principal Component Analysis (PCA), which worked by creating a database of synthetic Stokes profiles by means of an SVD technique. In such a database, given \emph{eigenprofiles} are obtained that are later used as a basis for expanding the observed Stokes profiles. Hence, the description of observations can be made with the help of a few coefficients, thus speeding up the inversion process. One year later, LILIA (LTE Inversion based on Lorien Iterative Algorithm), a code with similar properties as SIR, was presented by \citet{2001ASPC..236..487S} and FATIMA (Fast Analysis Technique for the Inversion of Magnetic Atmospheres), a PCA code, was introduced by \citet{2001ApJ...553..949S}. A different technique was proposed by \citet[][see also \citeauthor{2003NN.....16..355S} \citeyear{2003NN.....16..355S}]{2001ASPC..236..511C} that used artificial neural networks (ANNs) whereby the system was trained with a set of synthetic Stokes profiles. The structure obtained therefrom finds the solution for the free parameters by interpolating among the known ones. Although the training can be slow, the inversion of observational data is very fast. In practice, both the synthetic training set of ANNs and the synthetic database of PCA have employed ME profiles to keep the implementation feasible. Otherwise, the number of free parameters would render the two techniques impracticable. A PCA code to analyze the Hanle effect in the He~{\sc i}~D$_{3}$ line was developed by \citet[][see also \citeauthor{2005ApJ...622.1265C}, \citeyear{2005ApJ...622.1265C}]{2003ApJ...582L..51L}. 

A substantial modification of the original SIR code, called SIRGAUSS, was presented by \citet{2003ASPC..307..301B} in which the physical scenario included the coexistence of an inclined flux tube ---that is pierced twice by the LOS--- within a background. Such a scenario is used to describe an uncombed field model of sunspot penumbrae \citep{solanki+montavon1993}. An evolution of this inversion code, called SIRJUMP, was later used by \citet{2009ApJ...704L..29L} that was able to infer possible discontinuities in the physical quantities along the LOS. A further code presented by \cite{2004PhDULL...A} was able to deal with the Zeeman effect in molecular lines. The very same year, \citet{2004A&amp;A...414.1109L} published {\sc HeLIx}, an ME inversion code that dealt with the Hanle and the Zeeman effect in the He {\sc i} line at 1083 nm. Another ME inversion code was presented by \citet{2007A&A...462.1137O} with the helpful feature that was written in IDL, so that it is easily manipulated by relatively inexperienced users and employed as a routine in high-level programming pipelines. Also in 2007, \citeauthor{2007A&amp;A...464..323B} took over the \cite{1984SoPh...93..269L} method and extended it to include unresolved magnetic structures. Unfortunately, they fail to obtain the magnetic field strength and the filling factor separately; only their product is reliable. Self-consistent levels of confidence in the ME inversion results were estimated through the code proposed by \citet{2007A&amp;A...476..959A} using Bayesian techniques. A rigorous treatment of optical pumping, atomic level polarization, level crossings and repulsions, Zeeman, Paschen-Back, and Hanle effects on a magnetized slab was included in HAZEL \citep{2008ApJ...683..542A}, with which analysis of the He~{\sc i} D$_3$ and the multiplet at 1083 nm can be carried out. 

Oriented to its extensive use with the data coming from the Helioseismic and Magnetic Imager \citep{2003ASPC..307..131G} aboard the Solar Dynamics Observatory, \citet{2011SoPh..273..267B} presented VFISV (Very Fast Inversion of the Stokes Vector), a new ME code but with several further approximations and simplifying assumptions to make it significantly faster than other available codes. \citet{2011A&amp;A...535A..45M} presented an alternative inversion code in which, with a significant number of simplifying assumptions on top of the ME approximation (such Stokes $I$ profiles being Gaussians and magneto-optical effects being almost negligible), some moments of the Stokes profiles are used to retrieve the vector magnetic field and the LOS velocity. In 2012, a significant step forward was provided by \citeauthor{2012A&amp;A...548A...5V}, who combined spectral information with the known spatial degradation effects on two-dimensional maps to obtain a consistent restoration of the atmosphere across the whole field of view. An aim similar to van Noort's is followed by \citet{2013A&amp;A...549L...4R}, who, by means of a regularized method (indeed based on PCA), deconvolve the spectropolarimetric data that are later inverted with SIR. Based on the concept of sparsity,  \citet{2015A&amp;A...577A.140A} have proposed a novel technique that allows the inversion of two-dimensional (potentially three-dimensional) maps at once.

The interested reader can complement this chronological overview with the reviews by \citet{1995INV-DelToroRuizCobo, 1996SoPh..164..169D, 1997INV-DelToroRuizCobo}, \citet{2001ASPC..236..487S}, \citet{2003AN....324..383D}, \citet{2006ASPC..358..107B}, and \citet{2012ApJ...748...83A} and the didactical introductions and discussions by \citet{1994KAP...book...S}, \citet{2003isp..book.....D}, and \cite{2004ASSL..307.....L}. A critical discussion on the different techniques and the specific implementations will be developed throught the paper, which is structured as follows: the basic assumptions of radiative transfer are discussed in Section~\ref{section:assumptions}; the following two Sections discuss the approximations used for the model atmospheres and the Stokes profiles; an analysis of the forward problem, namely the synthesis of the Stokes spectrum, is presented in Section~\ref{section:synthesis}, which is followed by an analysis of the sensitivities of spectral lines to physical quantities (Section~\ref{section:response}); the basics of inversion techniques are  analyzed in Section~\ref{section:techniques} and a discussion on inversion results presented in Section~\ref{section:discussion}; finally, Section~\ref{section:conclusions} summarizes the conclusions. An appendix proposes an optimum way of initializing the inversion codes through the use of classical estimates.

\newpage


\section{Radiative transfer assumptions}
\label{section:assumptions}

The propagation of electromagnetic energy through a stellar atmosphere ---and its eventual release from it--- is a significantly complex, non-linear, three-dimensional, and time-dependent problem where the properties of the whole atmosphere are involved. From deep layers up to the stellar surface, the coupling between the radiation field and the atmospheric matter implies non-local effects that can connect different parts of the atmosphere. In other words, the state of matter and radiation at a given depth may depend on that at the other layers: light emitted at one point can be absorbed or scattered at another to release part or all of its energy.

The description of the whole system, matter plus radiation field, needs to resort to the  solution of the coupled equations that describe the physical state of the atomic system and that of the radiation traveling through it. Therefore, we have to simultaneously solve the so-called statistical equilibrium equations and the radiative transfer equation. The first assumption we shall make is that radiative transfer is one dimensional; that is, that the transfer of radiative energy perpendicular to the line of sight can be neglected in the matter--radiation coupling. For most solar applications so far, this assumption has been seen to be valid. Since the purpose of this paper is not directly related to either of the two systems of equations, let us simply point out what their main characteristics and ingredients are, and how the whole problem can be simplified in different situations. We refer the interested reader to the book by \citet{2004ASSL..307.....L} for a full and rigorous account of all the details.


Most classical radiative transfer descriptions in the literature do not deal with polarization. They are typically qualified as radiative transfer studies for unpolarized light but the name is ill-chosen. Formally speaking, those analyses are for light traveling through homogeneous and isotropic media  \citep{2003isp..book.....D}. As a consequence of that heritage, the community is used to speak about atomic level populations either calculated through the Boltzmann and Saha equations (the LTE approximation; see Sect.\ \ref{sec:LTE}) or not (the non-LTE case; see Section\ \ref{sec:NLTE}). These isotropic descriptions of the transfer problem, however, are not valid when a physical agent such as a vector magnetic field establishes a preferential direction in the medium, hence breaking the isotropy. Moreover, the outer layers of a star are a clear source of symmetry breaking. The exponential density decrease with height makes the radiation field anisotropic: outward opacity is much smaller than inward opacity. This should also be the case with collisions between particles: they are more probable at the bottom than at the top of the atmosphere. In such a situation, the probability is not zero for the various degenerate levels of the atom (with respect to energy) to be not evenly populated and for non-zero coherences or phase relations between them to exist. The atomic system is then said to be polarized and its state is best described with the so-called density operator, $\vector{\rho}$, that provides the probabilities of the sublevels being populated (hence the populations) along with the possible \emph{correlations} or \emph{interferences} between every pair. In the standard representation that uses the eigenvectors of the total angular momentum, $\vector{J}^2$, and of its third component, $\vector{J}_z$, as a basis, the density matrix element
\begin{equation}
\label{eq:densmatelem}
\rho(\alpha j m, \alpha' j' m') = \langle\alpha j m | \rho | \alpha' j' m' \rangle
\end{equation}
represents the coherence or phase interference between the different magnetic sublevels characterized by their angular momentum quantum numbers. In Eq.\ (\ref{eq:densmatelem}), $\alpha$ and $\alpha'$ stand for supplementary quantum numbers relative to those operators that commute with $\vector{J}^2$ and $\vector{J}_z$. Certainly, the diagonal matrix elements $\rho_{\alpha} (j m, j m) \equiv \rho(\alpha j m, \alpha j m)$ represent the populations of the magnetic sublevels and the sum
\begin{equation}
\label{eq:pobla}
n_j = \sum_m \rho_{\alpha} (j m, j m) = \sum_{m=-j}^{j} \langle\alpha j m | \rho | \alpha j m \rangle
\end{equation}
accounts for the total population of the level characterized by the $j$ quantum number.

At all depths in the atmosphere, evolution equations for these density matrix elements have to be formulated that describe their time ($t$) variations due to the transport of radiation, on the one hand, and to collisions among particles on the other. All interactions with light ---namely, pure absorption ($A$), spontaneous emission ($E$), and stimulated emission ($S$)--- have to be considered. All kinds of collisions ---namely, inelastic ($I$), superelastic ($S$), and elastic ($E$) collisions--- have to be taken into account. Inelastic collisions induce transitions between any level $|\alpha jm\rangle$ and an upper level $|\alpha_u j_u m_u\rangle$ with a consequent loss in kinetic energy. Superelastic collisions induce transitions to a lower energy level $|\alpha_l j_lm_l\rangle$ with an increase in the kinetic energy of collision. Finally, elastic collisions induce transitions between degenerate levels $|\alpha jm\rangle$ and $|\alpha jm'\rangle$; in these, the colliding particle keeps its energy during the interaction. The statistical equilibrium equations (\ref{eq:equilradeqs}) and (\ref{eq:equilcoleqs}) that follow for radiative and collisional interactions, respectively, have slightly different application ranges. The former are valid for the multi-term atom representation and can even be used in the Paschen--Back regime, while the latter are only valid for the special case of the multi-level atom representation (although they can be generalized to the multi-term representation).\footnote{The concepts of multi-level or multi-term representation of an atomic system basically depend on the assumption or not, respectively, that coherences can be neglected among magnetic sub-levels that belong to levels characterized by different quantum numbers $\alpha$ and $j$. See \citet{2004ASSL..307.....L} for a detailed and rigorous description.} We make them explicit here for illustrative purposes only and refer the interested reader to the \citeauthor{2004ASSL..307.....L}'s (\citeyear{2004ASSL..307.....L}) monograph for details. According to that work, the radiative interaction equations in the magnetic field reference frame\footnote{Where the vector magnetic field marks the $Z$ direction.} can be written as
$${\displaystyle \frac{\rm d}{{\rm d} t}} \rho_{\alpha} (jm,j'm') = -2\pi {\rm i} \, \nu_{\alpha} (jm,j'm') \rho_{\alpha} (jm,j'm') +$$
$${\displaystyle \sum_{\alpha_l j_l m_l j'_l m'_l}^{\,}} \rho_{\alpha_l} (j_l m_l,j'_lm'_l) \, T_A (\alpha jmj'm',\alpha_l j_l m_lj'_lm'_l) +$$
\begin{equation}
\label{eq:equilradeqs} 
 {\displaystyle \sum_{\alpha_u j_u m_u j'_u m'_u}^{\,}} \rho_{\alpha_u} (j_um_u,j'_um'_u) \left[ T_E (\alpha jmj'm', \alpha_u j_u m_u j'_u m'_u) + T_S (\alpha jmj'm', \alpha_u j_u m_u j'_u m'_u) \right] - 
\end{equation}
$${\displaystyle \sum_{j''m''}^{\,}} \left\{ \rho_{\alpha} (jm,j''m'') \left[ R_A (\alpha j'm'j''m'') + R_E (\alpha j''m''j'm') + R_S (\alpha j''m''j'm') \right] \right. +$$
$$\rho_{\alpha} (j''m'',j'm') [ R_A (\alpha j''m''jm) + R_E (\alpha jmj''m'') + R_S (\alpha jmj''m'')] \}$$
where $\nu_{\alpha} (jm,j'm')$ is the frequency difference between the two sublevels and the $T$'s and $R$'s are radiative rates of coherence transfer and relaxation among the sublevels, respectively. Now, the collisional interactions give
$${\displaystyle \frac{\rm d}{{\rm d} t}} \rho_{\alpha} (jm,jm') = {\displaystyle \sum_{\alpha_l j_l m_l m'_l}^{\,}} C_I (\alpha j mm',\alpha_l j_l m_l m'_l) \rho_{\alpha_l} (j_lm_l,j_lm'_l) + $$
$${\displaystyle \sum_{\alpha_u j_u m_u m'_u}^{\,}} C_S (\alpha j mm',\alpha_u j_u m_u m'_u) \rho_{\alpha_u} (j_um_u,j_um'_u) +$$ 
\begin{equation}
\label{eq:equilcoleqs}
{\displaystyle \sum_{m'' m'''}^{\,}} C_E (\alpha j mm',\alpha j m'' m''') \rho_{\alpha} (jm'',jm''') -
\end{equation}
$${\displaystyle \sum_{m''}^{\,}} \left[ \frac{1}{2} X(\alpha jmm'm'') \rho_{\alpha}(jm,jm'') + \frac{1}{2} X(\alpha jm'mm'')^* \rho_{\alpha} (jm'',jm') \right. -$$
$$\left. \frac{1}{2} X_E(\alpha jmm'm'') \rho_{\alpha}(jm,jm'') + \frac{1}{2} X_E(\alpha jm'mm'')^* \rho_{\alpha} (jm'',jm') \right],$$
where the $C$'s are collisional transfer rates between levels and the $X$'s are relaxation rates. The indices refer to the corresponding type of collisions and the asterisk denotes the complex conjugate.

With the standard notation for the Stokes pseudo-vector $\vector{I} \equiv (I,Q,U,V)\trans$, where index {\scriptsize T} stands for the transpose, the radiative transfer equation can be written as \citep[e.g.,][]{2003isp..book.....D}
\begin{equation}
\label{eq:rte}
\frac{\rm d \vector{I}}{\rm d \tau_{\rm c}} = \matriz{K} (\vector{I} - \vector{S}),
\end{equation}
where $\tau_{\rm c}$ is the optical depth at the continuum wavelength, $\matriz{K}$ stands for the propagation matrix, and $\vector{S}$ is the so-called source function vector. Since the continuum spectrum of radiation can safely be assumed flat within the wavelength span of a spectral line and non-polarized as far as currently reachable polarimetric accuracies are concerned, the optical depth, defined as 
\begin{equation}
\label{eq:opticaldepth}
\tau_{\rm c} \equiv \int_s^{s_{\rm lim}} \chi_{\rm cont} \, {\rm d}s,
\end{equation}
is the natural length scale for radiative transfer. Note that the origin of optical depth ($\tau_{\rm c} = 0$) coincides with the outermost boundary of geometrical distances ($s_{\rm lim}$) and is taken where the observer is located so that $\tau_{\rm c}$'s are actual depths in the atmosphere. In Eq.\ (\ref{eq:opticaldepth}), $\chi_{\rm cont}$ is the continuum absorption coefficient (the fraction of incoming electromagnetic energy withdrawn from the radiation field per unit of length through continuum formation processes). The propagation matrix deals with \emph{absorption} (withdrawal of the same amount of energy from all polarization states), \emph{pleochroism} (differential absorption for the various polarization states), and \emph{dispersion} (transfer among the various polarization states). The product of \matriz{K} and $\vector{S}$ accounts for emission. The RTE can then be considered \emph{as a conservation equation}: the energy and polarization state of light at a given point in the atmosphere can only vary because of emission, absorption, pleochroism, and dispersion. Equation\ (\ref{eq:rte}) is strictly valid only under the assumption that the energy and polarization state of light are independent of time. To be more specific, we have assumed that the rate of change of the Stokes parameter profiles is much slower than the radiative and collisional relaxation time scales involved in the problem. 

A formal solution to the general RTE was proposed for the first time by \citet{1985SoPh...97..239L}, according to whom, the observed Stokes profiles at the observer's optical depth ($\tau_{\rm c}=0$) read 
\begin{equation}
\label{eq:rteformalsolution}
\vector{I}(0) = \int_{0}^{\infty} \matriz{O} (0,\tau_{\rm c}) \matriz{K} (\tau_{\rm c}) \vector{S} (\tau_{\rm c}) \rm d \tau_{\rm c},
\end{equation}
where $\matriz{O}$ is the so-called evolution operator, and a semi-infinite atmosphere has been assumed as usual. The solution is called \emph{formal} because it is not a \emph{real} solution as long as the evolution operator (and the propagation matrix and the source function vector) are not known. Unfortunately, no easy analytical expression can in general be found for $\matriz{O}$. Only in some particular cases, such as that in Sect.\ \ref{sec:milne}, can a compact form for the evolution operator and an analytic solution of the RTE be obtained. In all other cases, numerical evaluations of $\matriz{O}$ and solutions of the transfer equation are necessary. The emergent Stokes spectrum is obtained through an integral of a product of three terms all over the whole atmosphere. Claiming that some of the Stokes parameters are proportional to one of the matrix elements of $\matriz{K}$ is, at the very least, adventurous. This proportionality can only take place in very special circumstances (e.g., Sections \ref{sec:weakfield} and \ref{sec:weakatmosphere}).

\subsection{The non-local thermodynamic equilibrium problem}
\label{sec:NLTE}

Being a vector differential equation, the RTE should indeed be considered as a set of four \emph{coupled} differential equations. These can only be solved independently in specific media, either isotropic or very simplified ones. But the situation is far more complicated since both \matriz{K} and \vector{S} depend on the material properties described by $\rho_{\alpha} (jm,j'm')$, as well as on external fields such as a macroscopic velocity or a magnetic field. For their part, the radiative and collisional transfer and relaxation rates do depend on the radiation field. Therefore, Eqs.\ (\ref{eq:equilradeqs}), (\ref{eq:equilcoleqs}), and (\ref{eq:rte}) describe a very involved, non-local, non-linear problem, known as the \emph{non-local thermodynamic equilibrium}  (NLTE) problem and must be consistently solved altogether. The numerical solution of all those coupled equations requires iterative procedures that are summarized in Figure \ref{fig:NLTE}.

\epubtkImage{}{%
\begin{figure}[htbp]
    \centerline{\includegraphics[width=\textwidth]{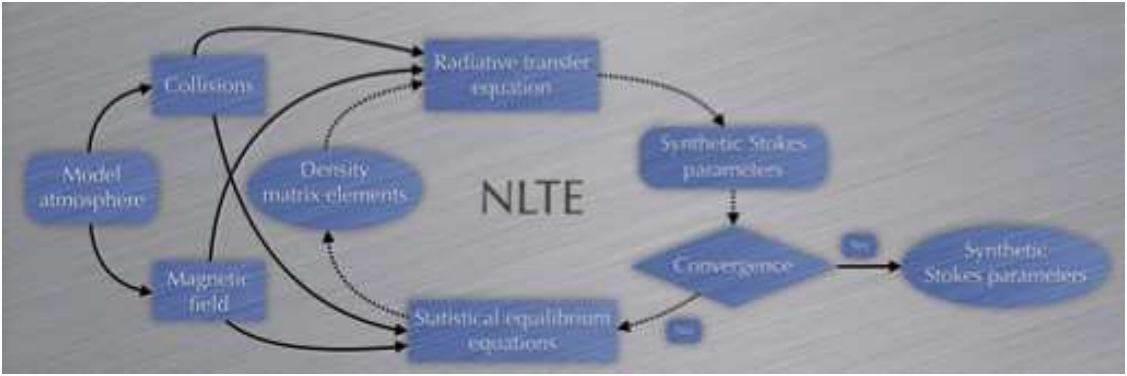}}
    \caption{Block diagram of the Stokes profile synthesis under NLTE conditions}
    \label{fig:NLTE}
\end{figure}}

By a model atmosphere we understand the set of thermodynamic variables (usually two, e.g., temperature and pressure, $T$ and $p$), dynamic (the macroscopic, bulk line-of-sight velocity field, $v_{\rm LOS}$), magnetic (the vector field \vector{B}, represented by $B$, the strength, $\gamma$, the inclination with respect to the LOS, and $\varphi$, the azimuth), and possibly some other, \emph{ad hoc} variables (such as the micro- and macroturbulence velocities, $\xi_{\rm mic}$ and $\xi_{mac}$, the filling factor, $f$ ---the area fraction of the resolution pixel that is filled with the unknown atmosphere--- and so forth). All these variables have to be specified as functions of the optical depth. Numerically, that model can be represented by a vector \vector{x} of $np+r$ components, $n$ being the number of depth grid points throughout the atmosphere, $p$ the number of physical quantities varying with depth, and $r$ the number of quantities that are assumed constant throughout the LOS. For example, one such model atmosphere would look like
\begin{equation}
\label{eq:modelatmos} \begin{array}{rcl}
\vector{x} & \equiv & [T(\tau_1), T(\tau_2), \ldots, T(\tau_n), p(\tau_1), p(\tau_2), \ldots, p(\tau_n), B(\tau_1), B(\tau_2), \ldots, B(\tau_n),\\ 
 & & \gamma(\tau_1), \gamma(\tau_2), \ldots, \gamma(\tau_n), \varphi(\tau_1),\varphi (\tau_2), \ldots, \varphi(\tau_n), \\ 
& & v_{\rm LOS}(\tau_1), v_{\rm LOS}(\tau_2), \ldots, v_{\rm LOS}(\tau_n), \xi_{\rm mic}, \xi_{\rm mac}, f]\trans, \end{array}
\end{equation}
where we have assumed specifically that both micro- and macroturbulence (as well as the filling factor) are constant with depth. This assumption is based on the fact that experience teaches that the increase in spatial resolution reached with new instruments makes less and less necessary the use of such \emph{ad hoc} parameters.

Once this model atmosphere is set, the necessary ingredients for the RTE and the statistical equilibrium equations can be calculated. The solution of the RTE has to be compared with that coming from it after modification driven by the new density matrix elements resulting from the solution of the statistical equations. If the differences are considered small compared with a given threshold, then a new synthetic set of Stokes parameters has been found. If not, the equilibrium equations have to be modified in order to iterate the procedure until convergence is reached. The direct problem of obtaining the Stokes spectrum of a given line coming out from a given model atmosphere then turns out to be very complex. It cannot always be computed with the necessary speed and accuracy. Approximations are, thus, in order. 

\subsection{The local thermodynamic equilibrium approximation}
\label{sec:LTE}

Imagine now that coherences among the Zeeman sublevels can be neglected, and that all of them are evenly populated. That is, assume that
\begin{equation}
\label{eq:ltecondition}
\rho(\alpha jm,\alpha' j'm') = \delta_{\alpha \alpha'} \delta_{jj'} \delta_{mm'} \rho_{\alpha j},
\end{equation}
where $\delta$ is Kronecker's delta. In such conditions, $n_{j} = (2j+1) \rho_{\alpha j}$. Assume also that $n_{j}$ and the population of other ionic species can be evaluated through the equations of thermodynamic equilibrium at the local temperature (the Boltzmann and Saha laws; e.g., \citeauthor{1992oasp.book.....G}, \citeyear{1992oasp.book.....G}). This assumption will be valid only in the case that the photon mean free path ($\ell = 1/\chi_{\rm cont}$)\footnote{For instance, at the bottom of the photosphere, $\ell \simeq 100$ km.} is small compared to the scale of variation of the physical quantities, i.e., when the atomic populations depend only upon the values of the local physical quantities. Besides, it can be shown that if Kirchoff's law is further assumed, \citep[e.g.,][]{2004ASSL..307.....L} the source function vector reduces to 
\begin{equation}
\label{eq:sourcefunction}
\vector{S} = (B_{\nu} (T), 0, 0, 0)\trans,
\end{equation}
where, $B_{\nu} (T)$ is the Planck function at the local temperature. These are the conditions of the so-called \emph{local thermodynamic equilibrium} approximation (LTE) and have automatically decoupled the RTE from the material equations. Then, if LTE can be supposed for a given spectral line, the synthesis of its Stokes profiles simplifies significantly because iterative procedures are no longer needed. This is graphically explained in Figure\ \ref{fig:LTE}.

\epubtkImage{}{%
\begin{figure}[htbp]
    \centerline{\includegraphics[width=\textwidth]{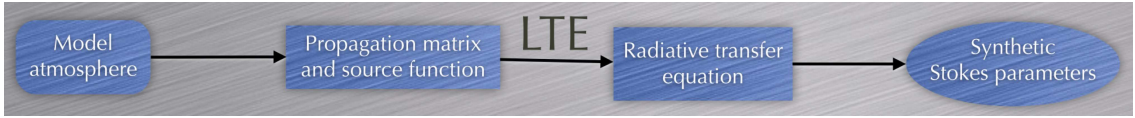}}
    \caption{Block diagram of the Stokes profile synthesis under LTE conditions.}
    \label{fig:LTE}
\end{figure}}

In some circumstances, it may be useful to relax the fulfillment of the Boltzman law and, instead, admit that $\rho_{\alpha j}$ deviate from the LTE values, $\hat{\rho}_{\alpha j}$, so that 
\begin{equation}
\label{eq:depcoeff}
\beta_j = \frac{\rho_{\alpha j}}{\hat{\rho}_{\alpha j}}
\end{equation}
are \emph{departure} coefficients that measure how far the conditions are from LTE. Thus, although radiative transfer remains with the LTE scheme sketched in Fig.\ \ref{fig:LTE}, the second block is affected by Eq.\ (\ref{eq:depcoeff}) and the $\beta$'s are needed to calculate the level populations. As we are going to see, this departure-coefficient approximation can be very useful for formulating NLTE inversion procedures (see Section\ \ref{sec:non-lte}).

\epubtkImage{}{%
\begin{figure}[htbp]
    \centerline{\includegraphics[width=0.75\textwidth]{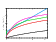}}
    \caption{Stokes $I$, LTE source function for various atmospheric models: the umbral model E by \citet[][black line]{1986ApJ...306..284M}, the penumbral model by \citet[][red line]{1994ApJ...436..400D}, the plage model by  \citet[][blue line]{solanki1986}, and the quiet-Sun models by \citet[][purple line]{gingerich+etal1971} and \citet[][green line]{vernazza+etal1981}.}
    \label{fig:source}
\end{figure}}

\subsection{The Milne--Eddington approximation}
\label{sec:milne}

An even more simplified approximation is obtained by further assuming that thermodynamics is sufficiently described with a source function that depends linearly on the continuum optical depth,
\begin{equation}
\label{eq:sourcelin}
\vector{S} = (S_{0} + S_{1}\tau_{\rm c}) \, \vector{e}_{0},
\end{equation}
where $\vector{e}_{0} \equiv (1,0,0,0)\trans$, and that the other physical quantities ($\vector{B}$, $v_{\rm LOS}$, etc.) in the model are constant throughout the atmosphere, hence defining a constant \matriz{K}. Figure \ref{fig:source} shows the LTE source function (the first component of the vector in Eq.\ \ref{eq:sourcefunction}) at 525 nm for several realistic model atmospheres, namely, the umbral model E by \citet[][black line]{1986ApJ...306..284M}, the penumbral model by \citet[][red line]{1994ApJ...436..400D}, the plage model by  \citet[][blue line]{solanki1986}, and the quiet-Sun models by \citet[][yellow line]{gingerich+etal1971} and \citet[][green line]{vernazza+etal1981}. The hypothesis of linearity does not seem very accurate for all the models. Nevertheless, in spite of its seemingly unrealistic nature, when we are dealing with a weak spectral line, the optical depth interval at which the line is sensitive to the atmospheric quantities is usually small enough to consider that a linear source function is not a bad approximation. There is wide experience in showing how useful the ME approximation is for inferring average values of the magnetic field vector and the LOS velocity, starting with the paper by 
\citet[][for a check with other approaches see \citeauthor{1998ApJ...494..453W}, \citeyear{1998ApJ...494..453W}]{skumanich+lites1987}. The key point is that the RTE has an analytic solution (Stokes $\vector{I}$ at $\tau_{\rm c} = 0$) under these assumptions \citep[e.g.,][]{2003isp..book.....D}:
\begin{equation}
\label{eq:milnesolution}
\vector{I} (0) = (S_0 + \matriz{K}^{-1} S_1) \, \vector{e}_0.
\end{equation}
The analytic character of the solution helps in grasping many of the relevant features in line formation; it cannot reproduce Stokes line asymmetries,\footnote{By Stokes line asymmetries or Stokes profile asymmetries we mean deviations from the even (Stokes $I$, $Q$, and $U$) or odd (Stokes $V$) functional shape about the central wavelength of the line. This is commented on in several places in this review, e.g., Secs.\ \ref{sec:misma}, \ref{sec:meatmosphere} and \ref{sec:varying}, and discussed in Section\ \ref{section:synthesis}.} though \citep{auer+heasley1978}. Using this useful feature, \citet{1985SoPh...97..239L} had the clever idea of tailoring the functional shape of the source function so that it might be used to synthesize chromospheric line profiles while preserving an analytic solution because of the constancy with depth of the propagation matrix. Atomic polarization is neglected in this modeling. The so-called ``field-free approximation" is assumed. The latter grants substitution of the scalar components of the source function for those corresponding to the same atom in the absence of a magnetic field \citep{1969SoPh...10..268R}. Later on, \citet{1988ApJ...330..493L} elaborated \citeauthor{1985SoPh...97..239L}'s  idea and proposed a new source function that was incorporated into their inversion code  to interpret the observed profiles of the Mg~{\sc i}~b lines at 517.27 and 518.36~nm. Specifically, they wrote the RTE in terms of the line center optical depth, $\tau_{0}$, which remains the same as in Eq.\ (\ref{eq:rte}) but substituting \matriz{K} by $\matriz{K}' \equiv r_0 \matriz{K}$, where $r_0$ is the continuum-to-line absorption coefficient ratio and with a new source function $\vector{S}'$ that follows from two distinct continuum and line source functions given by
\begin{equation}
\label{eq:chromosmilne} \begin{array}{c}
\vector{S}_{\rm cont} = \vector{S}, \\
{\displaystyle \vector{S}_{\rm lin} = \vector{S} - \sum_{i=1}^2 A_i {\rm e}^{-\varepsilon_i \tau_0}}, \end{array}
\end{equation}
where $\vector{S}$ is defined in Eq.\ (\ref{eq:sourcelin}). The exponential shape of the last two terms in $\vector{S}_{\rm lin}$ tries to mimic the consequences in the source function of the actual chromospheric rise of temperature. The $A$'s and $\varepsilon$'s are free parameters that can be tuned to fit the observed profiles. With this formulation, the analytic solution of the transfer equation (at $\tau_{0} = 0$) turns out to be
\begin{equation}
\label{eq:chrmossolution}
\vector{I}(0) = \left[ S_0 + \matriz{K}'^{-1} S_1 - \sum_{i=1}^{2} A_i (\matriz{K}' + \varepsilon_i \bbbone)(\matriz{K}' - r_0 \bbbone) \right] \vector{e}_0,
\end{equation}
where $\bbbone$ stands for the identity $4\times4$ matrix.\footnote{Note that this is not a non-LTE inversion technique but a phenomenological approach that can help in fitting the profiles of chromospheric lines that are indeed formed under conditions far from local thermodynamic equilibrium.}

Further exploiting the analytic character of the 
Milne--Eddington solution, slight modifications in the assumptions were also suggested by \cite{landolfi+landi1996} to deal with small velocity gradients and even with discontinuities along the LOS. In summary, we can say that approximations to the RTE predicated on keeping the \matriz{K} matrix constant or almost constant are useful and still a field for exploitation in observational work.

\subsection{The weak-field approximation}
\label{sec:weakfield}

A further simplification of radiative transfer is sometimes used. When the magnetic field can be assumed constant with depth and weak enough, the resulting Stokes $V$ profile of many lines turns out to be proportional to the longitudinal component of the field, regardless of the remaining physical quantities (see Section\ \ref{sec:weakatmosphere}). Under this assumption (and for not extremely weak fields since linear polarization is zero to first order approximation), the ratio between Stokes $U$ and $Q$ is proportional to the tangent of twice the field azimuth. The weakness of the field is guaranteed provided that \citep[e.g.,][]{2004ASSL..307.....L}
\begin{equation}
\label{eq:weakfield}
g_{\rm eff} \frac{\Delta \lambda_{\rm B}}{\Delta \lambda_{\rm D}} \ll 1,
\end{equation}
where $g_{\rm eff}$ is the effective Land\'e factor of the line, $\Delta \lambda_{\rm B}$ is the Zeeman splitting, and $\Delta \lambda_{\rm D}$ is the Doppler width of the line. The effective Land\'e factor is given by
\begin{equation}
\label{eq:lande}
g_{\rm eff} = \frac{1}{2} (g_u + g_l) + \frac{1}{4} (g_u-g_l) [j_u(j_u + 1) - j_l(j_l+1)],
\end {equation}
where $g_u$ and $g_l$ are the Land\'e factors of the upper and lower level of the transition, respectively. In $LS$ coupling, those factors are functions of the quantum numbers:
\begin{equation}
\label{eq:landelevel}
g = \frac{3}{2} + \frac{s(s+1)-l(l+1)}{2j(j+1)}.
\end{equation}
The Zeeman splitting is given by 
\begin{equation}
\label{eq:zeemansplitting}
\Delta \lambda_{\rm B} = \frac{\lambda_0^2 e_0 B}{4\pi mc^{2}},
\end{equation}
where $\lambda_0$ is the central, rest wavelength of the line, $e_0$ and $m$ are the charge and mass of the electron, $B$ is the magnetic field strength, and $c$ stands for the speed of light. For its part, the Doppler width is given by
\begin{equation}
\label{eq:dopplervel}
\Delta \lambda_{\rm D} = \frac{\lambda_0}{c} \sqrt{\frac{2kT}{m_a} + \xi_{\rm mic}^2},
\end{equation}
where $T$ is the temperature, $k$ is the Boltzmann constant, and $m_a$ is the mass of the atom.

From a formal point of view, Eq.\ (\ref{eq:weakfield}) is a good conditioning inequality. However, in practical terms, one should establish what is meant by \emph{much less than} 1. This is addressed in Section\ \ref{sec:weakatmosphere} but we can be sure that the wider the line, the more the weak-field approximation applies. Hence, broad chromospheric lines are good candidates for using it. One of the first attempts at measuring a magnetic field with a chromospheric line, known to the authors of this review, was carried out as early as  \citeyear{1990ApJ...361L..81M} by \citeauthor{1990ApJ...361L..81M} who (photographically) observed Stokes $I$ and $V$ profiles of the Ca {\sc ii} H line and interpreted them in terms of the weak-field approximation. This approach remains useful as interest in the chromosphere increases \citep[e.g.,][]{2013A&amp;A...556A.115D}.

\subsection{The MISMA hypothesis}
\label{sec:misma}

Driven by the ubiquitous appearance of Stokes profile asymmetries in observations, \citet{1994ssm..work...29L} suggested considering the atmospheric physical quantities, instead of deterministic stratifications, to have stochastic distributions about mean values \emph{with possible correlation effects} among them. Assuming that the source function nevertheless varies linearly with depth through the whole atmosphere and that the propagation matrix stays constant at the spatial scale of each of the realizations of such a common stochastic distribution, he found an analytic solution for the transfer equation. Certainly inspired by the \citeauthor{1994ssm..work...29L}'s proposal, \citet{1996ApJ...466..537S} put forward a new approach. Realizing that the wavelength symmetries in the propagation matrix elements do indeed avoid such Stokes profile asymmetries in the absence of LOS velocity gradients in the regular formulation of the transfer problem \citep{1992soti.book...71L}, they proposed that the solar atmosphere may be pervaded by MIcro-Structured Magnetic Atmospheres (MISMAs). The hypothesis implies a highly inhomogeneous atmosphere at scales much smaller than the photon mean free path whereby the integration of Eq.\ (\ref{eq:rte}) turns out to be very difficult. An alternative formulation is thus in order by locally averaging the propagation matrix and the emission vector. The resulting equation reads
\begin{equation}
\label{eq:rtemisma}
\frac{{\rm d} \vector{I}}{{\rm d} s} = - \left< \matriz{K}' \right> (\vector{I} -\vector{S}').
\end{equation}
It formally looks very much like the regular RTE but is formulated in terms of geometrical distances, $s$; $\matriz{K}' = \chi_{{\rm cont}} \matriz{K}$; 
\begin{equation}
\label{eq:sprimamisma}
\vector{S}' \equiv \left< \matriz{K}' \right>^{-1} \left< \matriz{K}' \vector{S} \right>;
\end{equation}
and the averages are taken over a distance $\Delta s$ that may vary along the optical path. The distance $\Delta s$ is supposed  to be still smaller than $\ell$ for Stokes $I$ to be assumed constant within its range. In addition, the averages are considered to vary smoothly along the line of sight. 

With all these assumptions, Eq.\ (\ref{eq:rtemisma}) is formally the same as Eq.\ (\ref{eq:rte}). All the mathematical tools developed to solve the latter can be used to find a solution to the former. This is so despite the (numerically) inconvenient formulation in terms of geometrical distances: it requires either non-equally-spaced grid points or an increase in computation time. The good news is that, since correlations may exist among the physical parameters of the microstructures, the symmetry properties of matrix $\left<\matriz{K}'\right>$ are automatically destroyed. Hence, asymmetric Stokes profiles can appear naturally.

\newpage


\section{Degrees of approximation in the model atmospheres}
\label{sec:approxmod}

Provided that physical atmospheric quantities are bounded functions of the optical depth, we can safely expect that they are either continuous or have some jump (Heaviside-like) discontinuities throughout the line formation region. Therefore, except for the discontinuity points, a Taylor expansion approximation seems simple and sensible. The good feature of Taylor expansions is that you can keep them at a given order of approximation that can be subsequently increased if needed. The sequential approach is of great help in following the principle of Occam's razor ---\emph{lex parsimoniae}--- which, in our opinion, should prevail in the interpretational work. The question arises as to whether an order of approximation is useful or whether it should be increased to give account of the observations. The answer must be found in the degree of accuracy with which we are trying to reproduce the observables. Hence, it has to do with the balance between the signal and the noise: if the next order of approximation only introduces variations that are below, say, three times  the rms noise, $\sigma$, then its use is discouraged. If, on the contrary, the difference between the observed and synthetic profiles is greater than $3\sigma$, its use may be advisable.\footnote{By adopting $\sigma$ as a measure of noise we are assuming that the noise statistics is Gaussian and this seems a common and sensible assumption as well. Requiring signals to be larger than $3\sigma$, therefore, implies more than 99.7~\% certainty in the detection. We refer the reader to \citet{2012ApJS..201...22D} for a discussion on polarimetric accuracy and signal-to-noise ratio. For Bayesian selection among model atmospheres, see \citet{2012ApJ...748...83A}.} Let us postpone the discussion to the following sections and present here the various atmospheres we are considering. We start with the zeroth order approximation and assume that physical quantities are constant with depth to continue with gradients, higher order variations, and jumps or discontinuities.

\epubtkImage{}{%
\begin{figure}[htbp]
    \centerline{\includegraphics[width=0.8\textwidth]{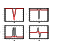}}
    \caption{Examples of ME Stokes profiles of the Fe~{\sc i} line at 617.3 nm as observed with an instrument whose Gaussian spectral PSF has a FWHM of 6 pm. Two model atmospheres are used that differ only in the magnetic field strength: $B=1200$~G for the black lines and 200~G for the red ones.}
    \label{fig:meprofile}
\end{figure}}

\subsection{Constant physical quantities}
\label{sec:constantquantities}

Let us distinguish among three possibilities, namely, the Milne--Eddington approximation, the weak-field approximation, and an atmosphere where $\vector{B}$ and $v_{{\rm LOS}}$ are constant but where thermodynamics is properly accounted for with a realistic stratification of temperature.\footnote{Indeed, two variables are needed for specifying the thermodynamical state of the medium. However, most of the spectral lines used in typical observations present a very low dependence, if any, on pressure. Therefore, we shall assume that pressure is stratified according to hydrostatic equilibrium throughout the paper.}

\subsubsection{The Milne--Eddington atmosphere}
\label{sec:meatmosphere}

As commented on in Sect.\ \ref{sec:milne}, a Milne--Eddington atmosphere provides an analytic solution to the RTE. With nine parameters, the Stokes profiles of a spectral line can be synthesized. The model parameters are the three components of the magnetic field, $B$, $\gamma$, and $\varphi$, the LOS velocity, $v_{\rm LOS}$, and the so-called thermodynamic parameters: the line-to-continuum absorption coefficient, $\eta_0$ ($=1/r_0$), the Doppler width of the line, $\Delta\lambda_{\rm D}$, the damping parameter, $a$, and the two coefficients for the source function, $S_0$ and $S_1$. The actual values of $\eta_0$, $\Delta\lambda_{\rm D}$, and $a$ may vary significantly throughout the atmosphere. Therefore, assigning one single value for each may be, say, risky. Experience, however, indicates that this is possible. Reasonable fits to actual data can be obtained with this approximation and we can even understand the relationship between the single-valued parameters and their actual stratification \citep{1998ApJ...494..453W}. Only Stokes profiles with definite symmetry properties can be formed in an ME atmosphere. Stokes $I$, $Q$, and $U$ are even functions of wavelength while Stokes $V$ is odd. This is a consequence of the absence of velocity gradients \citep{auer+heasley1978} and will be discussed later in Section\ \ref{section:synthesis}. Figure\ \ref{fig:meprofile} shows two examples of ME profiles corresponding to the Fe~{\sc i} line at 617.3 nm as observed with an instrument whose (Gaussian) spectral profile (point spread function, PSF) has a full width at half maximum (FWHM) of 6 pm. The thermodynamic model parameters are $\eta_0 = 5.06$, $\Delta\lambda_{\rm D} = 2.6$ pm, $a= 0.22$, $S_{0} = 0.1$, and $S_{1} = 0.9$; they come from a fit to the FTS spectrum \citep{1984sfat.book.....K, 1987ftp...book...B}. The magnetic inclination and azimuth are both equal to 30$^{\circ}$; $B=1200$~G for the black lines and 200~G for the red ones.

\subsubsection{The weak-field atmosphere}
\label{sec:weakatmosphere}

As stated in Sect.\ \ref{sec:weakfield}, when $B$ is constant with depth and very weak, then the Stokes $V$ profile turns out to be proportional to the longitudinal component of the magnetic field independently of the remaining quantities. It can be shown \citep[e.g.,][]{2004ASSL..307.....L} that
\begin{equation}
\label{eq:vpropmag}
V(\lambda) \simeq - g_{\rm eff} \, \Delta\lambda_{\rm B} \cos\gamma \, \frac{\partial I_{\rm nm}}{\partial \lambda},
\end{equation}
where $I_{\rm nm}$ is the non-magnetic Stokes $I$ profile, corresponding to the line in the absence of a magnetic field. Equation\ (\ref{eq:vpropmag}) has been key for many magnetic inferences. In fact, written as $V = C B_{\parallel}$, it is known as the magnetographic equation since it provides a calibration of the magnetographic signal. When magnetographs used only one or two wavelength samples of the circular polarization, the magnetographic equation was indeed the only means of obtaining estimates of the component of the magnetic field along the line of sight. Nowadays, with modern magnetographs providing more samples in all four Stokes parameters, that equation is still useful for morphological, qualitative estimates but cannot be trusted everywhere and under all circumstances. The modern way to evaluate $C$ indeed implies some radiative transfer calculations in given model atmospheres \citep[e.g.,][]{2011SoPh...268...57M}, and these calculations readily show that the approximation saturates at low magnetic field strengths. In the left panel of Fig.\ \ref{fig:maximav}, we plot the maximum of the Stokes $V$ profile as a function of the field strength (the field is along the LOS, $\gamma=0^{\circ}$) with an instrumental profile FWHM of 6 pm (asterisks) and of 8.8 pm (diamonds). In solid lines, the linear (red) and quadratic (blue) fits are also shown. Only strengths up to 600~G are plotted because the relationship is evidently nonlinear above that threshold. For weaker fields, it is apparent that the instrumental broadening of the profiles helps linearity to hold as differences between the linear and quadratic fits are smaller for the broader PSF. Those differences are for most of the points above $3 \cdot 10^{-3} I_{\rm c}$; that is, more than $3\sigma$, with $\sigma$ being the noise level of the polarization continuum signal of typical observations. Such differences are clearly detectable by current means. Hence, the approximation loses validity for yet weak fields. Deviations from linearity are even clearer if one sees the green lines in the figure, which correspond to linear fits including only data points for which $B$ is less than 200~G. In our example, the weak field approximation for the Stokes $V$ peaks breaks down at fields stronger than 300~G with a FWHM of 6 pm and stronger than 400~G with a FWHM of 8.8 pm. Certainly, if the instrument has a narrower spectral PSF or if the noise is smaller, the approximation fails earlier. The approximation clearly worked better for older instruments.  

\epubtkImage{}{%
\begin{figure}[htbp]
    \centerline{\includegraphics[width=0.5\textwidth]{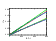}\includegraphics[width=0.5\textwidth]{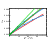}}
    \caption{Maximum of the Stokes $V$ profile as a function of the magnetic field strength for a longitudinal field (left panel). Maximum of the Stokes $Q$ profile as a function of the square magnetic field strength (right panel). Asterisks correspond to an instrumental profile FWHM of 6 pm and diamonds to a FWHM of 8.8 pm. Red lines represent linear fits to the points; blue lines display quadratic fits; green lines correspond to fits for fields weaker than 200~G.}
    \label{fig:maximav}
\end{figure}}

Further arguments can be supplied for the user to be cautious about weak field assumptions with typical, visible photospheric lines. The first one is that Eq.\ (\ref{eq:vpropmag}) is hardly applicable, as shown in Fig.\ \ref{fig:weakfieldapprox}, not only because Stokes $V$ does not follow it but because Stokes $I$ deviates from $I_{{\rm nm}}$ even sooner \citep[and, up to first order, $I=I_{{\rm nm}}$ must hold for Eq.\ (\ref{eq:vpropmag}) to be valid; e.g.,][]{2004ASSL..307.....L}. In the left column of the figure, the differences between the left-hand and the right-hand members of the equation are plotted. Colors correspond to 600~G (black), 500~G (red), 400~G (blue), 300~G (green), 200~G (purple), and 100~G (dark green). The dashed, horizontal purple lines mark the $3\sigma$ level. The upper rows are for a FWHM of 6 pm and the bottom row is for a FWHM of 8.8 pm. The plots in the left column are of course consistent with the results from Figure\ \ref{fig:maximav}. Those in the right column are illustrative of how Stokes $I$ varies with the magnetic field strength. Differences between the various profiles can easily be discerned above the $3\sigma$ level. When the profiles themselves are affected by noise, unlike in these plots, detecting the differences may be more difficult but the message is clear: {\bf contrary to the common belief}, the Stokes $V$ profile is not the only tool for estimating the longitudinal component of weak magnetic fields; Stokes $I$ helps a lot and should not be forgotten. 

The second argument concerns the diagnostic capability for typical lines to disentangle $B$ from $\gamma$ in the weak-field regime. Most statements about the only accurate retrieval to be the longitudinal magnetic field component are based on Eq.\ (\ref{eq:vpropmag}), as if it were the only available tool from radiative transfer. Stokes profiles other than $V$ are often obliterated. It is easy to understand \citep[e.g.,][]{2004ASSL..307.....L}, however, that the mere deviations between $I$ and $I_{\rm nm}$ we have seen in Fig.\ \ref{fig:weakfieldapprox} should imply the appearance of linear polarization signals (provided that the inclination is different from zero): such Stokes $I$ deviations from $I_{\rm nm}$ are second order terms in an expansion of all four Stokes profiles.\footnote{The expansion is in terms of powers of a dimensionless parameter that scales the vector magnetic field, and that is only valid when $\Delta\lambda_{\rm B} \rightarrow 0$.} At second order, Stokes $Q$ and $U$ are no longer zero (or below the noise) either and start to provide additional information. It can also be proven \citep[e.g.,][]{2004ASSL..307.....L} that $Q \propto B^2 \sin^2 \gamma$, as shown in the right panel of Fig.\ \ref{fig:maximav}, where the maximum of Stokes $Q$ is plotted against $B^2$ for a field that is inclined $45^{\circ}$ with respect to the vertical.\footnote{Stokes $Q$ is assumed to be defined here in the reference frame where Stokes $U$ is zero (constant magnetic azimuth).} Here, deviations between linear and quadratic fits are smaller than for the $V$ case (note that the $Y$ scale is an order of magnitude smaller) but the interesting point is that, above $B=200$ G, linear polarization signals begin to be larger than $3\sigma$ and, hence, detectable.

\epubtkImage{}{%
\begin{figure}[htbp]
    \centerline{\includegraphics[width=0.9\textwidth]{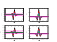}}
    \caption{Differences between the Stokes $V$ profile and its weak-field approximation (left column) and differences between the Stokes $I$ profile and that for a zero field strength. Colors indicate values of the longitudinal component of the field. The dashed horizontal lines mark the $3\sigma$ level of typical, modern observations. The upper row is for a FWHM of 6 pm and the bottom one for a FWHM of 8.8 pm. Colors correspond to 600~G (black), 500~G (red), 400~G (blue), 300~G (green), 200~G (purple), and 100~G (dark green).}
    \label{fig:weakfieldapprox}
\end{figure}}

A third argument we want to bring to the reader's attention is related to the common belief that weak fields are hardly distinguished from strong fields (say above 1 kG) with a filling factor significantly smaller than 1. We will return to this issue in Secs.\ \ref{sec:analyticrfs} and \ref{sec:weakretreival}, as the problem has already been discussed in the literature \citep[e.g.,][]{2010ApJ...711..312D}. Let us only mention here that the loss of linearity of Stokes $V$ above, say, 400~G and, most importantly, the behavior of Stokes $I$ are reasons enough for the two types of atmospheres to be distinguished by observational means.

If Eq.\ (\ref{eq:vpropmag}) were universally accepted, then it would indicate that the RTE is almost useless since the emergent profile is proportional to one of the matrix elements of $\matriz{K}$. Elementary mathematics readily explain that this is not possible except for, perhaps, a small value range of fields. In summary, we must acknowledge that {\bf Stokes $V$ is not proportional to the longitudinal component of the magnetic field}.

\subsubsection{Constant vector magnetic field and LOS velocity}
\label{sec:constantbv}

There is still a third option to deal with constant $\vector{B}$ and $v_{\rm LOS}$. Imagine that the atmosphere is a regular one as far as thermodynamics is concerned but where the magnetic and dynamic quantities do not vary with depth. Since the propagation matrix is no longer constant, no analytic solution of the RTE is available.\footnote{The clause is not very rigorous but is true in practical terms. Indeed, one can conceive other \matriz{K} stratifications that still allow an analytic solution of the RTE \citep{1985SoPh...97..239L}.} One is then led to use numerical techniques to synthesize the spectrum. The atmosphere, however, is greatly simplified since the number of parameters is reduced. This can be very helpful for quicker analyses of the data or as a makeshift for more elaborate subsequent approaches that include variations of $\vector{B}$ and $v_{\rm LOS}$ with the optical depth. This is the approximation used for the first version of the SPINOR code \citep{1992A&amp;A...263..339S} or as an option in the SIR code \citep{1992ApJ...398..375R}.

\subsection{Physical quantities varying with depth}
\label{sec:varying}

The community has gathered a great deal of evidence about variations of $\vector{B}$ and $v_{\rm LOS}$ along the optical path everywhere over the solar disk. In addition, physical laws such as those of magnetic flux and mass conservations demand that these quantities vary with optical depth in a number of structures. The approximations in the former subsections cannot then be considered but as first-step approaches or simplified descriptions of reality. In any case, we can safely assume that stratifications of the physical quantities  are bounded functions of $\tau_{\rm c}$ (or whichever variable parameterizing the optical path), as we admitted in the beginning of this Section.

A historical landmark for the full acknowledgement of LOS velocity gradients from an observational point of view was established by the discovery by \citet{1974A&amp;A....31..179M} and \citet{1974A&amp;A....35..327I,1974A&amp;A....37...97I,1975A&amp;A....41..183I} of a broadband circular polarization in sunspots. The true explanation was already suggested in the last of those papers, although schematically founded on the assumption of two slabs with \emph{different velocity} and magnetic field strengths. The broadband observations were soon related to spectral line net circular polarization (the integral of the Stokes $V$ profile over the wavelength span of the line):  \citet{1975SoPh...42...21G} computed all four Stokes profiles in the presence of an LOS velocity gradient and certainly obtained  asymmetric profiles; later on, \citet{auer+heasley1978} demonstrated that a necessary and sufficient condition for such a net circular polarization had to be found in velocity gradients along the line of sight, although they were neglecting magneto--optical effects. Rigorous derivations (including dispersion effects) have later been obtained and can be found, for example, in the elegant work by \citet{1981NCimB..62....1L}. The symmetry properties of the propagation matrix elements predict no net circular polarization (or Stokes $V$ area asymmetry) in the absence of an LOS velocity gradient. Other mechanisms such as  insufficient spatial resolution that implies mixtures of individual atmospheres within a pixel, may produce asymmetries in the peaks (the so-called amplitude asymmetries) but the integral of $V$ will remain zero. Therefore, any net circular polarization is unambiguous observational evidence for the presence of velocity gradients. And Stokes $V$ area asymmetries are observed practically everywhere. Unfortunately, no such unambiguous evidence exists for the presence of magnetic field gradients, although we know on physical grounds there are plenty of them, such as those through magnetic canopies where a magnetic layer is overlaying a non-magnetic one. 

\subsubsection{Parameterizing the stratifications}
\label{sec:parameterizing}

Among the numerical codes relevant to this review (see Sect.\ \ref{section:techniques}) there are some that acknowledge variations of $\vector{B}$ and $v_{\rm LOS}$. We deal here with what might be called ``normal'' or ``regular'' stratifications, such as those employed by \citet{1992ApJ...398..375R}, \citet{1998A&amp;A...336L..65F}, \citet{2000ApJ...530..977S}, and \citet{2001ASPC..236..487S}, and leave some others, devoted to specific solar features, to the following paragraphs.

Since the number of depth grid points used for the numerical integration of the RTE can be high, it may be advisable to reduce the degrees of freedom of the variations with depth of the physical quantities. As commented on above, a reasonable approach would be to follow higher order polynomials in a stepwise form. From constant values to linear, parabolic, third-order polynomial dependences, and so on. Then, if we assume, for instance, that $v_{\rm LOS}$ is linear with $\tau_{\rm c}$, we only need to specify the velocity at two grid points (\emph{nodes} in SIR's terminology) and three if it is parabolic, hence reducing the number of free parameters of the model. We do not need to specify $T$, $\vector{B}$, and $v_{\rm LOS}$ at every single point we use for solving the RTE but only at a few of them. We shall see in Sect.\ \ref{section:techniques} that one can go even further with this kind of approach and consider more involved optical depth dependences if necessary.

\subsubsection{The MISMA atmosphere}
\label{sec:mismaatmos}

As we explained in Sect.\ \ref{sec:misma}, the MISMA hypothesis guarantees the appearance of Stokes profile asymmetries but at the expense of introducing a significant number of extra free parameters. In fact, even in the simplest MISMA atmosphere \citep{1996SoPh..164..203S}, where all the micro-structures are described by ME atmospheres, one has in principle as many as ten free parameters per  needed component (also known as micro-structure). To the nine regular ME parameters, the volume occupation fraction for each micro-structure must be added. In more complicated MISMAs, the number of parameters is even higher \citep{1997ApJ...491..993S}. Moreover, in spite of the very detailed physical description where equilibrium equations are required for slender flux tubes, the inclination and azimuth of the magnetic field are kept constant throughout the whole atmosphere, which does not seem very realistic. (Canopies are found almost everywhere owing to the fanning out of magnetic field lines with height.) Last, but not least, {\bf when the structuring of the atmosphere is established at sizes comparable to $\ell$, the average propagation matrix does not result in an RTE as in Eq.\ (\ref{eq:rtemisma}), which is no longer valid}. Modern observations with continuously increasing spatial resolution do indeed show this kind of structuring both in quiet and active regions and sunspots. For example, single magnetic flux tubes of approximately  150 km size have been fully resolved by \citet{2010ApJ...723L.164L}; their evolution followed for half an hour by \citet{2014ApJ...789....6R}; and the internal structure of network magnetic structures revealed \citep{2012ApJ...758L..40M} with {\sc Sunrise}/IMaX observations (\citeauthor{2011SoPh...268...57M}, \citeyear{2011SoPh...268...57M}; \citeauthor{2011SoPh..268....1B}, \citeyear{2011SoPh..268....1B}). In our opinion, the MISMA hypothesis, being a clever idea for producing asymmetries, is advisable as a ``when-all-else-fails" atmosphere but  there are yet conventional radiative transfer treatments that provide reasonable interpretation of the observations.

\subsubsection{Other special atmospheres}
\label{sec:interlaced}

This subsection is devoted to three special cases where the physical scenario envisaged to explain the observations requires a specific configuration that is not intended to be universally valid. Those specific configurations, however, help in interpreting the Stokes profiles emerging from given solar features.

\paragraph{Interlaced atmospheres}

Imagine that you can assume that your line of sight is piercing a number $n$ of alternate boundaries $\left\{ s_{i} \right\}_{i=1,\ldots,n} (s_{1}<s_{2}<\ldots<s_{n})$ between two distinct atmospheres, as when observing from a side two identical thin flux tubes that are close but not stuck to each other. In such a scenario, the structuring of the atmosphere is comparable in size with $\ell$ and, therefore, the MISMA hypothesis does not hold. If you happen to know the solution, $\vector{I}_{\pm 1}$, of the RTE in each of the two atmospheres, labeled $\pm 1$, \citet{1995A&A...294..855D} found out that the formal solution to the problem is 
\begin{equation}
\label{eq:interlaced}
\vector{I} (s) = \vector{I}_{+1} (s) + \sum_{i=1}^{n} (-1)^{n-i} \left[ \prod_{j=i}^{n} \matriz{O}_{(-1)^{n-j}} (s_{j+1}, s_{j}) \right] \Delta \vector{I} (s_{i}), 
\end{equation}
for any $s \in [s_{n}, s_{\rm lim}]$, where the $+1$ atmosphere is assumed to be the outermost one, $\Delta \vector{I} (s_{i}) \equiv \vector{I}_{-1} (s_{i}) - \vector{I}_{+1} (s_{i})$, and $\matriz{O}_{\pm 1}$ are the evolution operators for both atmospheres \citep[e.g.][]{1985SoPh...97..239L}. Equation (\ref{eq:interlaced}) is at the root of the flux-tube inversion code by \citet[][see \citeauthor{1996A&A...306..960B}, \citeyear{1996A&A...306..960B} as well]{1997ApJ...478L..45B}. A different treatment of discontinuities along the line of sight was proposed by \citet{2003ASPC..286..235B} where the density of depth grid points is increased in the discontinuity neighborhood.

\paragraph{Atmospheres with Gaussian profiles}

The existence of net circular polarization in the penumbrae of sunspots was also the driver for \citet{2003ASPC..307..301B} to propose an implementation of the uncombed model by  \citet{solanki+montavon1993}. The scenario is based on two components; namely, a magnetic component and a penumbral magnetic flux tube, the latter occupying a fractional area of the resolution element. The model parameters of the penumbral tube are built by Gaussian modifications (in depth) of those in the background. All the Gaussians have the same width and are located at the same depth, but their amplitudes depend on the specific model parameter, of course. With these premises, the SIR code was modified into the so-called SIRGAUS code, which has been used, among others by \citet{2007PASJ...59S.601J}, \citet{2008A&A...481L..17J}, \citet{2010ApJ...713.1310I}, and \citet{2014A&amp;A...566A.139Q}.

\paragraph{Atmospheres with jump discontinuities}

Discontinuities can be treated numerically by decreasing the depth grid step or by using Eq.\ (\ref{eq:interlaced}). A specific implementation of such discontinuities was first used by \citet{2009ApJ...704L..29L} for an analysis of sunspot light bridges. Like SIRGAUS, it is based on a modification of the SIR code to take this particular scenario into account. In it, two magnetic atmospheres coexist in the resolution pixel: a background atmosphere whereby $\vector{B}$ and $v_{\rm LOS}$ are constant with depth, and another magnetic atmosphere where those quantities have a Heaviside-like discontinuity. This code (called SIRJUMP) has also been used in practice, e.g., by \citet{2012ApJ...758L..40M} and \citet{2012ApJ...748...38S}.

\newpage


\section{Degrees of approximation in the Stokes profiles}
\label{sec:approxprof}

Since the ultimate goal of inversions is the \emph{bona fide} reproduction of observed profiles, an analysis of the properties of Stokes spectra as functions of the wavelength is in order. Such an analysis should be aimed at finding the most conspicuous characteristics of the profiles in order for these characteristics to be the best reproduced among all the features. In other words, if, for instance, a given Stokes $I (\lambda)$ profile shows only small deviations from a Gaussian, we should aim to obtain the Gaussian that best simulates the profile and identify the model parameters responsible for this bulk behavior. In some cases we may be satisfied just with this ``coarse'', or not very detailed, description and leave small deviations or nuances to further, in-depth analysis that might even be carried out separately. As we are going to see, this approximation of incremental complexity for the profiles is well in line with the successive approximations we have described for the model atmospheres in Section \ref{sec:approxmod}.

The Stokes $Q$, $U$, and $V$ profiles and Stokes $I$ in line depression; that is,
\begin{equation}
\label{eq:intensityld}
I_{\rm d} \equiv 1 - \frac{I}{I_{\rm c}},
\end{equation}
as functions of $x \equiv \lambda-\lambda_{0}$ (where $\lambda_{0}$ is the central wavelength of the line), can be decomposed as sums of even and odd functions of $x$, as any other function defined over $\bbbr$.\footnote{Strictly speaking, we should take a mean LOS velocity wavelength shift into account as well, if it is large enough.} Specifically, if we call $S (x)$ any one of the profiles, then
\begin{equation}
\label{eq:s+s-}
S (x) = S_{+} (x) + S_{-} (x),
\end{equation}
where 
\begin{equation}
\label{eq:splussminus}
S_{+} (x) \equiv \frac{S(x) + S(-x)}{2} \,\,\, {\rm and} \,\,\, S_{-} (x) \equiv \frac{S(x) - S(-x)}{2}.
\end{equation}
By construction, $S_{+}$ is even and $S_{-}$ is odd.\footnote{The property is also valid for Stokes $I$. We have, however, chosen $I_{\rm d}$ for reasons that will become clear a little later in this Section.}

\epubtkImage{}{%
\begin{figure}[htbp]
    \centerline{\includegraphics[width=0.6\textwidth]{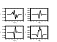}}
    \caption{Differences between the Stokes profiles of the Fe~{\sc i} line at 617.3 nm as synthesized in two model atmospheres that differ in the LOS velocity. See text for details.}
    \label{fig:approxprofiles}
\end{figure}}

This parity property is very interesting because, as we have seen in former Sections, the Stokes profiles of any line formed in the absence of velocity gradients have definite symmetry (parity) properties. Since asymmetries in regular profiles are relatively small, that is, the profiles usually display a predominant parity character (even for Stokes $I$, $Q$, and $U$, and odd for Stokes $V$), a sum of even and odd profiles may give account of the observed spectra as if the opposite parity component was indeed a \emph{perturbation} related to velocity gradients. This can explain the success of ME inversion codes for fitting many observations \citep[cf. \citeauthor{1998ApJ...494..453W}, \citeyear{1998ApJ...494..453W};][]{2010A&A...518A...2O}. The ME atmosphere accounts for  the main bulk of the observed Stokes profiles. In Fig. \ref{fig:approxprofiles} we plot the differences among the Stokes profiles of the Fe~{\sc i} line at 617.3 nm as synthesized in two model atmospheres. Both have the HSRA \citep{gingerich+etal1971} stratification of temperature with $B = 1500$ G and $\gamma=\varphi = 30$\degree. One of the models has a constant $v_{\rm LOS} = 1.87$ \kms and the other a small gradient from $v_{\rm LOS} = 2$ \kms at the bottom of the atmosphere through 1.75 \kms at the top. Both have $\xi_{\rm mac} = 1$ \kms and have been convolved with the IMaX PSF. Note that these differential profiles display almost the opposite parity character to their corresponding Stokes profiles.
\epubtkImage{}{%
\begin{figure}[htbp]
    \centerline{\includegraphics[width=\textwidth]{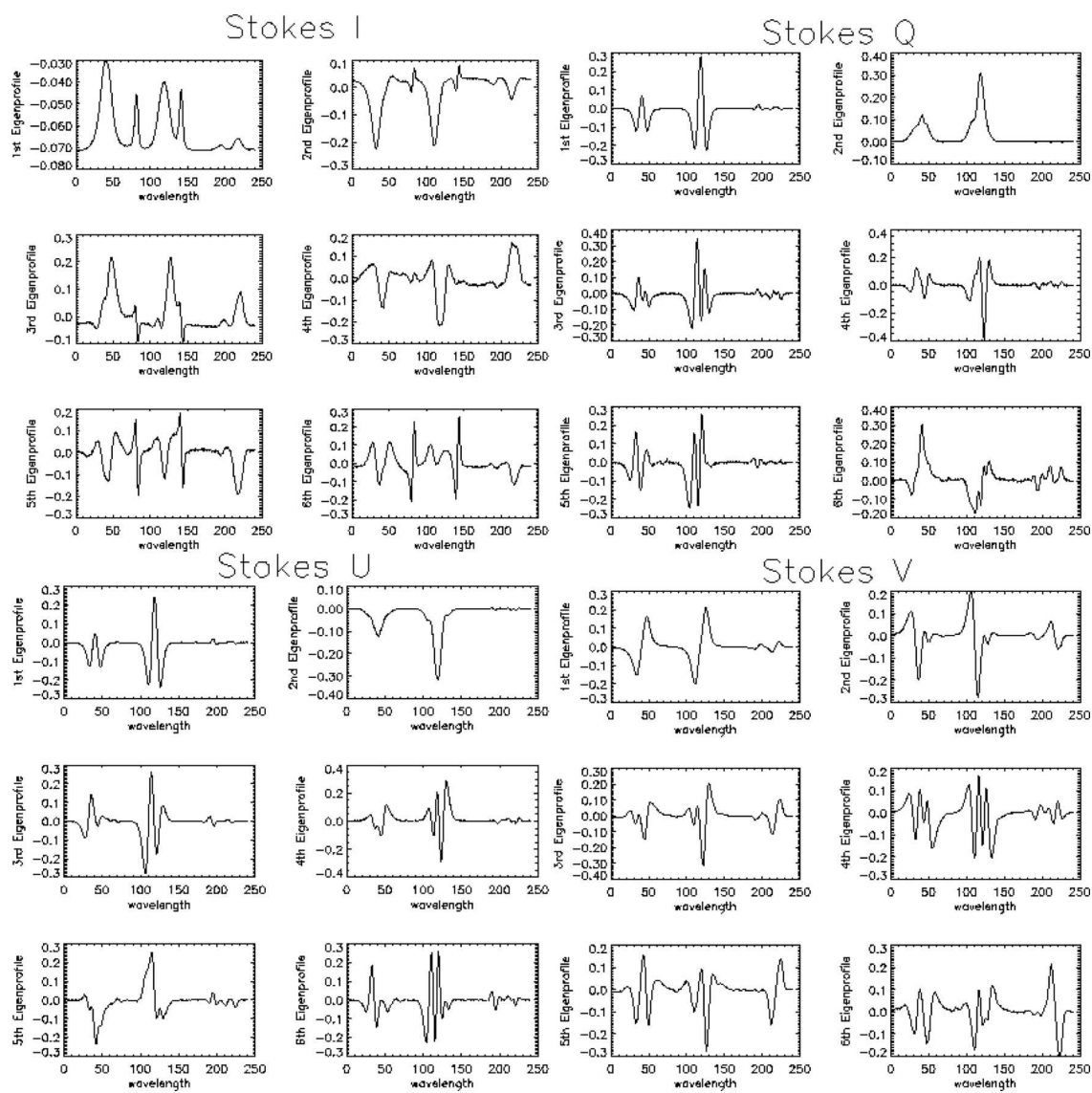}}
    \caption{First six eigenprofiles for Stokes $I$, $Q$, $U$, and $V$. They are obtained from observations of the Fe~{\sc i} line pair at 630.1 and 603.2 nm. Adapted from \citet{rees+etal2000}.}
    \label{fig:eigenprofiles}
\end{figure}}
A description with only $S_{+}$ or $S_{-}$ can thus provide a first approach analysis to a large number of observations. Indeed, $S_{+}$ should be good for $I$, $Q$, and $U$, and $S_{-}$ for $V$ as the differences are smaller than our ``nominal'' noise of $10^{-3} \, I_{\rm c}$. This, of course, cannot always be the case. Very peculiar Stokes profiles are often observed as our polarization accuracy increases. For instance, \citet{sigwarth+etal1999} first reported the observation of one-lobed $V$ profiles that were later studied in detail by \citet{2000A&amp;A...357..351G} and \citet{sigwarth2001}. Most of these profiles are found in the internetwork \citep[e.g.,][]{2012ApJ...748...38S}. 

A different description of the Stokes profiles as functions of wavelength was proposed by \citet{rees+etal2000} who suggested that they can be described as sums of given \emph{principal components} or \emph{eigenprofiles}. If those eigenprofiles are contained in a database and are properly selected, they can increasingly give account of the profile shapes just by increasing the number of principal components in the expansion. An example of such eigenprofiles is given in Fig.\ \ref{fig:eigenprofiles}. By adding these components properly weighted, the corresponding Stokes profiles are synthesized. This is the basis for all the PCA inversion techniques presented so far and the concept is fairly simple.

A similar approach to that of PCA was proposed by \citet{2003A&A...412..875D}, based on the fact that Stokes $I_{\rm d}$, $Q$, $U$, and $V$ belong to $\bbbl^2$, the space of square integrable functions over $\bbbr$. Since $\bbbl^2$ is a Hilbert space with a well defined scalar product, an exact, infinite expansion of the profiles is possible in terms of any of the several bases of the space. Among those basis systems, \citeauthor{2003A&A...412..875D} selected the family of Hermite functions, $h_n (x)$, because of the similarity between the shapes of the first few elements of the family and the Stokes profiles (see Figure\ \ref{fig:hermite}). Somehow, the Hermite functions (see the aforementioned paper for a definition) provide a suitable basis for approximating the observed profiles with finite expansions of a few terms. Apart from their possible use in inversion codes that has not been investigated so far, the expansion of Stokes profiles in terms of Hermite functions has been used by \citet{2012SoPh..276..415T} for compressing observed Stokes profiles and by \cite{2012ApJS..203....7H} for an automatic solar active region detection. The first authors, after expansion of the Stokes profiles, only keep the coefficients compressed with a conventional algorithm. This way they reduce the storage space by a factor 20 while keeping most of the information virtually noise free. The latter author discriminates the different active regions after looking at the complexity of the emerging Stokes profiles as described by their Hermite-function expansion coefficients.

\epubtkImage{}{%
\begin{figure}[htbp]
    \centerline{\includegraphics[width=0.85\textwidth]{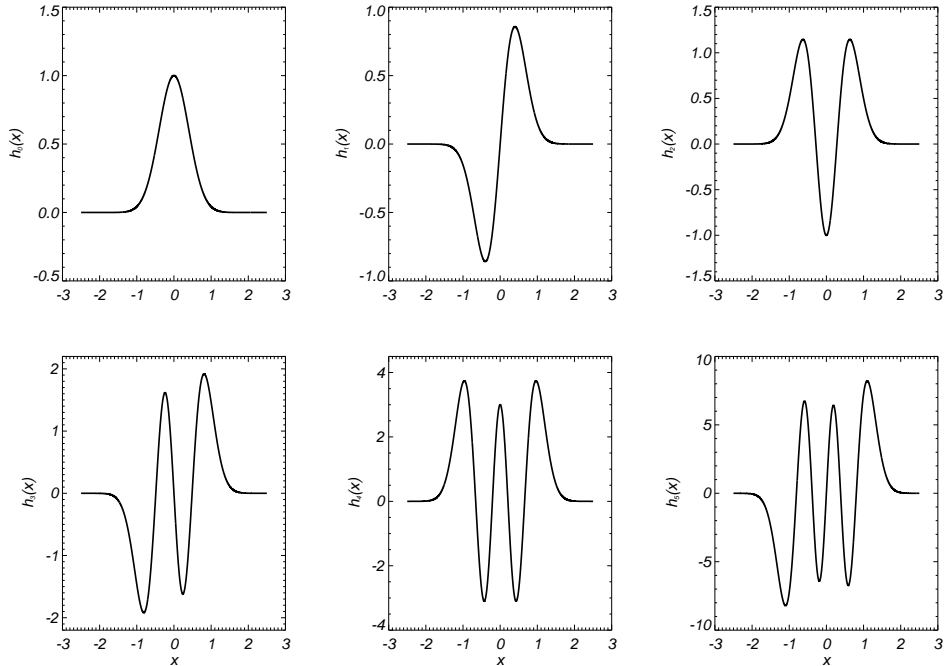}}
    \caption{First six Hermite functions. The abscissa has to be understood as a normalized wavelength (by the Doppler width of the line, for instance).}
    \label{fig:hermite}
\end{figure}}

\newpage


\section{A synthesis approach}
\label{section:synthesis}

As described in the introduction, an approximate knowledge as to how the Stokes profiles react to the various model parameters is advisable as it helps to select the adequate observables as ``orthogonal'' as possible. The ideal way to explore the diagnostic capabilities of Stokes profiles is by means of response functions (see Sect.\ \ref{section:response}). Tackling the problem head-on, that is, synthesizing the profiles in different model atmospheres, may help grasp basic ideas on the Stokes profile behavior, though. The idea is to study how the Stokes profiles vary when the model parameters are modified. This is the aim of this section.

\subsection{Constant atmospheres}
\label{sec:constantatmos}

As we have been doing in the two previous sections, let us start with the easiest case of atmospheres that do not vary with optical depth and, specifically, with ME atmospheres, since their analytic solution of the RTE enables a quick numerical overview of the space of model parameters. Figure~\ref{fig:iqvgrowthgamma} shows Stokes $I$, $Q$, and $V$ for the Fe~{\sc i} line at 617.3 nm with the same thermodynamic parameters used in Figs.\ \ref{fig:meprofile} and \ref{fig:weakfieldapprox}. Since the linear polarization $L^{2} \equiv Q^{2} + U^{2}$ is rotationally invariant and $\varphi$ is constant throughout the ME atmosphere, we are assuming to have selected the preferred reference frame where $U$ is identically zero, so that $L=Q$. In practical terms we have selected $\varphi = 0^{\circ}$ for all the profiles. The magnetic field strength is $B=300$, 500, 900, and 1200~G for the four rows from top to bottom. The magnetic inclination is encoded in color: $\gamma=0$ (dark green), 15 (purple), 30 (pink), 45 (green), 60 (blue), 75 (red), and $90^{\circ}$ (black). If we again assume a typical noise $\sigma=10^{-3} I_{\rm c}$, then the small differences in the core of Stokes $I$ for $B=300$ may not be detected and a neat distinction between $\gamma = 0$ and $15^{\circ}$ or $\gamma = 90$ and $75^{\circ}$ may hardly be reachable with Stokes $Q$ and $V$ at a $3\sigma$ level. Most certainly, however, there should not be any problem to distinguish between 0, 30, 60, and $90^{\circ}$. Of course, the dependences of $I$ and $V$ on $\gamma$ are significant enough when the field is stronger. One should not be restricted to longitudinal components even for this small field strength, and the situation improves further for lower noise levels. This is a well-known issue: in polarimetric observations we are photon starved, indeed as much as night-time astronomers may be for detecting very faint objects. 

\epubtkImage{}{%
\begin{figure}[htbp]
    \centerline{\includegraphics[width=\textwidth]{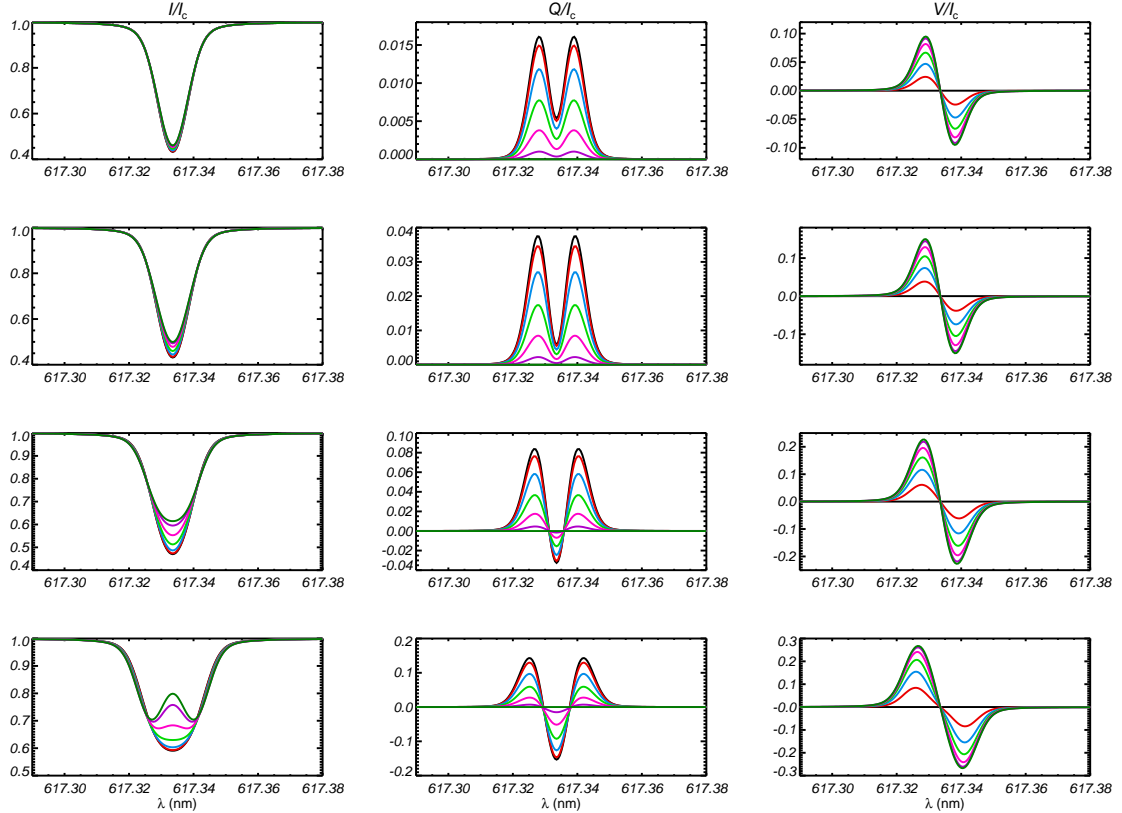}}
    \caption{Stokes $I$, $Q$, and $V$ as functions of the magnetic field inclination: dark green is for $\gamma=0$, purple for 15, pink for 30, green for 45, blue for 60, red for 75, and black for $90^{\circ}$. The magnetic field strength is different for the various rows: from top to bottom, $B=300$, 500, 900, and 1200~G. The magnetic azimuth is identically zero.}
    \label{fig:iqvgrowthgamma}
\end{figure}}

\epubtkImage{}{%
\begin{figure}[htbp]
    \centerline{\includegraphics[width=\textwidth]{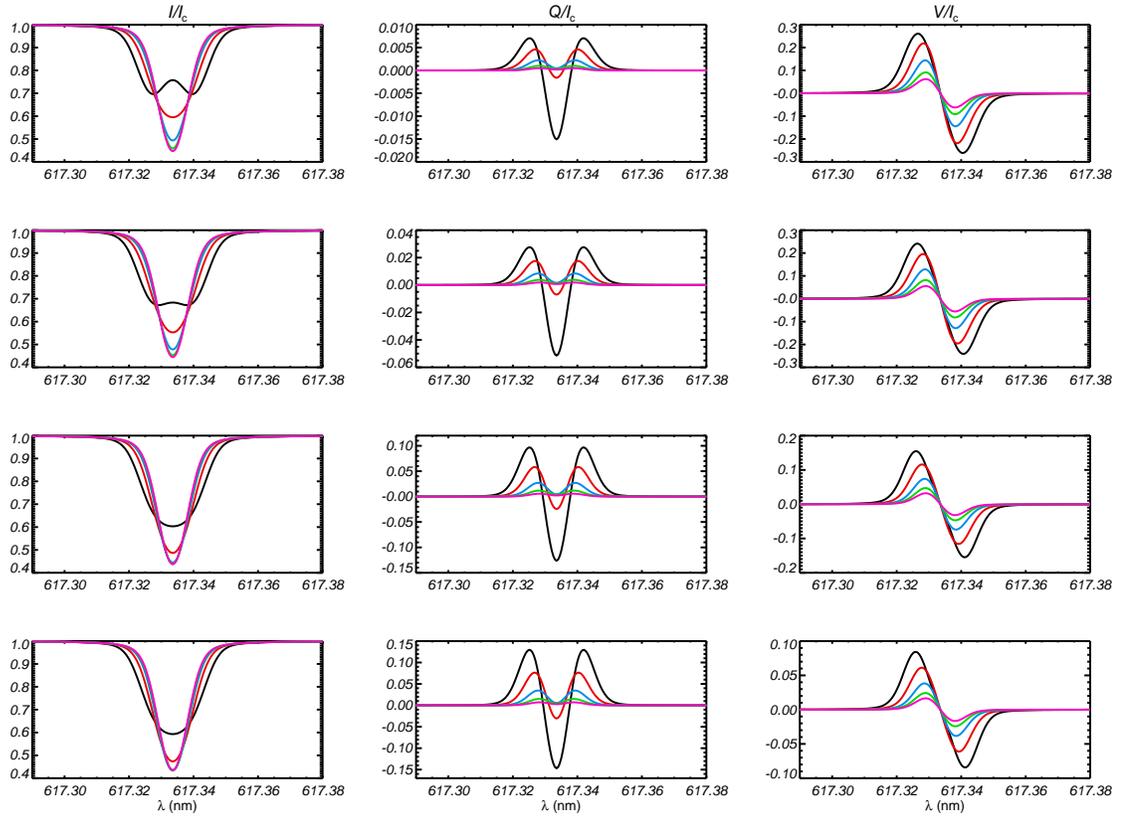}}
    \caption{Stokes $I$, $Q$, and $V$ as functions of the magnetic field strength: black is for $B=1200$, red for 900, blue for 500, green for 300, and purple for 200G. The magnetic field inclination is different for the various rows: from top to bottom, $\gamma=15$, 30, 60, and $75^{\circ}$. The magnetic azimuth is identically zero.}
    \label{fig:iqvgrowthstregth}
\end{figure}}

The alternative way to gauge the sensitivity of the Stokes profiles to $B$ and $\gamma$ by synthesizing the profiles is shown in Figure~\ref{fig:iqvgrowthstregth}. The rows correspond to different inclinations: $\gamma=15$, 30, 60, and $75^{\circ}$ from top to bottom. The magnetic field strength is this time encoded in colors: $B=1200$ (black), 900 (red), 500 (blue), 300 (green), and 200~G (yellow). This complementary view shows that the dependence on $B$ is in fact stronger than on $\gamma$. Therefore, properly sampled Stokes profiles with enough polarimetric accuracy should be able to provide the required information to infer the magnetic field strength and inclination separately for most of the strength spectrum. Weaker fields will have bigger uncertainties for sure, but they should not imply a theoretical inability. As a matter of fact, the weaker the fields we want to explore, the smaller the noise we need in our observations, but this is somehow obvious.

\subsection{Depth-dependent atmospheres}
\label{sec:depthdependent}

In the solar atmospheres, physical quantities do vary with depth. Acknowledging such variations almost always implies resorting to numerical solutions of the transfer equation. That was first done by \citet{1969SoPh....9..372B,1969SoPh...10..262B}. Numerical results by \citet{1969SoPh....8..264S,1970SoPh...12...84S} and by \citet[][see Fig.\ \ref{fig:wittmann}]{1971SoPh...20..365W} soon appeared and numerical codes were described \citep[e.g.][\citeauthor{1976A&amp;AS...25..379L}, \citeyear{1976A&amp;AS...25..379L}]{1974SoPh...35...11W}. Those first numerical codes capable of synthesizing the Stokes profiles were based on the fourth-order Runge--Kutta algorithm that is very accurate at the price of being very computationally expensive. A generalization of the method by \citet{1964CR....258.3189F} to polarized light was proposed by \citet{auer+etal1977} and later modified by \citet{1987nrt..book..241R} in order to take magneto-optical effects into account. A fast solution of the RTE, after being reformulated as an integral Volterra equation (first suggested by \citeauthor{1969SoPh....8..264S}, \citeyear{1969SoPh....8..264S}), was formulated by \citet{rees+etal1989} with their so-called DELO method.

\epubtkImage{}{%
\begin{figure}[htbp]
    \centerline{\includegraphics[width=0.8\textwidth]{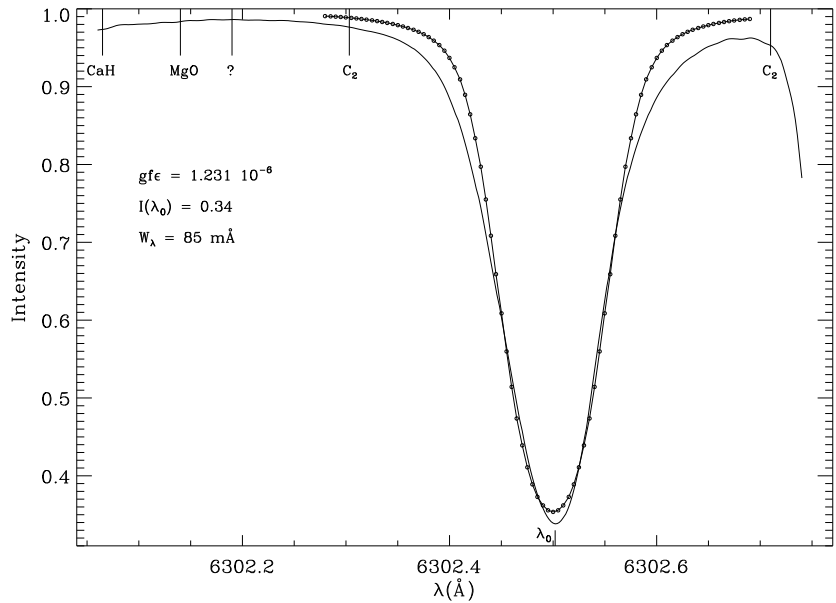}}
    \caption{Comparison between the observed and computed Stokes $I$ profile of the Fe~{\sc i} line at 630.25~nm. Adapted from \citet{1971SoPh...20..365W}.}
    \label{fig:wittmann}
\end{figure}}

An improvement in accuracy and computational speed was obtained by \citet{1998ApJ...506..805B} with an Hermitian method based on developing the Stokes vector as a fourth-order polynomial with depth. In Fig.\ \ref{fig:bellotrubio} we show an example of a synthesis of the same line as in Fig.\ \ref{fig:wittmann}, where clear asymmetries in wavelength in the Stokes profiles can be seen. According to Sections \ref{sec:varying} and \ref{sec:approxprof}, such asymmetries have been produced by the variation with depth of physical quantities. 

Learning how the various profile features of the four Stokes parameters depend on the many model parameters is certainly difficult and cannot be summarized in this paper. Experience, however, can train a researcher to be able to deduce ---many times after a quick glance (the \emph{art})--- a specific stronger $v_{\rm LOS}$ or $B$ here or there in the atmosphere. The situation is therefore much more complicated than for the ME case and one should rely upon inversions.

\epubtkImage{}{%
\begin{figure}[htbp]
    \centerline{\includegraphics[width=0.8\textwidth]{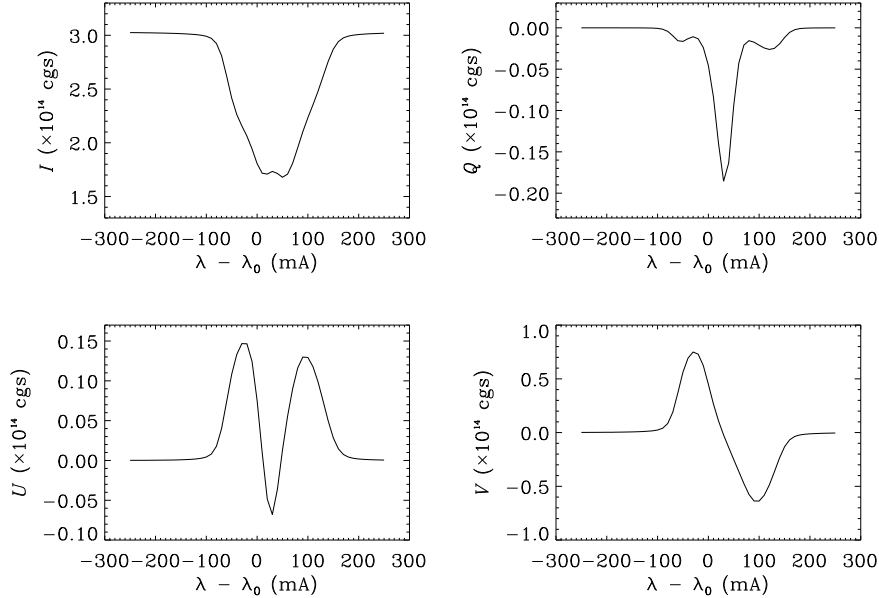}}
    \caption{Stokes profiles of Fe~{\sc i} line at 630.25 nm as synthesized with an Hermitian method. Adapted from \citet{1998ApJ...506..805B}.}
    \label{fig:bellotrubio}
\end{figure}}

\subsection{MHD simulations}
\label{sec:MHD}

The advent of magnetohydrodynamic (MHD) simulations such as  those by \citet{2003PhD...Voegler...A}, \citet{voegler+etal2005}, and \citet{2012ApJ...750...62R} has opened a new window on the exploration of the solar photosphere. They have enabled calculations that may help to envision what is expected from observations and to interpret them. The simulations also provide predictions that can be confronted with them. The enrichment has been remarkable because the realistic atmospheres resulting from the simulations have physical quantities varying along the optical path without any \emph{a priori} assumptions and may be closer to the actual Sun than other simplified atmospheres. MHD simulations have been used to test the reliability of inversion techniques \citep{2010A&A...518A...2O,2012A&amp;A...543A..34D}. In the first of these works, a confirmation of the predictions by \citet{1996A&A...314..295S} was found: if your inference technique assumes magnetic fields and velocities constant with depth and you use it on data coming from an atmosphere where these quantities are depth dependent, the result is just the average of the actual stratification weighted with the generalized response function to perturbations of that quantity. In the second of these papers, the non-LTE inversion code called NICOLE is tested. Here we report on preliminary results by \citet{2015ApJ...inprep...H} to illustrate the role of simulations as a tool to determine the optimum wavelength sample of Stokes profiles. This is a particularly interesting topic that is very relevant to the observational (and hence interpretational) work. Are the available samples enough for capturing all the information encoded in the Stokes profiles? What is the optimum sampling one should use with a new instrument under development depending on the goals such an instrument aims to fulfill?

Figure\ \ref{fig:stokespower} shows the Stokes $I$ (in depression), $Q$, $U$, and $V$ profiles of the Fe~{\sc i} line at 630.25 nm across a slit over a sunspot simulation by \citet{2012ApJ...750...62R} (left column panels) and their corresponding power spectra (right column panels). The simulation contains the transition from the quiet Sun (at both sides of the $X$ dimension) through the penumbra and the umbra of a sunspot. The power spectra of Stokes $V$, $Q$, and $U$ are wider than that of Stokes $I$ as a natural consequence of their shapes. Therefore, a cut-off frequency is better found in the polarization profiles. In this example, Shanon's critical sampling interval is around 1.25 pm/pixel with the remarkable fact that no convolution with an instrumental PSF has been applied to the profiles. This value is coarser than that provided by several ground-based spectrographs with resolutions about $R\simeq 10^{6}$ that would be considered too fine for the required diagnostics. As soon as the profiles are observed by an instrument with a finite width PSF, the Nyquist frequency will shrink to smaller values. Hence, the critical sampling will be coarser. These kinds of calculations can therefore help in designing new instruments and, as far as this paper is concerned, in deciding the adequate spectral sampling for any synthesis or inversion code: if the sampling is too fine, we are wasting computational time; if the sampling is too coarse, we are neglecting available information.

\epubtkImage{}{%
\begin{figure}[htbp]
    \centerline{\includegraphics[width=\textwidth]{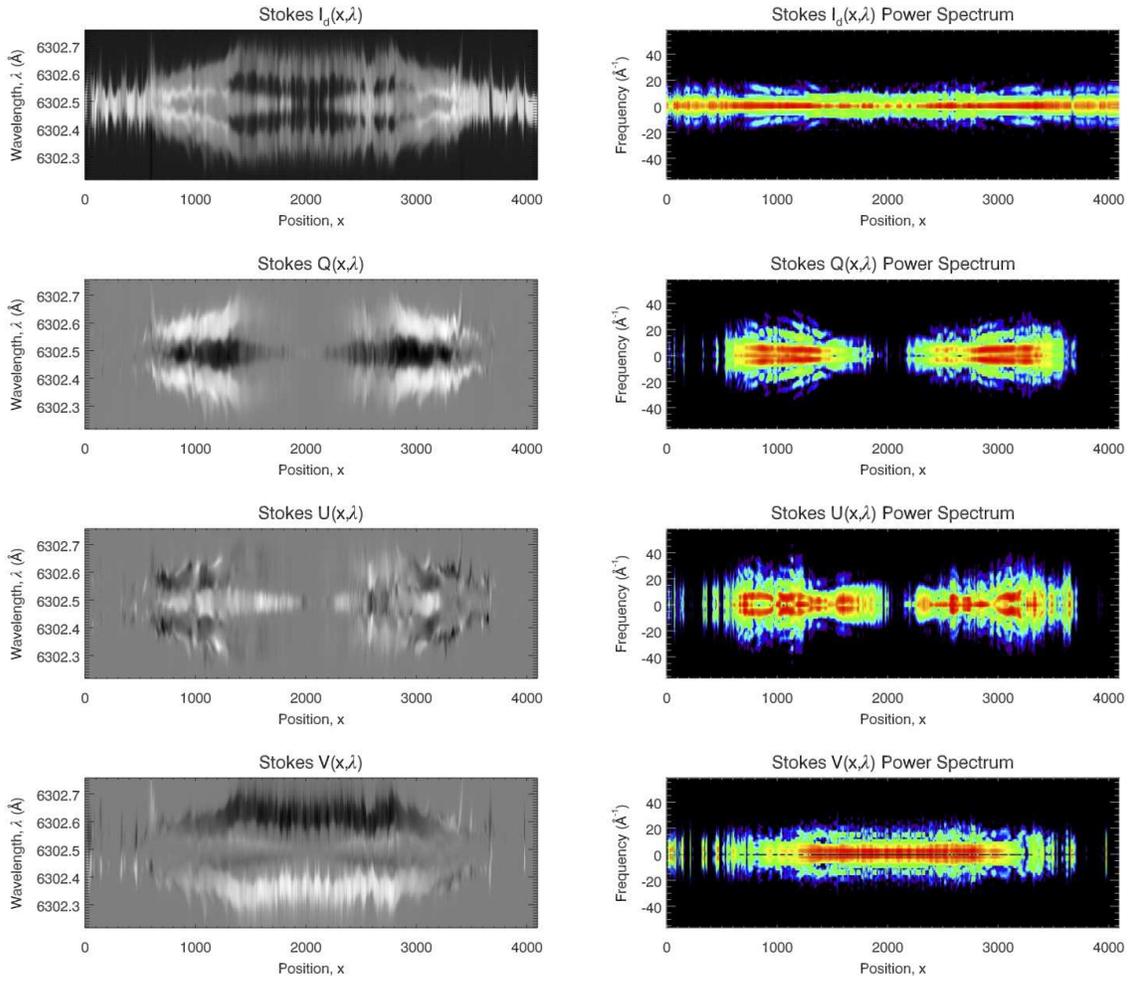}}
    \caption{Stokes profiles ($I$ is in line depression) across a sunspot simulation by \citet{2012ApJ...750...62R} (left column) and power spectra (right column). Adapted from \citet{2015ApJ...inprep...H}.}
    \label{fig:stokespower}
\end{figure}}

\newpage


\section{Response functions}
\label{section:response}

The needle in an old-fashioned ammeter is useful as far as it moves when a current circulates through the electric circuit. If, for instance, the needle does not move when a given very weak or very strong current passes through, then the ammeter is useless or \emph{insensitive} to those intensities. Keeping this analogy, our observables, the Stokes profiles are useful for inferring given physical quantities as long as they change when those physical quantities vary. Of course, to be detectable, the change should be larger than the noise. Therefore, the correct question to ask of a given spectropolarimetric proxy whether it senses, for example, $T$, $B$, or $v_{\rm LOS}$ is how much it modifies when $T$, $B$, or $v_{\rm LOS}$ change. The preceding section has been an attempt in that direction: we have been changing the various atmospheric quantities and checking the modifications in the Stokes profiles. We have proceeded in the \emph{direct} way, that is, through the solution of the \emph{differential} RTE. This direct approach is not very useful in practice, though, as we already recognized in Section \ref{sec:depthdependent}. Which quantity is to be modified first, at which optical depths, and by how much? If the problem is the simplest we talked about in the introduction (that of measuring velocities from the line core wavelength), then there may be some room for the direct approach. If not, the diagnostic capabilities of the Stokes profiles have to be further explored with the final goal of proceeding the \emph{inverse} way, that is, of solving the \emph{integral} equation known as the formal solution of the RTE (Eq.\ \ref{eq:rteformalsolution}). 

Here, we have a difficult problem where the observables depend nonlinearly on the unknowns. The nonlinear character is clear: the observables ---on the left-hand side of Eq.\ (\ref{eq:rteformalsolution})--- are equal to an integral of the product of three terms, each depending strongly non-linearly on the physical quantities that characterize the model atmosphere \citep[e.g.,][]{2003isp..book.....D}.  Changes in the Stokes spectrum are then very difficult to predict when modifications in the physical parameters occur. As in many other branches of physics, the diagnostic tools come out trough a linearization analysis. We can assume, for instance, that in a very special regime, when perturbations are small enough, changes occur linearly. These are the basics of linearization that, in the realm of solar physics were introduced for non-polarized light by \citet[][see also \citeauthor{1976SoPh...50..239C}, \citeyear{1976SoPh...50..239C}, and \citeauthor{1977A&amp;A....54..227C}, \citeyear{1977A&amp;A....54..227C}]{1971SoPh...20....3M} through the so-called weighting functions, although the name of \emph{response functions} (RFs) did not appear in the literature until the work by \citet{1975SoPh...43..289B}. Since polarization was not taken into account, those analyses were only strictly valid for isotropic media or, as far as we are concerned, for non-magnetic atmospheres. Response functions were introduced within polarized radiative transfer by the brothers \citet{1977A&A....56..111L} but RFs were paid little attention to until the works by  \citet{1982SoPh...77...13D,1983SoPh...87..221L}, \citet{1988A&amp;A...204..266G}, \citet{almeida1992}, and \citet{1992ApJ...398..375R,1994A&A...283..129R}. The latter found that the perturbations $\delta x_{i}$ exerted to the $p+r$ atmospheric quantities ($p$ of them varying with height and $r$ constant) induce modifications $\delta \vector{I} (0)$ to the observed Stokes profile given by 

\begin{equation}
\label{eq:deltai}
\delta \vector{I} (0) = \sum_{i=1}^{p+r} \int_0^{\infty} \vector{R}_i (\tau_{\rm c}) \, \delta x_i (\tau_{\rm c}) \, {\rm d}\tau_{\rm c},
\end{equation}
where 
\begin{equation}
\label{eq:responsefun}
\vector{R}_i (\tau_{\rm c}) \equiv \matriz{O}(0, \tau_{\rm c}) \, \left[ \matriz{K} (\tau_{\rm c}) \frac{\partial \vector{S}}{\partial x_i} - \frac{\partial \matriz{K}}{\partial x_i} (\vector{I} - \vector{S}) \right].
\end{equation}

Therefore, the modification of $\vector{I} (0)$ is given by a sum of terms, each related to one of the atmospheric quantities characterizing the medium. The terms are integrals over the whole atmosphere of the model atmospheric quantities weighted by the RFs. The physical meaning in Eq.\ (\ref{eq:deltai}) is straightforward: imagine that we change only a given quantity ($T$, $B$, $v_{\rm LOS}$, or any other) with a magnitude unity (i.e., 1 K, 1 G, 1 \kms, etc.) in the narrow surroundings of a given continuum optical depth $\tau_0$; then, the subsequent modification in the emergent Stokes spectrum is just the value of the corresponding RF at that optical depth:

\begin{equation}
\label{eq:rfexample}
\delta \vector{I} (0) = \vector{R}_i (\tau_0).
\end{equation}

Then, since the Stokes profiles are usually recorded normalized to some reference value (e.g., the average, ---unpolarized--- continuum intensity of the quiet Sun), units for RFs are inverse units of the corresponding quantity. That is, the response function to perturbations of temperature is measured in K$^{-1}$; the response to perturbations in the magnetic field strength is measured in G$^{-1}$; and so on. Thus, a response function can be defined as the modification that the Stokes spectrum experiences when the medium undergoes a unit perturbation of a given physical quantity at a given very narrow region in optical depth. Equation (\ref{eq:responsefun}) tells us that these modifications build upon the variations of the propagation matrix and the source function vector with respect to the physical quantities and their evolution through the atmosphere as driven by the evolution operator. The two variations have an opposite sign. This means that they are somehow competing as one could expect. While $\vector{S}$ represents the sources of photons, $\matriz{K}$ represents the sinks. Indeed we know that we do not only speak about photon removal but also about pleochroism and dispersion but, certainly, the propagation matrix role is somehow similar to that of a withdrawal. The counterbalancing between $\vector{S}$ and $\matriz{K}$ is very important to understanding radiative transfer because some analyses forget it and only account for the effects of $\matriz{K}$ (absorption in the non-polarized case). This was clearly pointed out and explained by \citet{1994A&A...283..129R}.

Equation (\ref{eq:deltai}) suggests that RFs play the role of partial derivatives of the observed Stokes profiles with respect to the atmospheric quantities once they have been discretized. This role is even more clear when we go down to the real world of a quadrature formula for that equation. Model atmospheres are usually described numerically by a grid of points that are spaced in logarithmic optical depth. Let $\Delta (\log \tau_{\rm c})$ be that spacing. If we call $x_{i,j} \equiv x_i (\tau_j)$ and $\vector{R}_{i,j} \equiv \vector{R}_i (\tau_j)$, then Eq.\ (\ref{eq:deltai}) can be written as

\begin{equation}
\label{eq:deltaiquad}
\delta\vector{I}(0) = \sum_{i=1}^p \sum_{j=1}^n a_j \vector{R}_{i,j} \, \delta x_{i,j} + \sum_{k=1}^r \vector{R}'_k \, \delta x_k,
\end{equation}
where $a_j = \Delta (\log \tau_{\rm c}) \ln 10 \, c_j \tau_j$, with $c_j$ being the quadrature coefficients. Therefore, if we include $a_j$ in the RFs, as one usually does in graphical representations, then Eq.\ (\ref{eq:deltaiquad}) shows the Stokes spectrum modifications as \emph{linear} expansions of the new variables $x_{i,j}$ and $x_k$. The first term on the right-hand side corresponds to those physical quantities that vary with depth; the second stands for those that are assumed to be constant.\footnote{For the specific meaning of the RF to perturbations of a constant quantity, $\vector{R}'$, see Sect.\ \ref{sec:rfproperties} below.} In summary, we can say that RFs are indeed partial derivatives of $\vector{I}(0)$ with respect to the (numerical) atmospheric parameters and, thus, they directly provide the sensitivities of the Stokes spectrum to perturbations of the physical conditions in the medium. Examples of these RFs are plotted in Figs.\ \ref{fig:figiresponse}, \ref{fig:figqresponse}, \ref{fig:figuresponse}, and \ref{fig:figvresponse}. They have been evaluated for Stokes $I$, $Q$, $U$, and $V$, respectively, of the Fe~{\sc i} line at 630.25 nm to perturbations of the temperature (top row panels), of the magnetic field strength (middle row panels), and of the LOS velocity (bottom row panels). The RF values are multiplied by $10^6\,$K$^{-1}$,  $10^6\,$G$^{-1}$, and $10^4\,(\!\kms)^{-1}$. The two columns correspond to two different model atmospheres. That in the left-hand columns has the temperature stratification of the HSRA model \citep{gingerich+etal1971}, a constant $B=2000$ G, $\gamma = 30\degree$, and $\varphi = 60\degree$; the plasma is at rest in this model. That in the right-hand columns has a 500 K cooler temperature, and a magnetic field 500~G weaker, 20\degree less inclined, and an azimuth of 10\degree; $v_{\rm LOS} = 1.58 + 0.3 \log \tau_{\rm c}$ ($\!$\kms). 

Equation (\ref{eq:deltaiquad}) hints at a way for calculating RFs through what could be called \emph{the brute force method}. This method is a four-step procedure: 1) synthesis of the Stokes spectrum in a given model atmosphere; 2) perturbation of just one of the (numerical) atmospheric parameters by a small amount and synthesis of the spectrum in the new model atmosphere; 3) calculation of the ratio between the difference of the two spectra and the perturbation; 4) repetition of steps 2) and 3) for each optical depth, for each wavelength sample, and for the remaining Stokes parameters. This is a formidable calculation as soon as the number of free parameters is large. Fortunately, Eq.\ (\ref{eq:responsefun}) provides a shortcut since the evolution operator, the propagations matrix, and the source function vector have to be calculated anyway in every synthesis of the spectrum. With only the added calculations of the derivatives, one can easily calculate RFs at the same time as  the RTE is solved. This property is extremely useful for inversion codes, as we shall see in Section\ \ref{section:techniques}.

Equation (\ref{eq:deltaiquad}) also offers an explicit explanation of the astrophysical ill conditioning we commented on in Section \ref{sec:introduction}: the same modification of $\vector{I}(0)$ may be produced by perturbations of different quantities or by perturbations of a single physical quantity but at several optical depths. That is, the effects of temperature can be similar to those of the magnetic field strength or the effects of perturbing $B$ at $\log \tau_{\rm c} = -0.5$ can be the same as those of perturbing $B$ at $\log \tau_{\rm c} = -3$. Therefore, we cannot say that the changes $\delta\vector{I}(0)$ are produced by perturbations of this physical parameter or that other without considering all of them at the same time. Cross-talk among some parameters may appear and, then, the retrieval of those parameters will be less reliable (see, e.g., Section \ref{sec:analyticrfs}). 

\subsection{Properties of response functions}
\label{sec:rfproperties}

A glance at Figs.\ \ref{fig:figiresponse}, \ref{fig:figqresponse}, \ref{fig:figuresponse}, and \ref{fig:figvresponse} readily tells us that some RFs are bigger than the others. This means that our line is more sensitive to some physical quantities than to others. However, the fact that RFs are measured in inverse units of those for their corresponding parameters makes it difficult to compare their relative sensitivity. In this regard, \emph{relative} RFs shed some light. If we consider relative perturbations $\delta x_{i,j} / x_{i,j}$, then we can define relative RFs $\tilde{\vector{R}}_{i,j} \equiv \vector{R}_{i,j} x_{i,j}$. Hence, $\tilde{\vector{R}}_{i,j}$ speak about the response of the Stokes spectrum to relative (i.e., dimensionless) perturbations. Experience shows that relative RFs to $T$ perturbations are the biggest at all depths and wavelengths, clearly indicating that temperature is the most important quantity in line formation. (Indeed, temperature is \emph{the} physical quantity that governs the thermodynamical state of the material medium because we assume that hydrostatic equilibrium prevails throughout our model atmospheres. After this assumption, pressure, the necessary second thermodynamic variable gets automatically prescribed.) Response functions to temperature perturbations start being different from zero at the deepest layers when compared to the remaining quantities. This is because the second term in the right-hand side of Eq.\ (\ref{eq:responsefun}) goes to zero as the continuum optical depth tends to infinity. This was explained by \citet{1994A&A...283..129R}. This physical fact implies that spectral lines tend to be insensitive at these low layers to the other physical quantities.

The Stokes profile wavelength symmetries are preserved in RFs: in the absence of velocity gradients, RFs of Stokes $I$, $Q$, and $U$ to any perturbation are even functions of wavelength and RFs of Stokes $V$ are odd. This means that, in fact, velocity gradients increase the diagnostic capabilities of spectral lines. In their absence, half of the profile is useless since the information they provide is exactly the same as the other half.\footnote{Indeed, such redundant information should help in decreasing the uncertainties in the retrievals by reducing the noise by a factor two.}

For given purposes, we can conceive constant perturbations with depth, in spite of the quantity being depth dependent. Owing to their nature, some physical quantities may be assumed constant with depth (e.g., macro- and microturbulent velocity, or any of the ME free parameters). In such cases, constant perturbations are in order. If so, then, Eq.\ (\ref{eq:deltai}) tells us that the resulting modification in the Stokes spectrum is given by the product of such a constant perturbation times the integral of the corresponding RF over the whole atmosphere. Hence, we can say that the RF to a constant perturbation, $\vector{R}'$ in Eq.\ (\ref{eq:deltaiquad}), is directly the integral of the regular response function or, in numerical terms,
\begin{equation}
\label{eq:rfcontantpertur}
\vector{R}'_k \equiv \sum_{j=1}^n a_j \vector{R}_{k,j}.
\end{equation}

As shown by \citet{1994A&A...283..129R}, RFs play the role of a PSF in the general theory of linear systems. Under this general theory, our system ---the Stokes spectrum--- experiences an input (the perturbation) and provides an output, $\delta\vector{I} (0)$. If the input is a Dirac delta, then the output is the corresponding value of the response function. If the input is harmonic throughout the atmosphere, then the response is the Fourier transform of the RF.

Response functions are model dependent. This property is extremely important in our understanding of spectral line sensitivities and, thus, in the inversion of the RTE. Instead of being a drawback, such a model dependence helps in disentangling the effects produced by the different quantities in distinct model atmospheres. Even in fixed model atmospheres, it is very difficult to discard one quantity or the other at once, however, and most of them have to be retrieved at the same time. Once this is carried out, one can theoretically understand the meaning of measurements \citep[][\citeauthor{2003isp..book.....D}, \citeyear{2003isp..book.....D}]{1996A&A...314..295S}.

\epubtkImage{}{%
\begin{figure}[htbp]
    \centerline{\includegraphics[width=\textwidth]{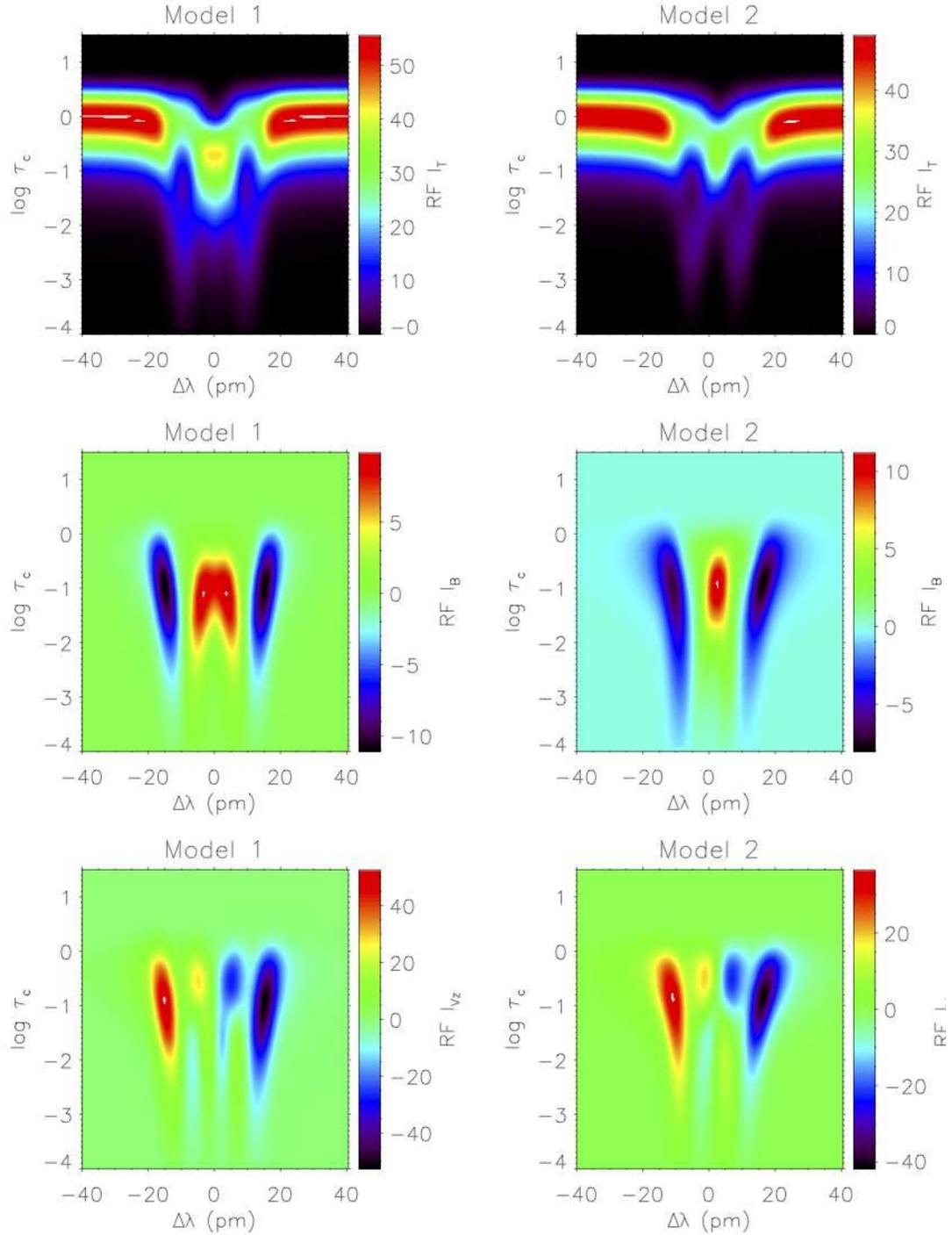}}
    \caption{RFs of Stokes $I$ to perturbations of $T$ (top panels), $B$ (middle panels), and $v_{\rm LOS}$ (bottom panels). Units are $10^{-6}\,$K$^{-1}$,  $10^{-6}\,$G$^{-1}$, and $10^{-4}\,(\!\kms)^{-1}$.The two columns correspond to two different model atmospheres. That in the left-hand column has the temperature stratification of the HSRA model \citep{gingerich+etal1971}, a constant $B=2000$ G, $\gamma = 30\degree$, and $\varphi = 60\degree$; the plasma is at rest in this model. That in the right-hand column has a 500 K cooler temperature, and a magnetic field 500~G weaker, 20\degree less inclined, and an azimuth of 10\degree; $v_{\rm LOS} = 1.58 + 0.3 \log \tau_{\rm c}$ ($\!$\kms). }
    \label{fig:figiresponse}
\end{figure}}

\epubtkImage{}{%
\begin{figure}[htbp]
    \centerline{\includegraphics[width=\textwidth]{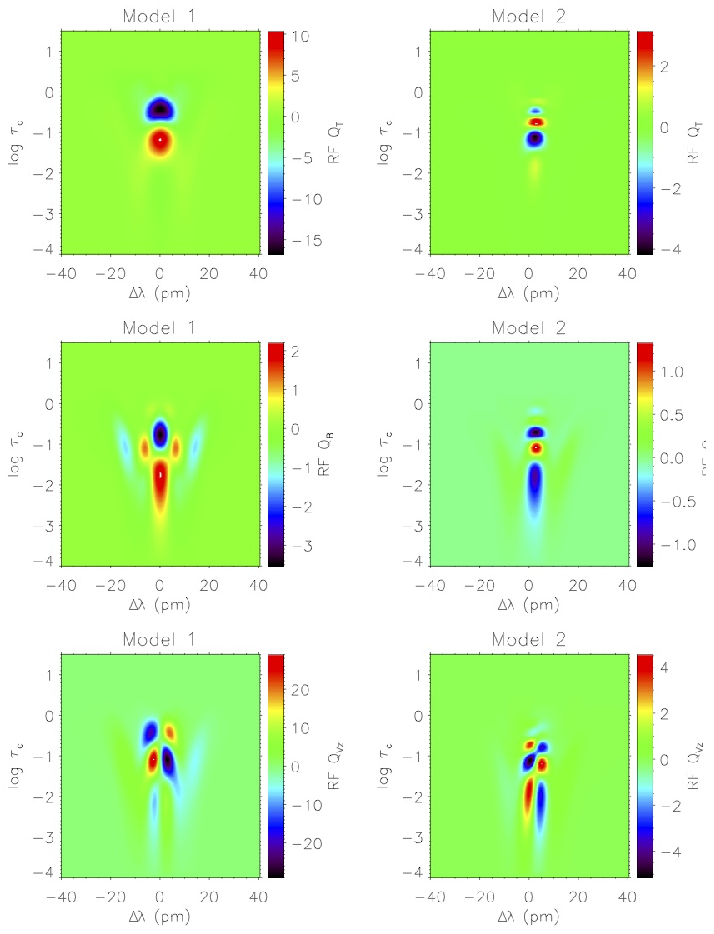}}
    \caption{Same as Fig.\ \ref{fig:figiresponse} for Stokes $Q$.}
    \label{fig:figqresponse}
\end{figure}}

\epubtkImage{}{%
\begin{figure}[htbp]
    \centerline{\includegraphics[width=\textwidth]{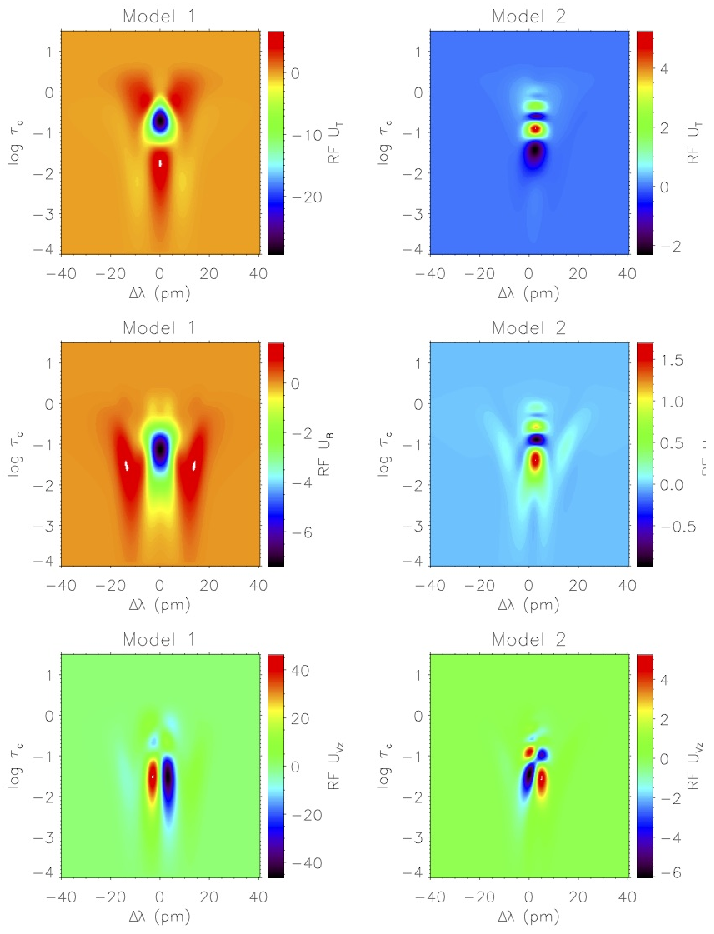}}
    \caption{Same as Fig.\ \ref{fig:figiresponse} for Stokes $U$.}
    \label{fig:figuresponse}
\end{figure}}

\epubtkImage{}{%
\begin{figure}[htbp]
    \centerline{\includegraphics[width=\textwidth]{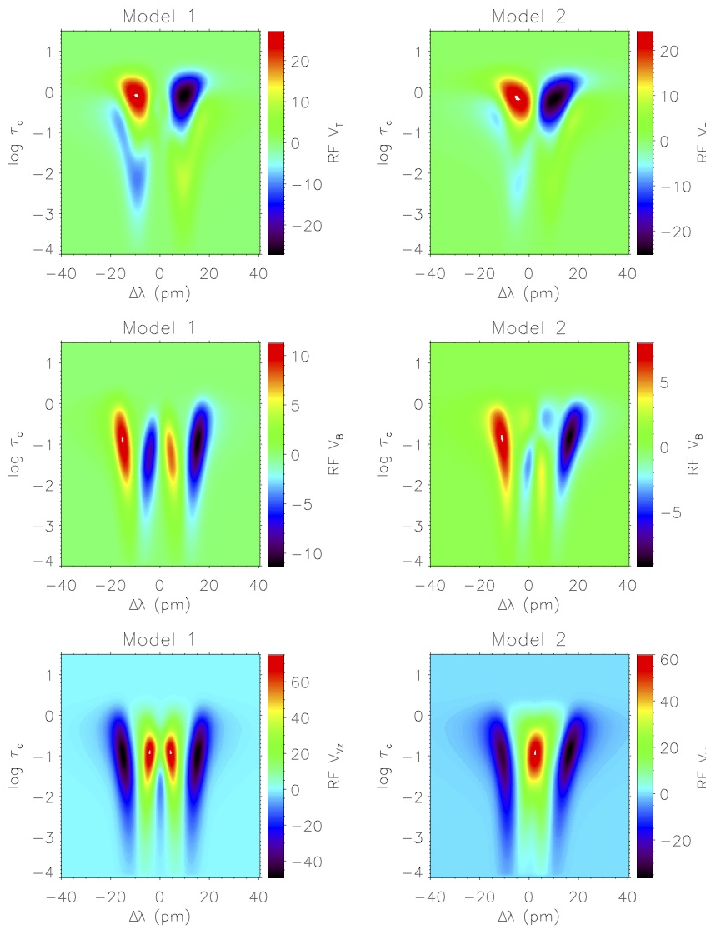}}
    \caption{Same as Fig.\ \ref{fig:figiresponse} for Stokes $V$.}
    \label{fig:figvresponse}
\end{figure}}

Last, but not least, the linear nature of RFs allows us to generalize them to any linear combination of Stokes profile wavelength samples. This property helps us understand what can be extracted from different proxies that are traditional in solar polarimetry but, most importantly, helps the astronomer in taking influence of the instrument into account. Since most instruments act as linear systems on light, the detected spectrum is a convolution of the actual spectrum with the instrument spectral PSF. Convolution is linear and, thus, one can easily conclude \citep{1994A&A...283..129R} that the RFs of the convolved spectrum are nothing but those of the original spectrum convolved as well with the PSF.\footnote{Convolutions with the spatial PSF of the instrument are also taken into account in modern inversions that take spatial degradation into account. See Section \ref{sec:spatialdegrad}.}

\subsection{Analytic response functions}
\label{sec:analyticrfs}

When all the terms on the right-hand side of Eq.\ (\ref{eq:responsefun}) can be calculated analytically (see Sect.\ \ref{sec:milne}), RFs are necessarily analytic and then we can use them to gain some physical insight into the diagnostic capabilities of the Stokes spectrum about the physical quantities that characterize the medium. This is the case of the ME approximation, where all the quantities are constant with depth. There, index $j$ in Eq.\ (\ref{eq:deltaiquad}) drops and (after inclusion of the coefficient into the RF) we can properly write:
\begin{equation}
\label{eq:rfme}
\vector{R}_i (\lambda) = \frac{\partial \vector{I}(\lambda)}{\partial x_i}.
\end{equation}
That is, that RFs are strict partial derivatives of the Stokes spectrum with respect to the free parameters of the problem \citep{2007A&A...462.1137O}. In Eq.\ (\ref{eq:rfme}) we have removed the $\tau_{\rm c} = 0$ indicator in the emergent Stokes spectrum and, rather, we have made explicit the dependence of RFs on wavelength. We have just seen in the former subsection that these RFs are indeed integrals over $\tau_{\rm c}$ and, hence, the dependence on it disappears. This analytic character is particularly important in the controversial discussion about the possibility of disentangling $B$, from $\alpha$ (the filling factor) in the case that our magnetic features are not fully spatially resolved. In Fig.\ \ref{fig:deltoroetal2010} we reproduce Fig. 3 from \citet{2010ApJ...711..312D}. It shows RFs to (constant) perturbations in those two quantities plus $v_{{\rm LOS}}$ in a ME atmosphere. As one can clearly see, both  Stokes $V$ RFs to $\alpha$ and to $B$ perturbations are almost proportional among themselves and to the Stokes $V$ profile itself. This can easily be traced back to the expected behavior from the weak field approximation, as expressed in Equation (\ref{eq:vpropmag}). From this proportionality we should conclude that it is indeed very difficult to discern the values of $B$ and $\alpha$ separately. If it were not for Stokes $I$, the widespread belief that only $\alpha B$ or $\alpha B \cos\gamma$ can be retrieved would be true. However, the Stokes $I$ RFs to $\alpha$ and $B$ are neatly different from each other and this necessarily implies that we have the means of inferring the two quantities independently. Stokes $Q$ and $U$ also help in disentangling the field strength and the filling factor for similar reasons as soon as they are above the noise level.

\epubtkImage{}{%
\begin{figure}[htbp]
    \centerline{\includegraphics[width=\textwidth]{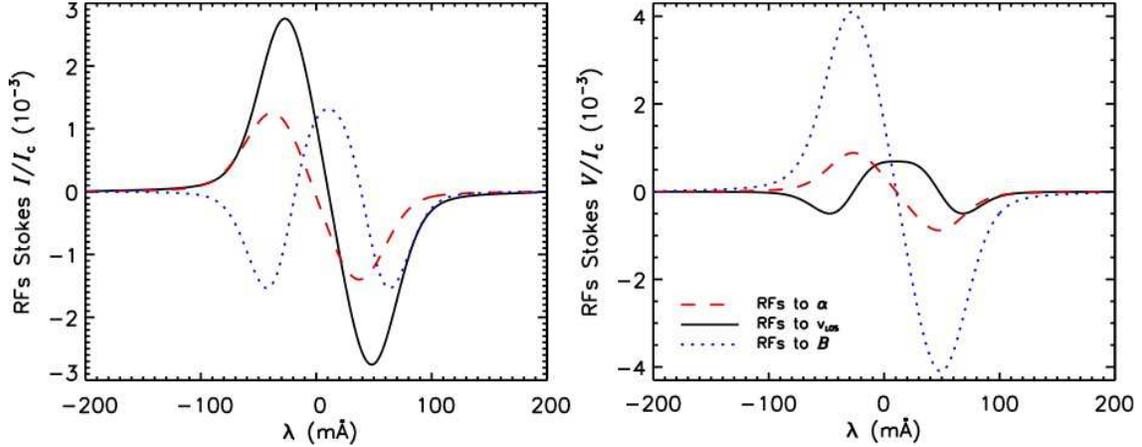}}
    \caption{Stokes I (left panel) and V (right panel) RFs to $v_{\rm LOS}$ (solid, black lines), to $B$ (dotted, blue lines), and to $\alpha$ (dashed, red lines) in the weak-field case. Perturbations of 10 \ms for $v_{\rm LOS}$, of 10~G for $B$, and of 0.1 for $\alpha$ have been assumed. Adapted from \citet{2010ApJ...711..312D}.}
    \label{fig:deltoroetal2010}
\end{figure}}

Among other features of RFs, the latter authors showed how the thermodynamical parameters of the ME atmosphere can have cross-talk among themselves: their RFs are fairly similar in shape, so that their effects can be misinterpreted by the inversion codes. Notably, the RFs to perturbations of $B$, $\gamma$, $\varphi$, and $v_{\rm LOS}$ are markedly different from one another and with respect to those of $\eta_0$, $\Delta\lambda_{\rm D}$, and $a$. This explains the good result of ME inversion codes in accurately retrieving the magnetic and dynamic parameters while the thermodynamic parameters are sometimes wrong. Our conclusion is also consistent with, and explains, the findings by \citet{2004A&amp;A...414.1109L} who decided to leave the damping parameter fixed with minor changes in the fitted magnetic and dynamic parameters while $\Delta\lambda_{\rm D}$ and $\eta_0$ were significantly affected. Linearity can also be useful for spatially coupled inversion techniques (see Sect.\ \ref{sec:coupled} below).

\newpage


\section{Inversion techniques}
\label{section:techniques}

Once we have discussed all the ingredients and assumptions, we can face the main problem in astrophysics, namely that of making theory and observations compatible. In other words, we can face the inversion problem by deriving the unknown physical quantities through comparison between observed and synthetic Stokes profiles.\footnote{The inversion problem could be thought of as the ``observational'' part of the compatibility game. The ``theoretical'' part involves the choice of hypotheses included in the physical scenario which, according to \citet{2010ApJ...711..312D}, defines the assumed model atmosphere. One has to decide whether the radiative transfer is LTE or NLTE, whether the physical quantities are dependent on the optical depth, whether macro- or microturbulence are needed, etc. All these assumptions settle the theoretical framework of the problem.} Figure \ref{fig:nonlteinversion} describes how the problem gets complicated as compared to the mere forward problem in Figure \ref{fig:NLTE}.

\epubtkImage{}{%
\begin{figure}[htbp]
    \centerline{\includegraphics[width=\textwidth]{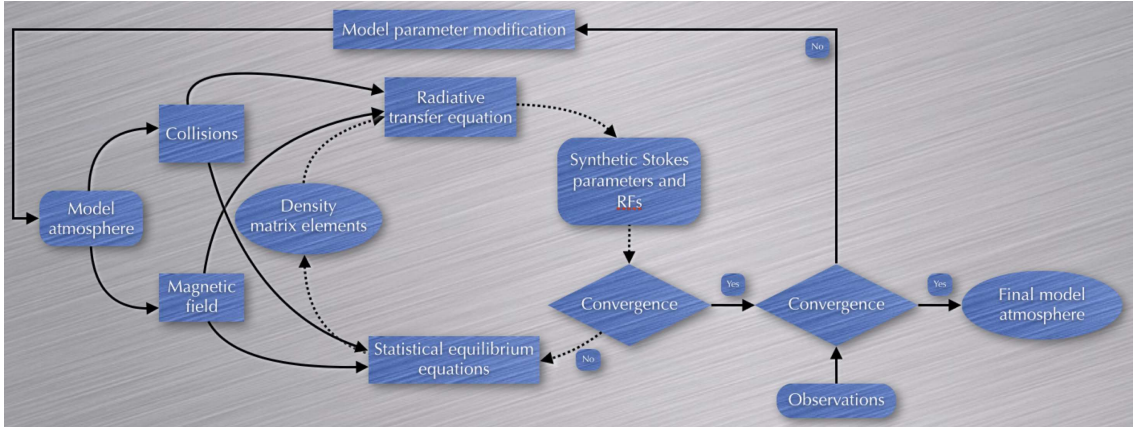}}
    \caption{Block diagram of the inversion problem under NLTE conditions.}
    \label{fig:nonlteinversion}
\end{figure}}

One can clearly see how a new overarching loop is present that indicates the needs for changing the model atmosphere if the synthetic Stokes spectra do not properly fit the observed ones. The problem turns out to be formidable and requires new, specific assumptions that make it tractable. In particular, some of the quantities have to be calculated from the strict NLTE conditions (see Section \ref{sec:non-lte}).

Even the simpler LTE problem gets complicated, and indeed becomes iterative, regardless of acknowledging the stratification in the physical quantities of the atmosphere (see Figure \ref{fig:lteinversion}). The needs for modifying such a model atmosphere according to the deviations between observed and synthetic profiles makes a loop necessary after calculating both the synthetic spectra and their derivatives with respect to the free parameters. Fortunately, we know how to calculate these derivatives through RFs at the same time as we synthesize the Stokes spectrum with little extra computational effort.

\epubtkImage{}{%
\begin{figure}[htbp]
    \centerline{\includegraphics[width=\textwidth]{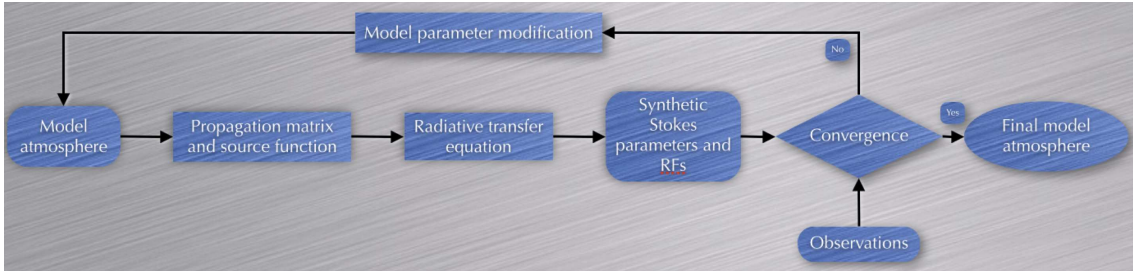}}
    \caption{Block diagram of the inversion problem under LTE conditions.}
    \label{fig:lteinversion}
\end{figure}}

Looking for convergence means measuring the distance between observed and synthetic profiles in the space of observables. Any inversion procedure must have a threshold below which the user can consider that convergence has been reached because the fit cannot be further improved within the current assumptions.

\subsection{Topology in the space of observables}
\label{sec:chisquare}

The topological problem depicted in the Introduction for the inversion problem (or any astrophysical inference) needs to be substantiated in minimizing a distance: a metric in the space of the observables. Since Stokes $I_{\rm d}$, $Q$, $U$, and $V$ belong to $\bbbl^2$, the quadratic norm of the difference turns out to be the natural distance between any two profiles. We want to approximate two sets of profiles, so that all four Stokes parameters should be taken into account. Therefore, when in practice we deal with discrete samples, the sought-for distance can be written as 
\begin{equation}
\label{eq:chi2}
\chi^2 (\vector{x}) \equiv \frac{1}{N_{\rm f}} \sum_{s=0}^3 \sum_{i=1}^q \left[ I_s^{\rm obs} (\lambda_i) - I_s^{\rm syn} (\lambda_i; \vector{x}) \right]^2 \, w_{s,i}^2,
\end{equation}
where index $s$ runs for the four Stokes parameters, we assume $q$ wavelength samples, and $N_{\rm f}$ stands for the number of degrees of freedom, that is, the difference between the number of observables ($4q$) and that of the free parameters (the number of elements in $\vector{x}$, $np+r$; see Section \ref{sec:NLTE} and Equation \ref{eq:modelatmos}). 

$\chi^2 (\vector{x})$ is a merit function of the atmospheric quantities that measures the distance between the observed and the synthetic profiles and has to be minimized in order to achieve a good fit. Having a normalized merit function to the degrees of freedom is useful to warn the user not to use an unreasonably large number of free parameters as compared with the number of observables. In such a case, $\chi^2$ would turn out to be always too big. The \emph{weights} $w_{s,i}$ can be used to favor some data more than the others. For instance, one can set them to the inverse of the measurement errors. For many applications they are simply kept at unity.

We can look at $\chi^2 (\vector{x})$ as a scalar field in an ($np+r$)-dimensional space. Since the number of dimensions may be too large, the minimization problem may turn out to be intractable. Before going to specific techniques that make it affordable, let us consider the paths through which we can look for the minimum of the merit function. That is, we have to find the derivatives of $\chi^2$ with respect to the atmospheric free parameters. \citet[][see \citeauthor{1994A&A...283..129R}, \citeyear{1994A&A...283..129R}, and \citeauthor{2003isp..book.....D}, \citeyear{2003isp..book.....D} as well]{1992ApJ...398..375R} showed that such derivatives are directly given by the RFs:
\begin{equation}
\label{eq:chiderivative}
\frac{\partial \chi^2}{\partial x_m} = \frac{2}{N_{\rm f}} \sum_{s=0}^3 \sum_{i=1}^q \left[ I_s^{\rm obs} (\lambda_i) - I_s^{\rm syn} (\lambda_i; \vector{x}) \right] \, w_{s,i}^2 \, R_{m,s} (\lambda_i),
\end{equation}
where, for the sake of a more compact notation, index $m$ runs from 1 to $np+r$ (including constant and variable physical quantities), the quadrature coefficients in Eq.\ (\ref{eq:deltaiquad}) are assumed to be included in the RFs when needed, and no distinction is made between $\vector{R}$'s and $\vector{R}'$'s. The same authors also demonstrated that the second derivatives can be approximated by
\begin{equation}
\label{eq:seconderivativechi}
\frac{\partial^2 \chi^2}{\partial x_m \partial x_k} \simeq \frac{2}{N_{\rm f}} \sum_{s=0}^3 \sum_{i=1}^q w_{s,i}^2 \, \left[ R_{m,s} (\lambda_i) \, R_{k,s} (\lambda_i) \right].
\end{equation}

Regardless of the way we approach its minimum in the hyperspace of parameters, that $\chi^2$ is the natural metric is reinforced by the fact that other metrics have been tried \citep[e.g.][]{2004A&amp;A...414.1109L} that have finally converged to almost the same formulation as in Eq.\ (\ref{eq:chi2}) \citep{2007A&amp;A...462.1147L}.

\subsection{Levenberg--Marquardt based inversions}
\label{sec:levenberg}

The process of profile fitting is nothing but the successive (and iterative) approximation of synthetic Stokes profiles until they reach a minimum distance to the observed ones. Hence, an initial guess model atmosphere is needed to start the procedure. Step by step, the model will be modified, so that the resulting synthetic Stokes profiles will approach more and more the observations. When we are close enough to the $\chi^{2}$ minimum, an approximate, parabolic motion may be useful:
\begin{equation}
\label{eq:parapprox}
\chi^{2} (\vector{x} + \delta\vector{x}) \simeq \chi^{2} (\vector{x}) + \delta\vector{x}\trans (\nabla \chi^{2} + \frac{1}{2} \matriz{H}' \delta\vector{x}),
\end{equation}
where the elements of the gradient are given by Eq.\ (\ref{eq:chiderivative}) and $\matriz{H}'$ is one half of the Hessian matrix, whose elements are given by Eq.\ (\ref{eq:seconderivativechi}), that is, $H'_{m,k} =  \partial^2 \chi^2/\partial x_m \partial x_k$. In Eq.\ (\ref{eq:parapprox}) a scalar product is understood between a transposed (row) vector and a regular (column) vector. When we are very near the minimum, it is clear that the second term in the right-hand side of Eq.\ (\ref{eq:parapprox}) should be zero, and this is done in the Levenberg--Marquardt (LM) algorithm \citep[e.g.,][]{press+etal1986} by requiring that
\begin{equation}
\label{eq:gradhessian}
\nabla \chi^{2} + \matriz{H} \delta\vector{x} = \vector{0},
\end{equation}
where the new matrix $\matriz{H}$ is defined by
\begin{equation}
\label{eq:hessian}
2 H_{ij} \equiv \left\{ \begin{array}{lll}
H'_{ij} (1 + \lambda), & \mbox{if} & i=j, \\
H'_{ij}, & \mbox{if} & i\neq j, \end{array} \right.
\end{equation}
where $\lambda$ is an \emph{ad-hoc} parameter that helps \emph{tuning} the algorithm for it to work as if the approximation is almost first order ($\lambda$ is large) or fully second order (when $\lambda$ is small). $\lambda$ is changed in every step in the iteration, depending on how far or close we are to the minimum as indicated by the variation of $\chi^2$.

At the end of the procedure we will most likely not find the true minimum but, hopefully, will be close enough to neglect the gradient term in Equation (\ref{eq:parapprox}). In such a case we can write
\begin{equation}
\label{eq:difend}
\Delta\chi^2 = \delta\vector{x}\trans \matriz{H}' \delta\vector{x}.
\end{equation}
The good news about this relationship is that, since the Hessian matrix is made up of RFs, one can finally obtain an expression for the inversion uncertainties in the physical quantities that are functions of the RFs \citep[see][]{2003isp..book.....D}.
\begin{equation}
\label{eq:uncertainties}
\sigma_m^2 \simeq \frac{2}{np+r} \frac{{\displaystyle \sum_{s=0}^3 \sum_{i=1}^q} \left[ I_s^{\rm obs} (\lambda_i) - I_s^{\rm syn} (\lambda_i; \vector{x}) \right]^2 \, w_{s,i}^2}{{\displaystyle \sum_{s=0}^3 \sum_{i=1}^q} R^2_{m,s} (\lambda_i) w_{s,i}^2}.
\end{equation}
Certainly, the larger the RFs, the smaller the uncertainties.

\subsubsection{Problems in practice}
\label{sec:sirstrategy}

\paragraph{Nodes and singular value decomposition}

With an LM algorithm, inversion of the RTE reduces in summary to solving Eq.\ (\ref{eq:gradhessian}), which implies the inversion of the modified Hessian matrix. One can certainly not expect the same practical problems when $\matriz{H}$ is built for an ME inversion or for a more general assumption where physical quantities vary with depth. Already in the ME case, $\matriz{H}$ has dimensions $9\times 9$ or $10 \times 10$ (if the filling factor is assumed to be different from unity). Inverting a $10 \times 10$ matrix is not difficult but, in the more general case, when the atmosphere is parameterized with a depth grid of 20 or 30 points, the Hessian may have several tens or even hundreds of elements in both dimensions. Inverting such matrices is by no means an easy numerical task.

A second problem can appear in practice as $\matriz{H}$ may be a quasi-singular (numerically singular) matrix because of the different sensitivities of the Stokes parameters to the various physical quantities that may vary even by orders of magnitude. One particular Stokes parameter of one specific spectral line may not be sensitive to a given physical quantity at given depths in the atmosphere. We already know, for instance, that, about or below $\log\tau_{\rm c} = 0$, only temperature leaves its fingerprints on the spectrum: the profiles are insensitive to the other quantities (see Section \ref{sec:rfproperties}). Hence, the corresponding matrix elements in $\matriz{H}$ will be close to zero, so that they hamper the Hessian matrix inversion. Here we report on the way SIR deals with these two problems. Other inversion techniques (e.g., LILIA, NICOLE, MILOS) apply similar procedures, although no much explicit information is available. The first problem can be circumvented by using several iteration cycles in each of which the number of free parameters is fixed and increased successively from cycle to cycle. The inversion of quasi-singular matrices is usually carried out through the singular value decomposition technique \citep[SVD; e.g.,][]{press+etal1986}.

\paragraph{Nodes and equivalent response functions}

Imagine that we only have one physical quantity to deal with in the inversion. Then, the number of free parameters is $n$, the number of depth grid points. Our Hessian is an $n \times n$ matrix. A practical way out of this involved numerical problem is found \citep{1992ApJ...398..375R} by assuming that all depth grid point perturbations are not free but bound by some interpolation formula. For example, we can use polynomial splines. This assumption allows to consider \emph{any} number $n^{\prime}$ of free parameters from $1$ through $n$. If such a number is 1, we assume we are applying a constant perturbation, whatever the original stratification is. The perturbation will be linear if the number is 2, parabolic if it is 3, and so on. As explained in \citet{2003isp..book.....D}, the use of nodes requires the evaluation of \emph{equivalent} RFs at the nodes, $\tilde{\vector{R}}$'s, in order to take information from the whole atmosphere into account. With this technique, the equivalent of Eq.\ (\ref{eq:deltaiquad}) in practice becomes
\begin{equation}
\label{eq:deltaisyn}
\delta \vector{I}^{\rm syn} (\lambda_l) = \sum_{m=1}^{n'p+r} \tilde{\vector{R}}_m (\lambda_l) \, \delta y_m,
\end{equation}
where $y_m$ is a new notation for the free parameters at the nodes.\footnote{Note that $n^{\prime}$ may be different from quantity to quantity, making the code more versatile.} Constant and depth-varying physical quantities are treated the same in Eq.\ (\ref{eq:deltaisyn}), and the quadrature coefficients are assumed to be included in the RF definitions. 

\paragraph{Different sensitivities to the various free parameters}

Since, by construction, the Hessian matrix is real and symmetric, its inversion is done through diagonalization. The quasi-singularity of the Hessian matrix shows up as some of the diagonal elements, $\gamma_k$, being too close to zero to be inverted with accuracy. This could be for two reasons. The first one is that, within the free parameters belonging to the same physical quantity, some depths are necessarily more important than others: as we have seen in Sect.\ \ref{section:response}, RFs tend to zero at some depths. This problem is numerically solved by setting $1/\gamma_k$ to zero, whenever $\gamma_k$ is considered too small (under a given threshold). By doing so, we are not (over)correcting a parameter that has no relevance at this time.\footnote{This is exactly the trick employed by SVD for its main application: inverting a quasi-singular matrix. According to  \citet{press+etal1986}, cancelling the inverse of the smallest eigenvalues provides the least-squares solution.} The second reason for singularity is that some physical quantities are more important than others. Therefore, the sensitivities of Stokes profiles to perturbations of those quantities can be larger (even by an order of magnitude) than perturbations to less significant quantities. This problem is overcome by using relative instead of absolute RFs as we explained in Section \ref{sec:rfproperties}. Nevertheless, and in order to make sure that all physical quantities are considered during each inversion cycle, the zeroing of the less significant diagonal elements is applied separately to each physical quantity. 

\epubtkImage{}{%
\begin{figure}[htbp]
    \centerline{\includegraphics[width=0.6\textwidth]{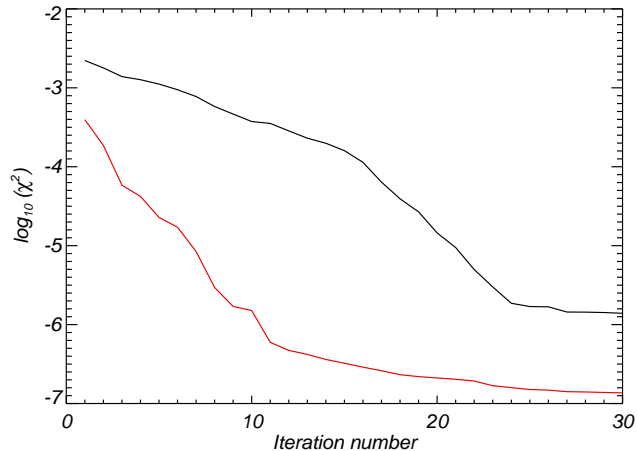}}
    \caption{Convergence rate (logarithm of the merit function in Eq. \ref{eq:chi2}) comparison between an inversion run by using fixed initial guesses for the physical quantities (black line) and the same by using the approximate estimates given in the Appendix (red line). One thousand Stokes profiles of the Fe~{\sc i} line at 617.3 nm have been used in the experiment and average results are plotted. They have been synthesized with the MELANIE code \citep{2001ASPC..236..487S} with uniformly distributed field strengths between 0 and 2000~G, inclinations and azimuths between 0\degree and 180\degree, and LOS velocities between -2 and 2 \kms. The profiles are sampled at 30 wavelengths regularly spaced every 2 pm. A C-programed version of MILOS \citep{2007A&A...462.1137O} has been used. No noise has been added to the profiles.}
    \label{fig:acceleration}
\end{figure}}

\paragraph{Initialization}

The lack of uniqueness we were discussing in the Introduction may be revealed in a dependence on the initialization parameters. The community has realized this fact for a long time and codes such as SIR or {\sc HeLIx} have been explicitly tested and showed robust against different initializations \citep{1992ApJ...398..375R, 2004A&amp;A...414.1109L}. Such robustness can nonetheless depend on the specific Stokes profiles and model atmosphere. Therefore, an advisable practice when doubts arise is to use several different initial guesses for estimating the uncertainties in the results for each physical quantity. Several attempts have been performed as well for finding an ideal initialization guess, including specific genetic algorithm procedures only for getting the initial guess (Skumanich, private communication). In our opinion, having initializations almost as complicated as the inversion itself does not make much sense. 

While preparing a given application, we discovered an outstanding, very economical way of making optimum initial guesses. Such an initialization is very much in the line we have been supporting throughout the paper, namely, the usefulness of a step-by-step approach. Using the classical center-of-gravity \citep{1979A&A....74....1R} and weak-field approximations of the Appendix, we have obtained a remarkable acceleration in the convergence, as shown in Figure\ \ref{fig:acceleration}.\footnote{The initialization for the inclination and azimuth angles are indeed very similar to those proposed in the paper by \citet{auer+etal1977}} According to \citet{2003ApJ...592.1225U}, the center of gravity technique has the remarkable property of being quite insensitive to the spectral resolution of the data. We can add that the results in Fig.\ \ref{fig:acceleration}, which have been obtained from profiles sampled at 30 wavelengths, are indeed fairly similar to those for six wavelength samples like those foreseen for the SO/PHI instrument (SO/PHI is the acronym for the Polarimetric and Helioseismic Imager for the ESA's \emph{Solar Orbiter} mission; see \citeauthor{2015IAUS..305..108S}, \citeyear{2015IAUS..305..108S}). 

\paragraph{Consistency among different ME implementations}

Milne--Eddington inversion techniques have become widely used to infer the vector magnetic field and the LOS velocity of the solar plasma. The physical interpretation of ME results was first investigated by \cite{1997ASPC..118..197W, 1998ApJ...494..453W} while comparing them to those obtained with SIR. A check of the theoretically predicted result and that obtained in practice was carried out in the latter paper for the HAO-ASP code \citep{1985NASCP2374..306S,1988ApJ...330..493L}. Predictions by \citet{1996A&A...314..295S} were confirmed: 1) measurements are essentially the result of averaging the actual parameter stratification with the corresponding generalized response function; 2) the so-called ME thermodynamic parameters had little correlation with the actual (quickly varying with optical depth) thermodynamic parameters. Further practice has usually shown that these ``thermodynamic'' parameters may not be very consistent among different runs for the same spectral line, while $\vector{B}$ and $v_{\rm LOS}$ are fairly accurate. This finding has been physically explained by \citet{2007A&A...462.1137O} who explored the shapes of ME response functions: RFs to perturbations in $\eta_{0}$, $\Delta \lambda_{D}$, and $a$ are fairly similar among themselves and, hence, cross-talk may appear between every two parameters; this is not the case, however, with RFs to perturbations in $\vector{B}$ and $v_{\rm LOS}$, which are neatly different. 

Although the consistency among different versions of the ME inversion is therefore guaranteed through physical analysis, the various implementations may have different numerical approximations and the technique can even be different (e.g., LM, genetic algorithms, PCA, etc.). Motivated by this fact, \citet{2014A&amp;A...572A..54B} have checked the consistency among the HAO-ASP, {\sc HeLIx}, and VFISV codes. They have found a positive confirmation of the previous results by using MHD simulations as a test bench instead of ME Stokes profiles.

\subsubsection{Automatic selection of nodes}
\label{sec:automatic}

In several specific problems, the optimum choice for parameterizing the atmosphere is actually an art, i.e., a skill that arises after the continuous exercise of intuitive skills. Codes like SIR help the user by enabling several node choices for each parameter.  Some different choices can yield similar fits; others can make it impossible to reach a convergent solution. As we have commented on in several places of the present paper, a generally good approach is \emph{lex parsimoniae}: between two solutions reaching a similar fit quality, that with fewest nodes should be selected. However, in practice, one cannot repeat the inversion several times in order to choose the optimum number of nodes for each parameter.

The current version of SIR includes an algorithm that automatically selects such a number of nodes for every parameter in each iteration. The algorithm is based on the quest for the roots or zeros of the partial derivative of $\chi^2$ with respect to each parameter, as written in Eq.\ (\ref{eq:chiderivative}). Let $a$ be one of the atmospheric quantities varying with optical depth and $a_p$ its value at $\tau_p$; that is, $a_p$ is one of the elements in the model atmosphere of Eq.\ (\ref{eq:modelatmos}). Let us call $d_{a_p} \equiv (\partial \chi^2/\partial a_p)$. Let us also suppose that we are only dealing with the intensity profile of one spectral line, and that, at a given iterative step,  $I^{\rm obs} > I^{\rm syn}$ for all wavelengths. If $R_{p,0} (\lambda_i)$ is positive for all wavelengths and all optical depths, it is then clear that $d_{a_p}$ will also be positive at all depths. To get a better fit, then, we will need to increase $a$ everywhere and just one node might be enough. Following this reasoning, it is easy to conclude that the number of nodes for a given physical quantity should be related to the number of times that the derivative $d_a$ changes its sign over the optical depth range. Obviously, as the derivative depends on the observational data, it is influenced by noise and, consequently, spurious zeros should be eliminated. Consequently, the algorithm determines the number of nodes after looking for positive relative maxima, and negative relative minima, larger in absolute value than a given threshold. An example of the behavior of this automatic selection feature in SIR is shown in Section \ref{sec:complexity}.

This automatic selection of the number of nodes can be considered a quantitative implementation of the principle of Occam's razor. Others are indeed possible. An alternative was presented by \citet{2006ApJ...646.1445A}. This author uses the minimum description length principle to effectively find the optimum number of expansion coefficients in PCA-based inversion techniques or the optimum number of nodes for the various atmospheric parameters in the SIR code. This problem is also addressed by \citet{2007ApJ...660.1690A}, who estimated the intrinsic dimensionality of spectropolarimetric data based on nearest neighbor considerations and applying the principle of maximum likelihood.

\subsubsection{A non-LTE inversion technique}
\label{sec:non-lte}

The only available non-LTE inversion technique so far is called NICOLE by \citet{2015A&amp;A...577A...7S}, which is an evolution of the IAC-NLTE code by \citet{2000ApJ...530..977S}. In turn, the latter was adapted from a previous one for non-polarized problems by \citet{1998ApJ...507..470S}. Since even the minute details of the code are extensively described in those papers, let us simply stress here the main assumptions underlying the code for those potential users to know the validity framework of the results. 

Regarding the minimization technique, NICOLE employs an LM algorithm that proceeds very similarly to SIR by using response functions. Since RFs cannot strictly be calculated in this specific non-linear, non-local problem, the \emph{fixed departure coefficients} (FDC) approximation is used to deal with the derivatives of the LTE atomic level populations once the $\beta$'s in Eq.\ (\ref{eq:depcoeff}) are fixed from a previous calculation \citep{1998ApJ...507..470S}. Although the approximation is not exact and, indeed, the authors show deviations from the correct values, FDC is good enough for the purposes of getting RFs that pave the way for the code to find the minimum distance between the observed and the synthetic profiles. The second important approximation in NICOLE is the field-free approximation \citep{1969SoPh...10..268R}, as we already mentioned in Section \ref{sec:milne}. It consists in obtaining the departure coefficients from an unpolarized, non-LTE code and uses them in a formal solution of the RTE. This way, the $\beta$'s are decoupled from the magnetic field. According to the authors, this approximation is valid because the actual level populations are governed by strong UV (weakly split) lines and those lines with large Zeeman splittings are weak enough not to have a significant influence on the statistical equilibrium equations. 

NICOLE only deals with polarization induced by the Zeeman effect. Hence, any polarization produced by scattering or depolarization through the Hanle effect are not taken into account. Other assumptions, such as the validity of \emph{complete frequency redistribution}, may have implications for applications in specific spectral lines. It has recently been used in the analysis of the Ca~{\sc ii} line at 854.2 nm \citep{2015ApJ...810..145D}.

\subsection{Database-search inversions}
\label{sec:pca}

The foundations of Principal Component Analysis inversion codes have already been explained in Sect.\ \ref{sec:approxprof}. The expansion of Stokes profiles as linear combinations of eigenprofiles is at the root of the technique. A set of synthetic Stokes profiles of a given spectral line, obtained with a large number of model atmospheres, is used as a training set to decompose each synthetic and observed profile into a sum of a small number of such eigenprofiles. The inversion topological problem, then, is reduced to a search in the low-dimensional space generated by the eigenprofiles or, more specifically, in the space of the expansion coefficients. The technique has proved to be efficient for quick inversions of the observations and looks very promising as a classification tool for profiles that can later be examined with more detailed techniques. We say so because no PCA code so far has been envisaged to go further than an ME or a slab atmosphere. This is natural in a way. One can build a database where a few constant parameters can get values within given ranges but the mere construction of such a database would be a formidable problem if variations with optical depth of the atmospheric physical quantities are considered. The technique, therefore, is unable to deal with gradients and the like, which ---we know--- populate a large fraction of the magnetic Sun. Nevertheless, as a first-order approach it is extremely useful everywhere and is eventually the only available tool to explore the behavior of some solar features \citep[e.g.,][]{2003ApJ...582L..51L,2005ApJ...622.1265C}. 

According to \citet{2002ApJ...570..379S}, the leading orders of the PCA expansion may have a direct (approximate) interpretation in terms of values for the physical quantities, specifically for the LOS velocity and the vector magnetic field. This result is in line with our discussion in Sects.\ \ref{sec:approxmod} and \ref{sec:approxprof} about successive approximations in the complexity of both the model atmospheres and the profiles. Since the profile database has to be created for each spectral line or group of lines, PCA is very well suited to analyze those spectral lines whose radiative transfer is particularly complicated, either because physical mechanisms other than the Zeeman effect are involved in their formation, or because the scenario is morphologically difficult, as in prominences, spicules, and other chromospheric structures \citep{2002ApJ...575..529L, 2005A&amp;A...436..325L, 2005ApJ...622.1265C, 2009ApJ...703..114C}. Extensions of the technique to stellar problems have also been proposed already \citep{2006ASPC..358..355S, 2006ASPC..358..405R, marian+etal2008b, 2012A&amp;A...544A...4P, 2015A&amp;A...573A..67P}.

A particularly interesting feature of the PCA technique is that once the observed Stokes profiles are expanded in terms of the eigenprofiles they become less noisy. This can be helpful for several applications \citep[e.g.,][]{marian+etal2008b, 2013A&amp;A...549L...4R}. 

\subsection{Other algorithm inversions}
\label{sec:otheralg}

\subsubsection{Artificial neural network inversions}
\label{sec:anns}

Artificial neural networks (ANNs) are systems through which a multidimensional input is translated into a multidimensional output by means of a non-linear mapping. The mapping (or, better, the parameters for the mapping) is (are) obtained by a previous process called training where the system is presented with inputs whose target outputs are already known. The process of training can be long and tedious but, once it is finished, the ANN can deal with new inputs with an extremely quick performance. Within the realm of solar physics, only multi-layer perceptrons have been proposed \citep[][\citeauthor{2003NN.....16..355S}, \citeyear{2003NN.....16..355S}]{2001A&amp;A...378..316C}. In these specific ANNs, the input, composed of $N$ \emph{neurons}, is sequentially ---layer by layer--- transformed into an output of $N$ neurons as well, of which a subset are the $M$ elements of the target. Following the notation by \citet{2005ApJ...621..545S}, the propagation rule between the input (layer 0) and the output (layer L) is given by 
\begin{equation}
\label{eq:annpropagation}
Y_n^l = f_{l} \left( \sum_{j=1}^{N} W_{n,j}^{l} Y_n^{l-1} + \beta_{n}^{l} \right),
\end{equation}
where $Y_n^l$ represents the contents of neuron $n$ in layer $l$, $W_{n,j}^{l}$ stands for the \emph{synaptic} weight connecting that neuron with neuron $j$ in layer $l-1$, and $\beta_{n}^{l}$ is a bias level. One or more layers may have a non-linear \emph{activation} function $f_{l}$ which depends on the specific implementation. In fact, the two above mentioned papers use a different $f$.

For ANNs, the topological problem is dealt with during the training process, while the $W$'s, the $\beta$'s, and the parameters defining $f$ are determined. It reduces to \citep{2001A&amp;A...378..316C} minimizing a quadratic distance similar to:
\begin{equation}
\label{eq:annmetric}
\xi^{2} = \sum_{i=1}^{P} \sum_{k=1}^{M} \left( Y_{k}^{L} - T_{k}^{i} \right)^{2},
\end{equation}
where $P$ is the total number of training input-target vector pairs and $T_{k}^{i}$ is the $k$th target value for the $i$th training pair.

Artificial neural networks have been used very seldom with actual data. In fact, as far as we know, only the two applications by \citet{2005ApJ...621..545S} and by \citet{carroll+kopf2008} have been published so far.

\subsubsection{Genetic algorithm inversions}
\label{sec:genetic}

A general description of a genetic algorithm (GA), along with an implementation of the so-called PIKAIA code, was given by \citet{1995ApJS..101..309C}. When compared to simple steepest ascent or descent techniques, these genetic algorithms are an alternative to, e.g., LM. The former can easily be stuck in local minima while GA and LM look and (eventually) find global minima of the multi-variable merit (or fitness) function. Some authors claim \citep[e.g.][]{2004A&amp;A...414.1109L} that GA techniques are more robust in finding global minima than LM, but no direct comparison is known to the authors of this review.

Devised for the specific problem of exploiting the chromospheric diagnostic capabilities of the He~{\sc i} multiplet at 1083 nm, \cite{2004A&amp;A...414.1109L} presented the so-called {\sc HeLIx} code. ({\sc HeLIx} is an acronym for Helium Line Information eXtraction.) The code is a direct adaptation of the PIKAIA routine to the He~{\sc i} multiplet formation. The line formation problem includes both the Zeeman and Hanle effects, and the presence of two blending photospheric lines of Si~{\sc i} and Ca~{\sc i}. Therefore, much care has to be taken with the analyzed wavelengths (some have to be weighted to zero) and, above all, with the complex, forward radiative transfer problem. The latter is dealt with for the photospheric Si~{\sc i} line by using the synthesis part of the SPINOR code (although no results from it are reported). The non-LTE effects in the He~{\sc i} triplet are neglected because the line is mostly optically thin. A simple ME atmosphere is assumed instead. The Hanle effect treatment is based on the oscillator model by \citet{2002Natur.415..403T}. The code later evolved (now it is called {\sc HeLIx$^+$}) to include the incomplete Paschen--Back effect  \citep{2006A&amp;A...456..367S} and to finally incorporate all the properties of the so-called constant property slab model by \citet{2005ApJ...619L.191T,2002Natur.415..403T} through the addition of the forward synthesis code by \citet{1982SoPh...79..291L} with extensions by \citet{2008PhD...UL...M}. Now, the code shares the synthesis calculation module with HAZEL \citep{2008ApJ...683..542A} and carrying out a direct comparison between the two codes with both numerical and actual observations would be a very interesting exercise, useful for the whole community. This would bring a gauge of pros and cons of GA versus LM algorithms. As a general rule of thumb we can say that genetic algorithm inversions become feasible methods whenever the evaluation of the merit function is extremely fast because this king of algorithms require the evaluation of the merit function a thousand or a million times.

\subsubsection{Bayesian inversions}
\label{sec:bayesian}

An alternative technique to the inversion problem that adds some extra statistical information on the results, namely confidence levels on the free parameters, has been proposed by \citet{2007A&amp;A...476..959A}, based on Bayes' theorem. According to that theorem, once the \emph{posterior} distribution $p(\vector{x}|\vector{I}^{\rm obs})$ is known, the position of its maximum indicates the most probable (a.k.a.\ optimum) combination of parameters that best fits the observations $\vector{I}^{\rm obs}$. The posterior (probability) distribution represents how much we know of the parameters once the observational data set is taken into account. As the reader may have already imagined, the Bayesian inversion is nothing but a maximization of $p(\vector{x}|\vector{I}^{\rm obs})$ instead of a minimization of the $\chi^2$ merit function. We are going to see that, indeed, the two optimization problems collapse to exactly the same result in given cases where the \emph{a priori} knowledge of the problem inserted in the calculations is the simplest (and ---probably--- the safest). Let us explain a little what we mean by this a priori knowledge.

According to the theorem, the posterior distribution is proportional to the product of a \emph{prior} distribution $p(\vector{x})$ and the \emph{likelihood} distribution, $p(\vector{I}^{\rm obs}|\vector{x})$:
\begin{equation}
\label{eq:bayes}
p(\vector{x}|\vector{I}^{\rm obs}) \propto p(\vector{x}) \, p(\vector{I}^{\rm obs}|\vector{x}).
\end{equation}

The likelihood distribution measures the probability that a given model atmosphere $\vector{x}$ can produce synthetic Stokes profiles $\vector{I}^{\rm syn}$ that fit the observed ones, $\vector{I}^{\rm obs}$. If the noise distributions are normal and typically independent of wavelength as it is usual to assume, then the likelihood is defined as \citep{2003itil.book.....M}
\begin{equation}
\label{eq:likelihood}
p(\vector{I}^{\rm obs}|\vector{x}) = {\rm e}^{-\frac{1}{2} \chi^2 (\vector{x})},
\end{equation}
where $ \chi^2 (\vector{x})$ is given in Equation (\ref{eq:chi2}). Imagine now, for a moment, that the prior is identically unity, $p(\vector{x}) = $ constant (within a reasonable range), which corresponds to a case where no a priori assumptions are made about the model parameters. (All possibilities are equally probable.) In such a case, Eq.\ (\ref{eq:bayes}) becomes
\begin{equation}
\label{eq:bayessimple}
p(\vector{x}|\vector{I}^{\rm obs}) \propto {\rm e}^{-\frac{1}{2} \chi^2 (\vector{x})}
\end{equation}
and indicates that maximizing the posterior distribution is {\bf exactly} the same as minimizing $\chi^2$. No matter what the optimization algorithm used for the inversion, introducing Eq.\ (\ref{eq:chi2}) into Eq.\ (\ref{eq:bayessimple}), we have the way for estimating confidence levels as given by the (multidimensional) posterior probability distribution. Two-dimensional cuts of $p(\vector{x}|\vector{I}^{\rm obs})$ allow one to explore the possible degeneracies or cross-talks among each pair of model physical quantities. As a matter of fact, and according to our discussions in Sect.\ \ref{section:response}, response functions and uncertainties derived from Eq.\ (\ref{eq:uncertainties}) provide qualitatively similar confidence levels although Bayes' theorem supplies a more graphical approach \citep[see figures in][]{2007A&amp;A...476..959A}. The only difficulty is then properly sampling the hyperspace of parameters (see below).

The prior distribution contains the information we may have of the model parameters without taking the observations into account. If all the model physical quantities are statistically independent, then 
\begin{equation}
\label{eq:prior}
p(\vector{x}) = \prod_i^{np+r} p(x_i).
\end{equation}
Unless other physical information is available, the typical assumptions one can make on the free parameters are the range of reliable values for each of them. Thus, a useful model for the prior distribution can be given by
\begin{equation}
\label{eq:modelprior}
p(x_i) = H(x_i, x_i^{\rm min}, x_i^{\rm max}),
\end{equation}
where $H(x,a,b)$ is the typical top-hat function
\begin{equation}
\label{eq:tophat}
H(x,a,b) = \left\{ \begin{array}{ll}
{\displaystyle \frac{1}{b-a}} & {\rm if} \,\, a \leq x \leq b, \\
0 & {\rm otherwise.} \end{array} \right.
\end{equation}
Establishing a prior, therefore, is analogous to the assumptions made by SIR on the (spline) smooth variations of the physical quantities along the atmosphere. The useful feature of $p(\vector{x})$ is that you can even consider correlations between the quantities and model them accordingly. This is in our opinion, however, a risky exercise because over fanciful correlations can be conceived that turn into an even more involved interpretation of the results. One has to make sure that the specific conditions of the problem enable this or that a priori assumption on parameter cross-talk.

In summary, if either $p(\vector{x}) = $ constant or is given by Eqs.\ (\ref{eq:prior}), (\ref{eq:modelprior}), and (\ref{eq:tophat}), the optimization problem is the same as that described in Sect.\ \ref{sec:chisquare} and, in principle, the LM algorithm could be used as well. The missing ingredient is the sampling of the free parameter hyperspace. An alternative method is then in order. Sampling the parameter space means repeating the synthesis of Stokes profiles many, many times. Typically, one needs of the order of $10^{np+r}$ samples of the posterior distribution. When the number of free parameters is high, such a brute-force method becomes impracticable.  \citet{2009ASPC..405..315A} propose using a ``not-so-brute-force'', Markov chain Monte Carlo method, where marginalized distributions of parameters can be obtained. The educated successive sampling grows linearly with the number of free parameters instead of exponentially. The decrease in computational cost has allowed the authors to deal both with ME atmospheres and with general LTE atmospheres where the physical quantities vary with depth \citep{2012ApJ...748...83A}.

\subsection{Inversions accounting for spatial degradation}
\label{sec:spatialdegrad}

A significant step forward has been adopted by three different techniques after acknowledging the spatial effects of non-ideal instruments \citep{2012A&amp;A...548A...5V,2013A&amp;A...549L...4R,2015A&amp;A...577A.140A}. While the spectral PSF of the instruments was soon incorporated into the inversion codes,\footnote{One typically convolves the synthetic Stokes profiles with the spectral PSF.} the spatial blurring could not be satisfactorily dealt with until these works. Note that, in spectropolarimetry, an extended spatial PSF not only degrades the quality and contrast of images but also introduces a spurious polarization signal that can be misinterpreted as magnetic fields and LOS velocities. Several attempts were proposed for mitigating (or circumventing) this spatial contamination, such as using a non-polarized, global quiet-Sun average \citep{skumanich+lites1987} or a non-polarized, local (1$^{\prime\prime}$ neighborhood) average \citep{2007ApJ...662L..31O}. In these examples, the magnetic structure is assumed to contribute with a filling factor $\alpha$. None of these can be considered fully consistent but they do provide an improvement in robustness and reliability of the results. So-called spatially-coupled inversions \citep{2012A&amp;A...548A...5V}, regularized deconvolution inversions \citep{2013A&amp;A...549L...4R}, and sparse inversions \citep{2015A&amp;A...577A.140A} attack the problem directly although through different means. The first technique uses the SPINOR code and the second employs SIR; a combination of the fast iterative shrinkage-thresholding algorithm \citep{2009ITIP...18.2419B} and the restarting scheme by \citet{2016FouMath...15.715O} is chosen for the third algorithm. 

\subsubsection{Spatially-coupled inversions}
\label{sec:coupled}

From a formal point of view, the gradient of the merit function and the Hessian matrix were very helpful in explaining the second order approximation that is behind the Levenberg--Marquardt algorithm. Instead of using $\nabla\chi^2$, let us think of the Jacobian matrix $\matriz{J}$ of the system, which is made up of the derivatives of all the data points (all wavelength samples of the four Stokes profiles) with respect to the free parameters. That is, the elements of $\matriz{J}$ are just the individual terms in the summation of Equation (\ref{eq:chiderivative}). It is then clear that the approximation in Eq.\ (\ref{eq:seconderivativechi}) is equivalent to saying that the Hessian matrix $\matriz{H}' \simeq \matriz{J}\trans \matriz{J}$ and that $\matriz{H} = \matriz{H}' - {\mbox{\bf diag}} \left[\matriz{H}' \right]$.
Consider now the inversion of the RTE for a whole spectropolarimetric image of $n\times m$ pixels individually (the uncoupled case). We can build a big (block-diagonal) new Jacobian $\matriz{J}$ with the ensemble of  $\matriz{J}_{k,l}$ of all the individual pixels:
\begin{equation}
\label{eq:ensemble}
\matriz{J} = \left( \begin{array}{ccccc}
\matriz{J}_{1,1} & \matriz{0} & \cdots & \cdots & \matriz{0} \\
\matriz{0} & \matriz{J}_{1,2} & \matriz{0} & \cdots & \matriz{0} \\
\vdots & \vdots & \ddots & \vdots & \vdots \\
\matriz{0} & \cdots & \cdots & \matriz{J}_{n,m-1} & \matriz{0} \\
\matriz{0} & \cdots & \cdots & \cdots & \matriz{J}_{n,m} \end{array} \right).
\end{equation}

\epubtkImage{}{%
\begin{figure}[htbp]
    \centerline{\includegraphics[width=\textwidth]{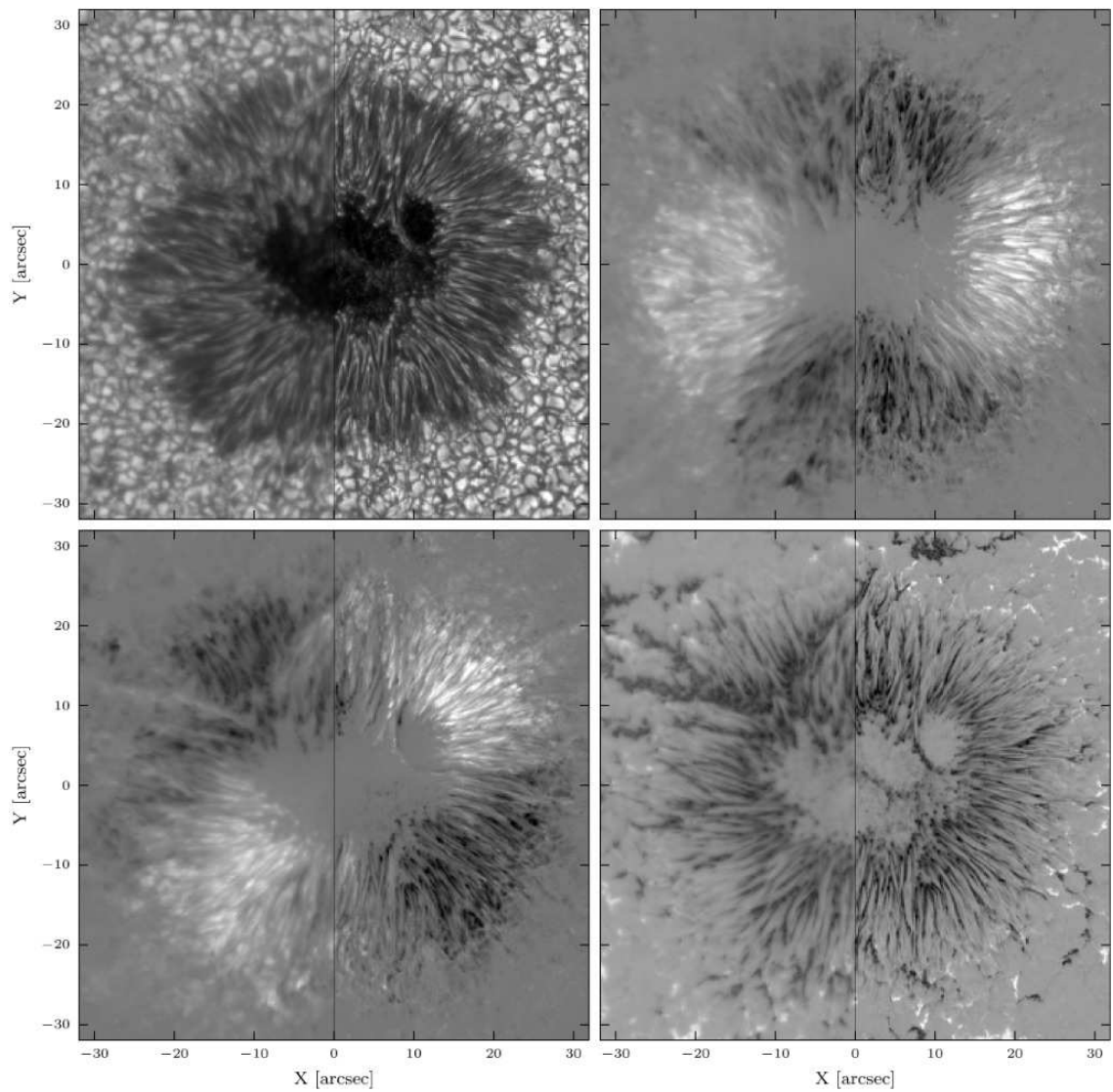}}
    \caption{Stokes images in the wing (+ 5.7 pm) Fe~{\sc i} line at 630.25 nm before (left half) and after (right half) spatially coupled inversion for the Stokes parameters I (top left), Q (top right), U (bottom left) and V (bottom right). Adapted from \citet{2012A&amp;A...548A...5V}.}
    \label{fig:vannoortfig}
\end{figure}}

The new (big) $\matriz{H}'$ readily becomes block diagonal, with the blocks being the $\matriz{H}'_{i,j}$ of all the individual pixels:
\begin{equation}
\label{eq:hensemble}
\matriz{H}' = \left( \begin{array}{ccccc}
\matriz{H}'_{1,1} & \matriz{0} & \cdots & \cdots & \matriz{0} \\
\matriz{0} & \matriz{H}'_{1,2} & \matriz{0} & \cdots & \matriz{0} \\
\vdots & \vdots & \ddots & \vdots & \vdots \\
\matriz{0} & \cdots & \cdots & \matriz{H}'_{n,m-1} & \matriz{0} \\
\matriz{0} & \cdots & \cdots & \cdots & \matriz{H}'_{n,m} \end{array} \right).
\end{equation}
The inverse of this matrix (or that of matrix $\matriz{H}$) is easy to obtain by individually inverting each of its block components.

Let us address now the spatially coupled inversion problem, where the effects of an assumed uniform PSF $\varphi(x,y)$ across the image are taken into account. The Jacobian can now be written as
\begin{equation}
\label{eq:bigjacobian}
\matriz{J} = \left( \begin{array}{rrcrr}
\varphi_{0,0} \matriz{J}_{1,1} & \varphi_{0,-1} \matriz{J}_{1,2} & \cdots & \varphi_{1-n,2-m} \matriz{J}_{n,m-1} & \varphi_{1-n,1-m} \matriz{J}_{n,m} \\
\varphi_{0,1} \matriz{J}_{1,1} & \varphi_{0,0} \matriz{J}_{1,2} & \cdots &  \varphi_{1-n,3-m} \matriz{J}_{n,m-1} & \varphi_{1-n,2-m} \matriz{J}_{n,m}\\
\vdots & \vdots & \ddots & \vdots & \vdots \\
\varphi_{n-1,m-2} \matriz{J}_{1,1} & \varphi_{n-1,m-3} \matriz{J}_{1,2} & \cdots & \varphi_{0,0} \matriz{J}_{n,m-1} & \varphi_{0,-1} \matriz{J}_{n,m} \\
\varphi_{n-1,m-1} \matriz{J}_{1,1} & \varphi_{n-1,m-2} \matriz{J}_{1,2} & \cdots & \varphi_{0,1} \matriz{J}_{n,m-1} & \varphi_{0,0} \matriz{J}_{n,m} \end{array} \right),
\end{equation}
which is no longer diagonal. Nevertheless, since the influence of the PSF should be relatively short and not involve all the image points, the resulting $\matriz{J}$ has a significant number of zero elements (it is sparse). If we call $Y \equiv \varphi \ast \varphi$ the autocorrelation function of the PSF, it can be shown that \citep{2012A&amp;A...548A...5V} 
\begin{equation}
\label{eq:bighessian}
\matriz{H}' = \left( \begin{array}{rrr}
Y_{0,0} \matriz{J}_{1,1}\trans \matriz{J}_{1,1} & \cdots & Y_{1-n,1-m} \matriz{J}_{1,1}\trans \matriz{J}_{n,m} \\
Y_{0,1} \matriz{J}_{1,2}\trans \matriz{J}_{1,1} & \cdots &  Y_{1-n,2-m} \matriz{J}_{1,2}\trans \matriz{J}_{n,m}\\
\vdots & \ddots & \vdots \\
Y_{n-1,m-2} \matriz{J}_{n,m-1}\trans \matriz{J}_{1,1} & \cdots & Y_{0,-1} \matriz{J}_{n,m-1}\trans \matriz{J}_{n,m} \\
Y_{n-1,m-1} \matriz{J}_{n,m}\trans \matriz{J}_{1,1} & \cdots & Y_{0,0} \matriz{J}_{n,m}\trans \matriz{J}_{n,m} \end{array} \right).
\end{equation}

Inverting matrix $\matriz{H}'$ is beyond our reach. However, the inversion of $\matriz{H}$ is affordable ---at least approximately--- because the linear system is sparse, although the number crunching problem is formidable. The authors propose strategies for the approximate $\matriz{H}'$ and recognize that human intervention (the \emph{artistic} part) is in the end more needed in these coupled inversions than in regular uncoupled ones, as expected. 

Although the improvement with respect to former techniques is clear and the procedure is opening a new avenue for physical inferences in the solar photosphere (see Figure \ref{fig:vannoortfig}), a few unsatisfactory oscillations appear here and there in the application to actual observations. Such oscillations show up a caveat of the technique: the possible amplification of high frequencies and, hence, of noise. It can be argued that the spatially coupled inversions carry out convolutions instead of deconvolutions, but it is also true that any spurious, high-frequency signal may be compatible with the inverted models, provided it is washed out by the convolution with the PSF. 

What is not clear either to the authors of the present review is the quite unexpected quantitative results in some applications. For example, \citet{2013A&amp;A...557A..24V} claim to have found magnetic field strengths of 7.5 kG at $\log\tau = 0$ in some parts of the penumbra. These values can easily break the observational paradigm where fields stronger of 4 kG have very seldom been observed. Since they may represent a new paradigm, they should be accompanied by an estimate of uncertainties that is not present in the paper. First of all, the fits obtained for the $V$ profile in their fig.\ 3, seem to be different from the observations by at least an order of magnitude greater than the noise at several wavelengths. Such a fit cannot be considered satisfactory. But even more important is the fact that the quoted field strength corresponds to very deep layers in the atmosphere. As explained by \citet{1994A&A...283..129R}, the second term in Eq.\ (\ref{eq:responsefun}) rapidly tends to zero at low layers because the difference between the Stokes profiles and the source function vector quickly vanishes at these layers. In these circumstances, it is easy to see that values for the magnetic quantities at $\log\tau = 0$ are extremely uncertain because the RFs go to zero (Equation \ref{eq:uncertainties}). Unless otherwise justified, those strong values at low layers cannot be interpreted but as (not-very-accurate) extrapolations of the magnetic field strength global stratification if SPINOR uses the equivalent response functions at the nodes of Section \ref{sec:sirstrategy}.\footnote{No explicit note on its usage is known to the authors.} If instead, the RFs are the regular ones, then we are afraid that the (uncoupled) inversion strategy should be modified by changing the nodes to other places in the atmosphere with greater sensitivity to perturbations in the magnetic and dynamic physical quantities.

\subsubsection{Regularized deconvolution inversions}
\label{sec:regularized}

A much simpler and computationally cheaper approach has been proposed by \citet{2013A&amp;A...549L...4R}. Based on the idea of Stokes profile expansion in terms of the principal components provided by a regular PCA technique, instead of deconvolving the Stokes profile images wavelength by wavelength, which is a very expensive and risky process,\footnote{Stokes $Q$, $U$, and $V$ images are almost noise for most wavelengths. Therefore, the risk of noise enhancement is high during any deconvolution.} deconvolution is applied to the PCA coefficient images. The resulting Stokes profiles after deconvolution are then inverted with SIR. The procedure is neat and simple, the uncertainties in the determination of physical quantities can be obtained through Eq.\ (\ref{eq:uncertainties}), and the possible overcorrections due to an excess in deconvolution can easily be controlled.

The idea is to assume that PCA expansions up to a degree $D$ are valid to describe the profiles fully before they reach the telescope. Under such an assumption, the observed Stokes profiles can be written as 
\begin{equation}
\label{eq:stokesregular}
\vector{I} (\lambda) = \sum_{i=1}^{D} (\vector{\omega}_i \ast \vector{P}) \, \phi_i (\lambda) + \vector{N},
\end{equation}
where $\vector{\omega}_i$ are the weights to the PCA eigenprofiles $\phi_i (\lambda)$, $\vector{P}$ stands for the spatial PSF of the instrument, and $N$ is the noise (assumed independent of wavelength). One of the interesting features of this regularized deconvolution  is that the noise contamination is largely minimized since the real signal is usually contained in the first few coefficients. Then, the $4\times N$ images $(\vector{\omega}_i \ast \vector{P}) + \vector{N}$ are deconvolved by a Richardson--Lucy algorithm \citep{1972JOSA...62...55R,1974AJ.....79..745L}. The resulting deconvolved Stokes profiles are then inverted with SIR. 

The main objection one may find to this technique is the same as that for PCA, namely that the most, say, \emph{peculiar} Stokes profiles, which indeed reveal very interesting physics, are not fully fit since one would need more PCA terms. The logical solution for such a problem would be to increase $D$ but this may not be advisable as the final computation time would be too long. A practical circumvention can be provided by classification techniques (e.g., \emph{k-means clustering}) that allow the grouping of the different Stokes profiles according to purely morphological criteria \citep{1967ProcFifth...1...281M,1957BullAcadPol...4...801S,1982IEEE...28...129L}.\footnote{As far as we know, clustering techniques were introduced in solar spectropolarimetry by  \citet[][see also \citeauthor{2011A&amp;A...530A..14V}, \citeyear{2011A&amp;A...530A..14V}]{2007ApJ...663.1386P}.} After such a classification, those peculiar profiles can be identified. One can then carry out the PCA expansion tailored to each group of profiles, hence optimizing the global performance. An increase in expansion terms may only be needed for small fractions of pixels in an image, thus keeping the whole inversion efficient. 

\subsubsection{Sparse inversions}
\label{sec:sparse}

An extremely interesting new generation of inversion methods has been proposed by \citet{2015A&amp;A...577A.140A}.
Based on the concept of sparsity or compressibility, this technique allows us to tackle the inversion of 2D maps ---and potentially 3D data sets--- all at once. The underlying idea of sparsity is the intrinsic redundancy of the data. That is, data can be projected to a parameter space where a reduced set of variables can fully describe that data set. Provided that a linear transformation exists between the data set and the new ---small sized--- parameter space, an affordable inversion of 2D maps can be carried out. In this first paper the authors present a 2D ME inversion based on a wavelet transformation of the model parameter space. They show that reducing the dimensionality of the model unknowns by a factor between three and five yields results comparable to ---or even better than--- pixel-to-pixel inversion.

Time saving is among the advantages of this kind of methods, along with their ability to 
easily compensate for the effects of the telescope PSF and the regularization of solutions introduced by
the sparsity hypothesis. Among the drawbacks we find the apparent impossibility of using LM methods ---since the Hessian matrix 
scales as the square of the number of free parameters. The authors suggest using proximal algorithms that can
increase the convergence speed of the standard gradient descent method that it is currently used.

\subsection{Summary of inversion techniques}
\label{sec:summaryits}

This section summarizes in Table \ref{tab:tableinversionsconstant} all the past and current inversion techniques that have been proposed or are in use for solar physics.  A distinction is made between those techniques that assume physical quantities that are constant with optical depth and those that allow the quantities to vary over the photosphere. LM stands for Levenberg--Marquardt, ANN for artificial neural networks, GA for genetic algorithm, B for Bayesian, and GD for gradient descent. The overwhelming majority of codes uses the Levenberg--Marquardt algorithm in order to find the minimum distance between the observed and the synthetic profiles.\footnote{The Florence code was modified by  \citet{2007A&amp;A...464..323B}.} \footnote{In the HAZEL code, the coarse approach to the minimum distance is made through a Lipschitzian method. See the paper.} \footnote{See \citet{2015A&amp;A...577A...7S} for details about NICOLE.}
\begin{savenotes}
\begin{table}[htbp]
  \caption{Inversion techniques. LM stands for Levenberg--Marquardt, ANN for artificial neural networks, GA for genetic algorithm, B for Bayesian, and GD for gradient descent.}
  \label{tab:tableinversionsconstant}
  \centering
  \begin{tabular}{llcc}
   \\
    \toprule
    Constant quantities\\
    \toprule
    Identifier & Reference & Method & In use \\
    \midrule
    KPNO & \mbox{\citet{1972lfpm.conf..227H}} & LM &  \\
    HAO-KPNO & \mbox{\citet{auer+etal1977}} & LM & \\
    Florence & \mbox{\citet{1984SoPh...93..269L}} & LM & $\checkmark$ \\
    HAO-ASP & \mbox{\citet{1985NASCP2374..306S}} & LM & $\checkmark$ \\
    IAC MISMA & \mbox{\citet{1997ApJ...491..993S}} & LM & $\checkmark$ \\
    CSIRO-Meudon & \mbox{\citet{rees+etal2000}} & PCA & $\checkmark$ \\
    HAO MELANIE & \mbox{\citet{2001ApJ...553..949S}} & LM & $\checkmark$ \\
    HAO FATIMA & \mbox{\citet{2001ApJ...553..949S}} & PCA & $\checkmark$ \\
    AIP ANN & \mbox{\citet{2001A&amp;A...378..316C}} & ANN &  \\
    HAO He {\sc i} D$_{3}$ & \mbox{\citet{2003ApJ...582L..51L}} & PCA & $\checkmark$ \\
    HAO ANN & \mbox{\citet{2003NN.....16..355S}} & ANN &  \\
    MPS {\sc HeLIx}	& \mbox{\citet{2004A&amp;A...414.1109L}} & GA & $\checkmark$ \\
    IAC Molecular & \mbox{\citet{2004PhDULL...A}} & LM & \\
    IAA MILOS & \mbox{\citet{2007A&A...462.1137O}} & LM & $\checkmark$ \\
    IAC HAZEL & \mbox{\citet{2008ApJ...683..542A}} & LM & $\checkmark$ \\
    HAO VFISV & \mbox{\citet{2011SoPh..273..267B}} & LM & $\checkmark$ \\
   IAC Sparse & \mbox{\citet{2015A&amp;A...577A.140A}} & GD & $\checkmark$ \\    
   \bottomrule
     \toprule
    Variable quantities\\
    \toprule
    Identifier & Reference & Method & In use \\
    \midrule
    ETH Flux tube & \mbox{\citet{1990A&amp;A...233..583K}} & LM &  \\
    IAC SIR & \mbox{\citet{1992ApJ...398..375R}} & LM & $\checkmark$ \\
    ETH IT & \mbox{\citet{1992A&amp;A...263..312S}} & LM & \\
    IAC Flux tube & \mbox{\citet{1997ApJ...478L..45B}} & LM & $\checkmark$ \\
    ETH SPINOR & \mbox{\citet{1998A&amp;A...336L..65F}} & LM & $\checkmark$ \\
    IAC NLTE & \mbox{\citet{2000ApJ...530..977S}} & LM & \\
    HAO LILIA & \mbox{\citet{2001ASPC..236..487S}} & LM & $\checkmark$ \\
    HAO-IAC NICOLE & \mbox{\citet{2001ASPC..236..487S}} & LM & $\checkmark$ \\
    KIS SIRGAUS & \mbox{\citet{2003ASPC..307..301B}} & LM & $\checkmark$ \\
    IAA SIRJUMP & \mbox{\citet{2009ApJ...704L..29L}} & LM & $\checkmark$ \\
    IAC Bayes & \mbox{\citet{2009ASPC..405..315A}} & B & $\checkmark$ \\
    MPS Spatially coupled & \mbox{\citet{2012A&amp;A...548A...5V}} & LM & $\checkmark$ \\
    IAC Regularization & \mbox{\citet{2013A&amp;A...549L...4R}} & LM & $\checkmark$ \\
    \bottomrule
  \end{tabular}
\end{table}
\end{savenotes}

\clearpage


\section{Discussion on inversion results}
\label{section:discussion}

\subsection{Increasing complexity in the model atmospheres}
\label{sec:complexity}

Following the approach of Sects.\ \ref{sec:approxmod} and \ref{sec:approxprof}, we want to discuss in this section how a given set of Stokes profiles can be fit with several assumptions about the stratification of the model atmosphere physical quantities. This discussion sheds light on the ill-conditioning issue we reported on in Sect.\ \ref{sec:introduction}: the same profile can be interpreted in several ways, depending on the complexity of the assumed model atmosphere; the only limiting factor for increasing such a complexity should be the noise in the observations and a reasonable dose of the principle of Occam's razor. To illustrate our discussion we shall be using SIR to carry out all the necessary calculations, in a similar way to that followed by \citet{2006ASPC..358..107B}. Given the SIR strategy explained in Sect.\ \ref{sec:sirstrategy}, the natural way to making a given physical quantity stratification more complex is by increasing the number of nodes and, hence, the polynomial degree of the spline that is assumed to describe such a stratification. Therefore, we shall be inverting the ``observed" profiles with various sets of nodes for each of the different atmospheric quantities. 

\epubtkImage{}{%
\begin{figure}[htbp]
    \centerline{\includegraphics[width=\textwidth]{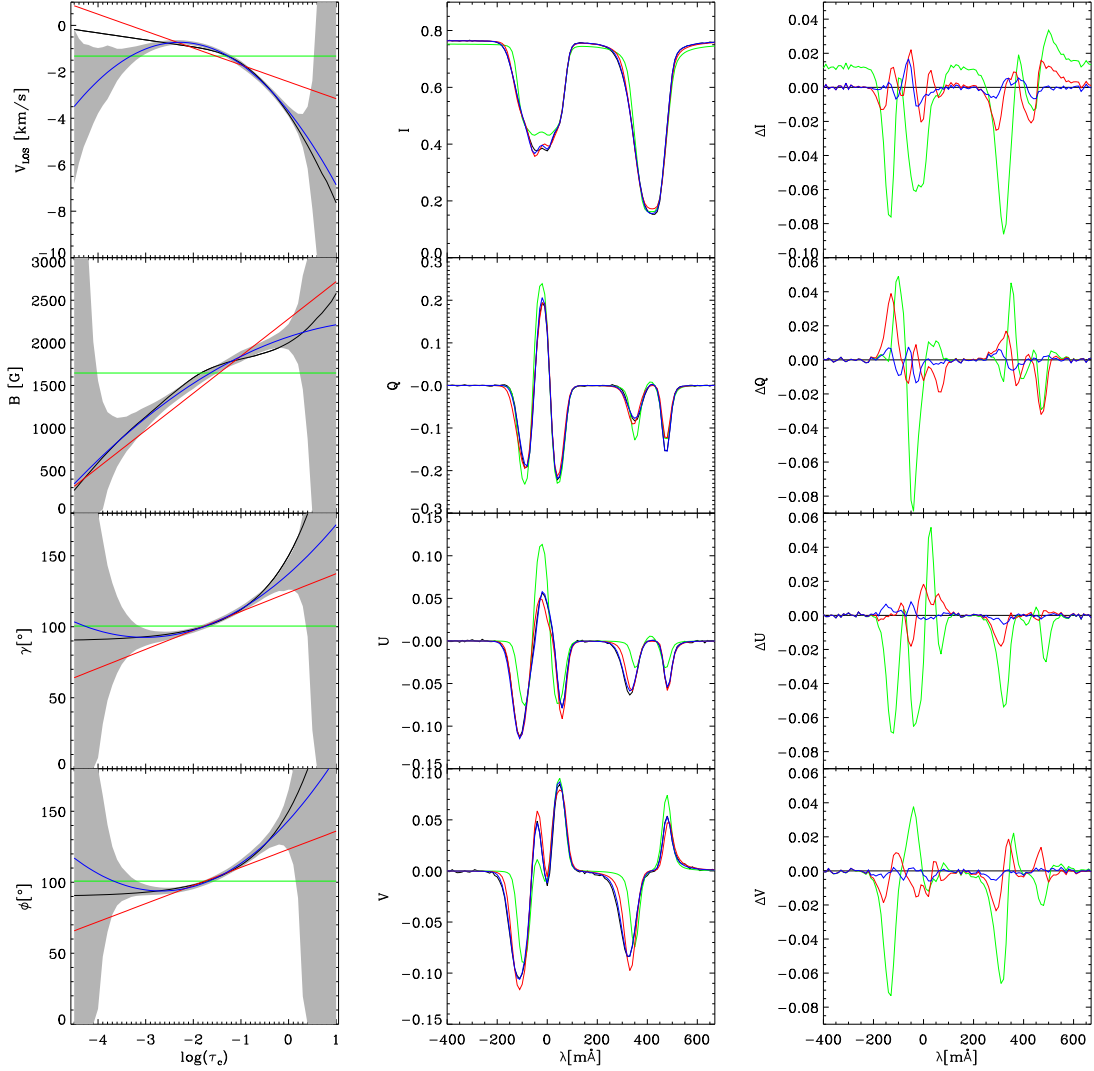}}
    \caption{Left column: $v_{\rm LOS}$, $B$, $\gamma$, and $\varphi$ stratifications with optical depth for the observed model atmosphere (black lines) and for the resulting models from the mode 1 (green lines), mode 2 (red lines), and mode 3 (blue lines) runs of SIR. Gray shaded areas cover the uncertainty region of the mode 3 solution. Middle column: the corresponding Stokes profiles. Right column: differences between the observed and inverted Stokes profiles. The abscissa for both the middle and right columns shows the wavelength centered at 630 nm.}
    \label{fig:Fig90ab}
\end{figure}}

Among all possible node combinations or modes, we have selected only six for the sake of simplicity. Each mode is characterized by the number of nodes, $n_{\rm B}$, used for $B$, $\gamma$, $\varphi$, and $v_{\rm LOS}$. The number of nodes for $T$, $n_{\rm T}$, is higher by two nodes than that for the magnetic and dynamic quantities, except for mode 1 where $n_{\rm T} = 2$. Mode 1 can be called ``\`a la ME" because it has just $n_{\rm B} = 1$. Since the starting guess model atmosphere has constant $\vector{B}$ and $v_{\rm LOS}$, only constant values for these quantities can result from this inversion. Mode 2 has $n_{\rm B} = 2$, so that linear stratifications are allowed in this mode. Mode 3 has $n_{\rm B} = 3$; hence, parabolic stratifications can come out from SIR in this mode. Mode 4 has $n_{\rm B} = 5$ and the stratification of the magnetic and dynamic quantities can be quartic. Mode 5 has $n_{\rm B} = 7$ and the stratifications can be of order 6. Finally, mode 6 uses the automatic node selection algorithm described in Section \ref{sec:automatic}. 

We have built a penumbral model atmosphere after making up a bit one of the resulting models from inversion of a \emph{Hinode} observation. We will call it hereafter \emph{the observed model}. Our choice is driven by the shape of the Stokes profiles emerging from such an atmosphere. They are far from being typical even and odd functions of wavelength. The stratifications for $v_{\rm LOS}$, $B$, $\gamma$, and $\varphi$ in the observed model are plotted (from top to bottom) with black lines in the left panels of Figs.\ \ref{fig:Fig90ab} and \ref{fig:Fig90cd}. With this model atmosphere, we have synthesized the two Fe~{\sc i} lines at 630.1 and 630.2 nm, convolved them with the \emph{Hinode} spectropolarimeter PSF, sampled with the instrument wavelength sampling interval, and finally added noise to a level of $10^{-3} \, I_{{\rm c}}$. The so-obtained Stokes profiles will be called \emph{the observed Stokes profiles} and are plotted in black lines in the middle panels of Figs.\ \ref{fig:Fig90ab} and \ref{fig:Fig90cd} (despite being barely discerned). Besides the observed model and profiles, both figures display the resulting models and fit profiles from the corresponding mode runs of SIR. Green, red, and blue lines correspond to modes 1, 2, and 3, respectively, in Fig.\ \ref{fig:Fig90ab}, and to modes 4, 5, and 6, respectively, in Figure \ref{fig:Fig90cd}. The right panels of the two figures show the differences between the observed and the fit profiles in the corresponding colored lines. These differences provide a direct measure of the fit quality.

\epubtkImage{}{%
\begin{figure}[htbp]
    \centerline{\includegraphics[width=\textwidth]{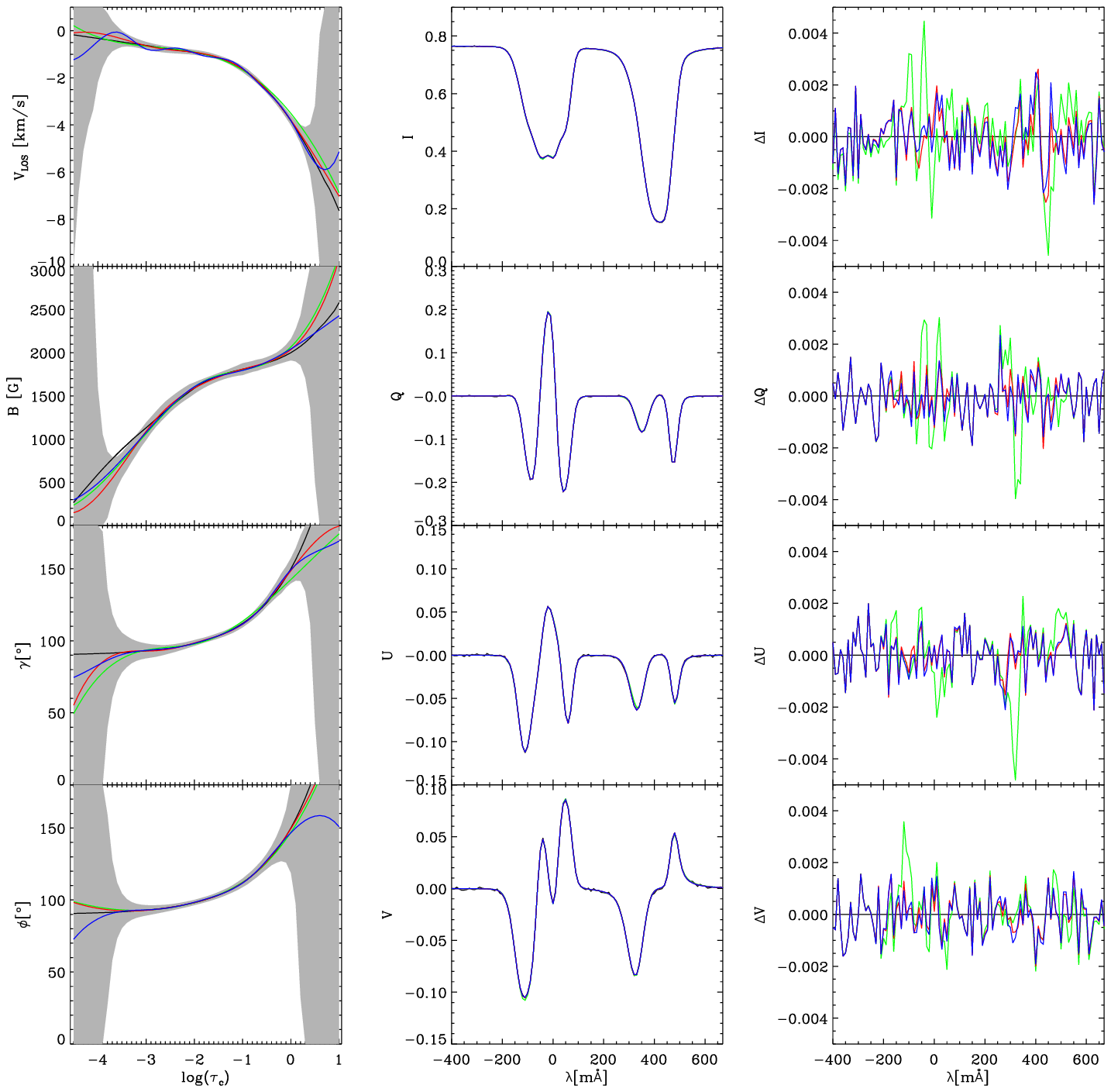}}
    \caption{Same as in Fig.\ \ref{fig:Fig90ab} but with red, green, and blue lines representing modes 4, 5, and 6, respectively.}
    \label{fig:Fig90cd}
\end{figure}}

The \`a-la-ME inversion yields a fairly good fit although, as expected, it is unable to reproduce the asymmetries in the profiles. The typical misfit is never larger than 10 \%. The results given by the \`a-la-ME inversion coincide with the actual values at $\log \tau_{\rm c} \simeq -1.5$. The exact coincidence takes place at different depths for each quantity but the important qualitative message to be extracted is that, in spite of its simplicity, the ME approximation is able to retrieve the atmospheric quantities at the mid-photosphere. Looking at the differences in the right panels of Fig.\ \ref{fig:Fig90ab}, the parity rules we commented on in Sect.\ \ref{sec:approxprof} seem not to operate but the reason is clear: the ME model is still too far from the observed model for linearity to hold. Moreover, given the strong asymmetry shown by the profiles, the specific wavelength around which we should symmetrize or anti-symmetrize the profiles has to be calculated because it is certainly different from the nominal rest wavelength of the spectral line. The linear stratification mode does a better job as it provides a mean gradient with which asymmetries start to be reproduced. The fits clearly improve with the parabolic mode. Note that, besides having reduced the misfits, the stratifications of $v_{\rm LOS}$, $B$, $\gamma$, and $\varphi$ are fairly well mimicked between $\log \tau_{\rm c} = -3$ and $\log \tau_{\rm c} = -0.5$. Note that the uncertainties evaluated with Eq.\ (\ref{eq:uncertainties}) ---and displayed in the figure with shaded gray areas--- indicate very well the range of reliability of the resulting stratifications. This is very important in practice where the observed model atmosphere is unknown. In our example, the uncertainties for the three components of the vector magnetic field are almost compatible with the linear stratification results. This is not the case for the LOS velocity where deviations are apparent and pave the road for more complex stratifications to improve the fits. In spite of this fact, the real gauge for deciding to proceed in the increase of nodes through the optical path is noise. As we have been discussing in many places in this paper, only if the differences between the observed and the synthetic profiles are larger than a few times the rms noise can we expect to obtain improvements with alternative model atmospheres. Since our noise still looks small enough when compared with the Stokes profile differences, we try with modes 4, 5, and 6. The retrieved model atmospheres are better than the former and, in particular, the uncertainty shaded areas in the top panels of Fig.\ \ref{fig:Fig90cd} (they correspond to mode 5) indicate that indeed the range of reliability has extended up to $\log \tau_{\rm c} = 0$. Notice that the size in the difference panels of the figure for mode 4 indicates that there is still some room to improve the fits, while mode 5 and 6 have profile differences compatible with the noise of the observations. Therefore, we cannot go any further (but indeed the fit quality is superb). 

\epubtkImage{}{%
\begin{figure}[htbp]
    \centerline{\includegraphics[width=0.8\textwidth]{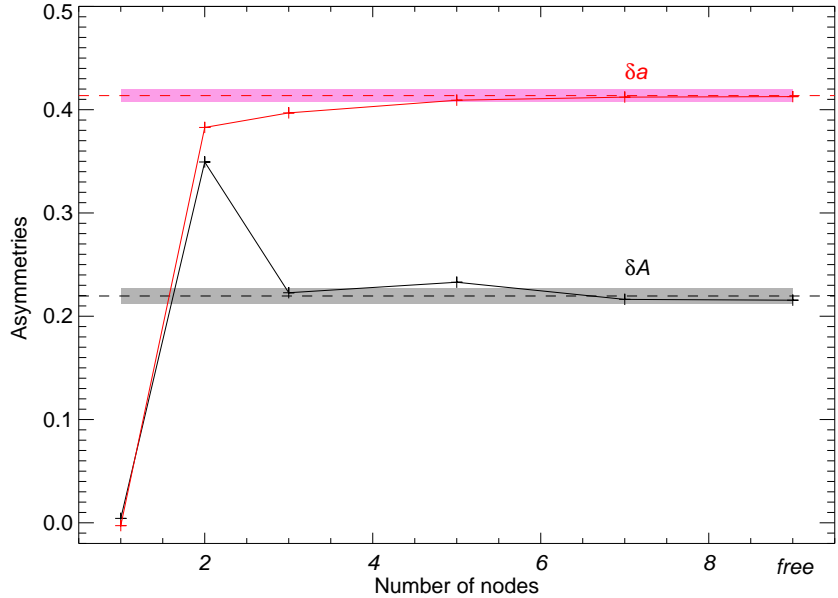}}
    \caption{Area ($\delta A$) and amplitude ($\delta a$) asymmetry of the Stokes $V$ profile as a function of the number of nodes for $B$. Dashed lines correspond to the asymmetries of the observed profile while shaded areas mark the uncertainties introduced by a noise of 10$^{-3}\, I_{\rm c}$.}
    \label{fig:asymmetries}
\end{figure}}

This example illustrates the ability we can have to retrieve very complex stratifications when both the noise is low and asymmetries are present. The latter feature is indeed important. On the one hand, if no asymmetries are present in the observed Stokes profiles we can readily discard part of the complexity: no variations of the LOS velocity with optical depth are present. On the other hand, and very remarkably, asymmetries increase the amount of available information: if profiles are symmetric (either even or odd), half of them can be thrown away although the retrievals will be noisier (see Section \ref{sec:rfproperties}). 

Since Stokes profile asymmetries have driven most of the evolution of concepts in inversion techniques and, in general, in radiative transfer, a further check on the way our numerical experiments have been able to reproduce those asymmetries is in order. Let us consider the typical definitions for the Stokes $V$ amplitude, $\delta a$, and area, $\delta A$, asymmetries:
\begin{equation}
\label{eq:asymmetries}
\delta a \equiv \frac{a_{\rm b} - a_{\rm r}}{a_{\rm b} + a_{\rm r}}, \,\,\, \delta A \equiv \frac{A_{\rm b} - A_{\rm r}}{A_{\rm b} + A_{\rm r}},
\end{equation}
where $a_{\rm b}$ and $a_{\rm r}$ stand for the amplitudes of the Stokes $V$ blue and red lobes, and $A_{\rm b}$ and $A_{\rm r}$ do for the unsigned areas of those lobes, respectively. Figure\ \ref{fig:asymmetries} describes the performance of the different modes in reproducing $\delta a$ and $\delta A$. Obviously, mode 1 shows zero asymmetries. Mode 2 approaches significantly the amplitude asymmetry but not the area one in this example. Mode 4 almost reproduce both quantities. The noise level (represented by the horizontal shaded areas) is so low, however, that only mode 5 and 6 provide an exact result.

\epubtkImage{}{%
\begin{figure}[htbp]
    \centerline{\includegraphics[width=\textwidth]{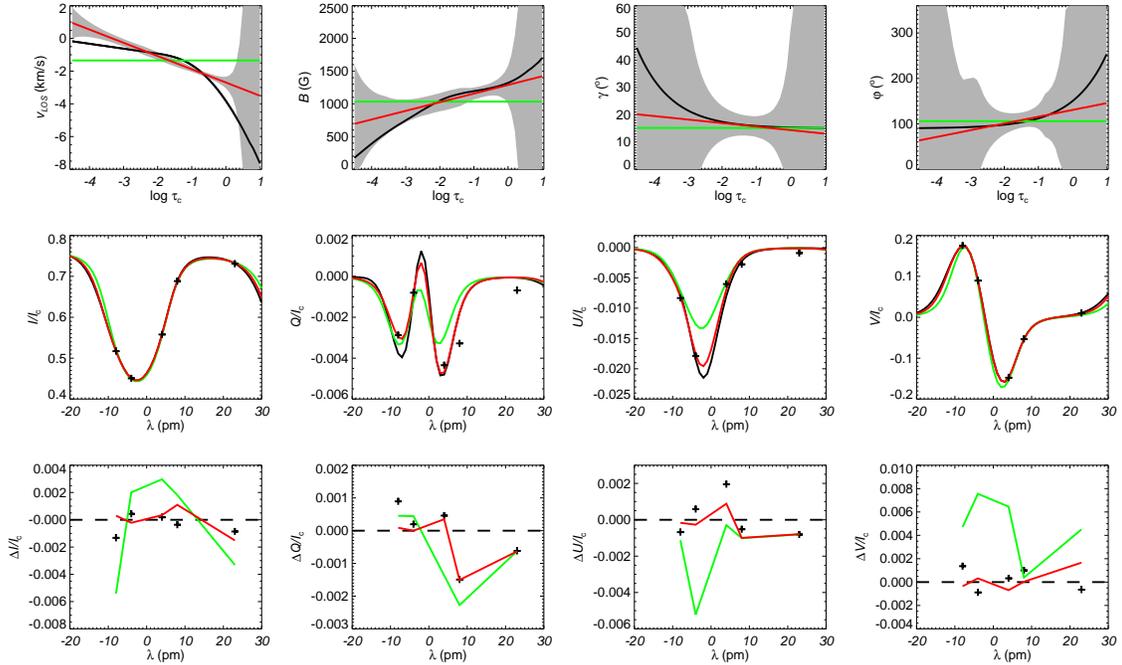}}
    \caption{Same as Figs.\ \ref{fig:Fig90ab} and \ref{fig:Fig90cd} but for a simulated IMaX observation.}
    \label{fig:FigIMAXV}
\end{figure}}

Certainly, if we decrease the amount of information by, for example, decreasing the number of wavelength samples and/or the polarization signal (because of the magnetic field being weaker) we cannot retrieve such complex model atmospheres any longer or, said otherwise, the range of reliability of the results will narrow down significantly so that low-order polynomial approximations are enough to give account of the observations. To exemplify this case, we have simulated a typical {\sc Sunrise}/IMaX observation in mode V5-6, that is, five wavelength samples and six accumulations. The IMaX Fe~{\sc i} line at 525.02 nm was synthesized in a quiet-sun model atmosphere (again taken from the actual observations), convolved with the IMaX PSF, sampled at -8, -4, 4, 8, and 22.7 pm from line center, and added noise at a level of 10$^{-3}\, I_{\rm c}$. Since $Q_{\rm c} = U_{\rm c} = V_{\rm c} = 0$ except perhaps in cases of very large LOS velocities, we just count in a total of $5\times4-3=17$ observables and, consequently, cannot afford to retrieve more than 17 unknowns. These 17 free parameters would cope with our mode 3 (five nodes for $T$ and three nodes for $v_{\rm LOS}$, $B$, $\gamma$, and $\varphi$). Indeed, this number is too high in practice and can only be reached in cases of strong asymmetries for the reasons we have just explained above: symmetries in the profiles reduce the degrees of freedom. Let us then consider mode 2 as the maximum achievable run and invert the observed profiles. Figure \ref{fig:FigIMAXV} is similar to Figs.\ \ref{fig:Fig90ab} and \ref{fig:Fig90cd} and the color codes are the same.\footnote{Panel arrangement is nevertheless rotated by 90$^{\circ}$.} The conclusions are clear, the \`a-la-ME mode provides fair values, and the linear approximation gives a reliable gradient on the physical quantities at around $\log \tau_{\rm c} = -1.5$.

\subsection{Inversion retrievals of weak fields}
\label{sec:weakretreival}

The reliability of inversion retrievals from zones with weak fields is a continuous matter of debate. Concerns are often published with different levels of arguments. As in any other aspect of life, criticism is always more prevalent than praise in any community. This is the case when discussing the ability of spectropolarimetric observations to distinguish weak fields and their inclinations. Most discussions are strongly biased by the fairly common misconception of Stokes $V$ being proportional to the longitudinal component of the magnetic field. We have shown in Sect.\ \ref{sec:weakatmosphere} that this approximation is valid only for a very limited range of values, and that the important observational parameter when dealing with weak signals is noise. Stokes profiles other than $V$ also provide information about $\vector{B}$. As long as the signal is not buried by noise, radiative transfer is powerful enough to provide sufficiently accurate magnetic quantities.

Sometimes the criticisms, are not correctly interpreted. An example is the evidence shown by \citet{marian+etal2006} that the pair of Fe~{\sc i} lines at 630 nm is not able to provide a single model for a scenario in which two depth-dependent atmospheres, one magnetic and another non-magnetic, fill a spatial resolution element of about 1$^{\prime\prime}$. This true result has often been interpreted as if the famous pair of lines were unable to provide a reliable inference of the vector magnetic field, and that only infrared lines were valid for such a diagnostic. \citet{2010ApJ...711..312D} explained that the scenario used by the former authors was perhaps too complicated for the available information. That is, that the visible line profiles are not enough to cope with the number of free parameters in a two-component, depth-dependent atmosphere. A simpler model atmosphere with just one magnetic component (and the other non-magnetic) may fit the profiles well enough. The latter authors gave both theoretical and observational arguments to defend the hypothesis that visible lines are reasonable diagnostics even for weak magnetic fields, in spite of infrared lines being more sensitive.

Later, \citet[][see also \citeyear{2012A&amp;A...547A..89B}, \citeyear{2013A&amp;A...550A..98B}]{2011A&amp;A...527A..29B} raised doubts about the retrievals of fairly inclined fields when the polarization signals are very weak because they come from quiet, internetwork regions \citep[e.g.,][\citeauthor{2008ApJ...672.1237L},\citeyear{2008ApJ...672.1237L}]{2007ApJ...670L..61O}. In our opinion, again, their claims are partial misinterpretations of the results, as we try to demonstrate with the calculations that follow.

Consider the pair of Fe~{\sc i} lines at 630 nm sampled as in the \emph{Hinode} \citep{kosugi+etal2007} spectropolarimeter \citep{2001ASPC..236...33L}. We have synthesized these lines in a quiet-Sun model for constant magnetic field strengths of 10, 15, 20, 25, 40, 50, 60, 75, 90, and 100~G. The inclination and azimuth may take any value but preserve an isotropic distribution, so that each of these values is equally probable. The LOS velocity has been set to zero for all the profiles. One thousand Stokes profile sets have been calculated for each value of $B$. Once synthesized, white noise has been added to the profiles with a standard deviation of $10^{-3}$ or $3.3 \cdot 10^{-4}  I_{\rm c}$, simulating ${\rm S/N} = 1000$ or $3000$.\footnote{We have followed here the customary procedure of adding equal rms noise to all four Stokes parameters. However, as stressed by \citet{2012ApJS..201...22D}, smaller noise should be added to Stokes $I$, which is always much better measured. This is so because almost all polarimeters are more efficient for Stokes $I$. In the optimum polarimeter case where Stokes $Q$, $U$, and $V$ have the same polarimetric efficiency, that of Stokes $I$ is higher by a factor of $\sqrt{3}$.} The synthetic profiles were then inverted with SIR \`a-la ME (two nodes in $T$ and one node for the rest of parameters). The results are summarized in Figure\ \ref{fig:histo_gamma6}. 

\epubtkImage{}{%
\begin{figure}[htbp]
    \centerline{\includegraphics[width=\textwidth]{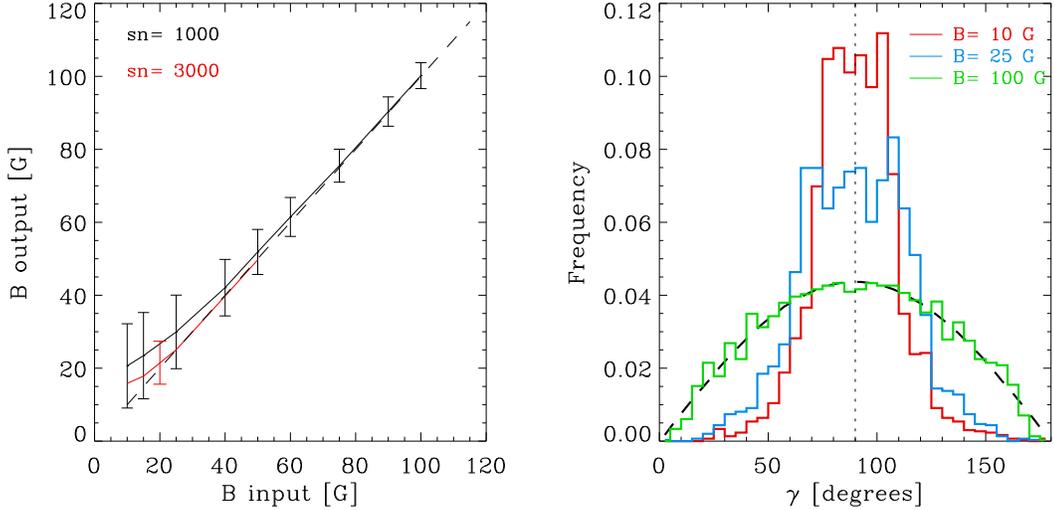}}
    \caption{Left panel: output magnetic field strength from the inversion as a function of input strength. Colors correspond to the two different values of the S/N. The dashed line marks the bisector of the first quadrant. Error bars represent rms values from the 1000 inversions. Right panel: distributions of the output field inclinations from the ${\rm S/N = 1000}$ inversions. Colors correspond to values of the field strength indicated in the inset. The input distribution of inclinations is represented by the dashed line.}
    \label{fig:histo_gamma6}
\end{figure}}

Fields weaker than 75 G (25 G) from the ${\rm S/N = 1000}$ (3000) inversions tend to be overestimated but in none of the cases is the excess output such that it would retrieve too strong a magnetic field. As a matter of fact, the results illustrate very well the fair reliability of the $B$ inference in practically all circumstances. Again, when noise decreases the inversion results improve. The magnetic inclination results, represented in the right panel of the figure, are also very illustrative of what is going on. Only results from the ${\rm S/N} = 1000$ experiments are plotted. One can clearly see that the weakest fields tend to result in an excess of inclined fields. {\bf This is,} however, {\bf an expected result that has nothing to do with any special inability of the visible lines or with the sought after  proportionality between Stokes $V$ and the longitudinal component of the magnetic field.} It is rather a consequence of the noise dominating the polarization signals. When the field is very weak, Stokes $V$ is very small, barely exceeding the noise level. At the same time, Stokes $Q$ and $U$ (which should theoretically be zero) simply show noise. Since the $V$ signals are not sufficiently larger than the linear polarization signals, the inversion code has no other option than to interpret the observations as very inclined magnetic fields: it is mostly fitting noise in $Q$ and $U$. The situation clearly improves as the field strength increases. The inclination distribution is well recovered for $B=100\,$G fields, even when S/N is only 1000.

Let us now consider a distribution of magnetic field strengths according to the probability density function (PDF) obtained by \citet{2007ApJ...670L..61O} from \emph{Hinode} observations. With these field strengths and an isotropic inclination distribution such as that for Fig.\ \ref{fig:histo_gamma6}, we have synthesized 10000 Stokes profiles to which white noise of rms amplitude of $\sigma = 10^{-3} \cdot I_{\rm c}$ was added. \`A-la ME inversions with SIR have been carried out. Both the inputs (black lines) and the results (red lines) are plotted in the upper panels of Figure \ref{fig:histo_campo}. Fields weaker than 20~G are slightly overestimated, but above 60 or 70~G the strength PDF is very nicely recovered. The inclination PDF shows an excess of horizontal fields in detriment of the more vertical ones. The same PDFs are shown in the lower panels for a selection of pixels where the maximum polarization signal (${\rm max} \{|V|,\sqrt{(Q^{2}+U^{2})} \}$) is greater than $4\sigma$. As one can clearly see, fields weaker than 10~G and a good portion of horizontal fields almost disappear. The inversions react very well. The underestimation of small inclinations and the excess of large ones can be attributed to insufficient S/N. When the experiments are repeated with higher signal-to-noise ratios, the PDFs agree accordingly better.

\epubtkImage{}{%
\begin{figure}[tbp]
    \centerline{\includegraphics[width=\textwidth]{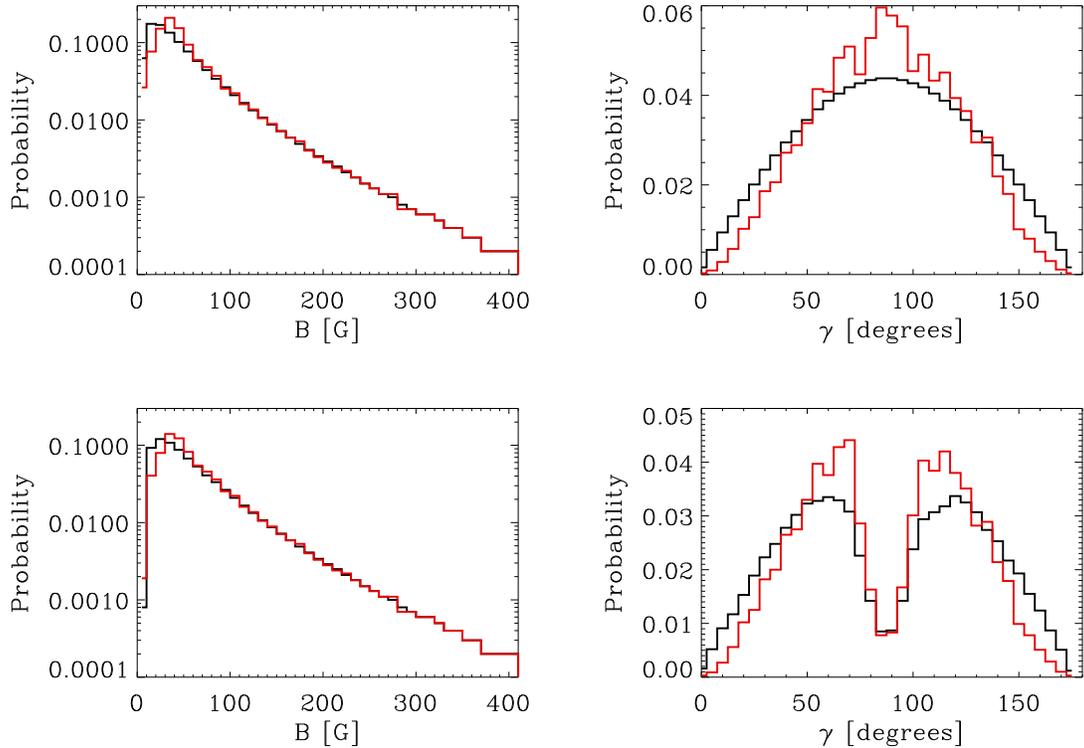}}
    \caption{Upper panels: magnetic field strength (left) and inclination (right) PDFs. $B$ follows the lognormal PDF from \citet{2007ApJ...670L..61O} with $B_{0} = 36.7$ and $\sigma = 1.2$. Inclinations follow the same random, isotropic distribution as for Figure \ref{fig:histo_gamma6}. Bottom panels: same as the upper ones but only for those points where the polarization signal is higher than a threshold (see text for details). Black lines correspond to the input and red lines to the output from \`a-la ME inversions.}
    \label{fig:histo_campo}
\end{figure}}

\newpage


\section{Conclusions}
\label{section:conclusions}

The inversion of the radiative transfer equation has been presented as a topological problem that maps the space of observables, the Stokes parameters, onto the space of the object physical quantities. The dependences of such a mapping on the definition of the two spaces implies a number of assumptions that are explicitly or implicitly made by any inference technique, regardless of it being called an inversion or not. Such assumptions determine to a great extent the uncertainties in the astronomical inferences, which depend on both the measurement errors and the analysis technique. 

In the observational space, one has to select the parameters to be measured and the level of noise with which such measurements are carried out. Signals (measured parameters) are useful insofar they vary after a modification in the object physical quantities. For the variation to be detectable it should be higher than the noise. If the signal does not change above noise levels after a perturbation in the physical quantities, then it is useless and must be discarded. In the object physical  space, the number of quantities and their assumed stratification with depth in the atmosphere are the key variables. If a physical quantity at a given depth in the atmosphere produces no measurable effect on the Stokes spectrum, then this quantity should not be looked for through the inversion process. The number of these physical quantities should not exceed the number of observables. 

The mapping between the two spaces is nothing but radiative transfer. Depending on the spectral line and the way we measure it (that is, the number of wavelength samples, the width of the spectral PSF, etc.) the transfer can be studied  through the full non-LTE problem, the LTE approximation or, rather, through further simplifications such as the Milne--Eddington approximation, the weak field approximation, etc. Strictly speaking, no available inversion technique deals with the full non-LTE problem and the only non-LTE code, NICOLE, relies on several approximations such as the fixed departure coefficient approximation or the field-free approximation in order to make the numerical problem tractable. 

We have provided arguments in favor of proceeding through a step-by-step approach in which the complexity of the problem increases sequentially until convergence has been reached. In this sense we strongly recommend initializing inversions with classical estimates of $B$ and $v_{\rm LOS}$ as provided by the center of gravity technique, and estimates of $\gamma$ and $\varphi$ as provided by the weak field approximation. The criterion for convergence has to be established in terms of noise: if the (rms) difference between observed and synthetic Stokes profiles is less than the typical noise of observations, increasing complexity in the object physical description adds no information. In this regard, we have given both conceptual and technical arguments for the MISMA hypothesis and inversion technique to be abandoned. At the other extreme of the ``complexity spectrum", the weak field approximation must only be used with much care and mainly for very broad chromospheric lines. Nothing in the transfer equation indicates that Stokes $V$ is proportional to $B \cos \gamma$. Only some matrix elements of \matriz{K} are. After integration through the atmosphere, the proportionality is most probably lost. Moreover, the information provided by the other Stokes parameters helps in disentangling the magnetic field strength from the inclination. In particular, Stokes $I$ very soon departs from the zero field conditions that are strictly necessary for the weak field approximation to apply. 

The step-by-step approach we suggest (and which is indeed implemented in the SIR inversion code) is to be preferred for two reasons: on the one hand, the atmospheric stratification of physical quantities can be described by a Taylor-expansion-like method where it is assumed to be first constant, then linear, later quadratic, and so on; on the other hand, the Stokes profiles ($I$ in line depression), as functions of the wavelength, belong to $\bbbl^2$ and, hence, can be expanded in terms of orthonormal bases such as those provided by the Hermite functions or those built for PCA techniques.

The various algorithms used for the inversion problem have been reviewed. They include the most widespread Levenberg--Marquardt algorithm, database search inversions, artificial neural networks, genetic algorithms, and Bayesian inferences. The most promising techniques for the near future, namely those which include spatial degradation by the telescope, are also discussed. Among these we find the so-called spatially-coupled inversions by  \citet{2012A&amp;A...548A...5V}, the regularized deconvolution inversions by \citet{2013A&amp;A...549L...4R}, and the sparse inversions by \citet{2015A&amp;A...577A.140A}. Suggestions for improving their performance and reliability are also given.

This paper ends with a discussion on a pair of topics that might seem controversial. The first one is a description of how the current implementation of SIR deals with a step-by-step approach. The sequential improvement on the fits is explicitly shown. The second topic discusses the reliability of weak field retrievals. The idea of a theoretical inability of Zeeman-sensitive spectral lines in the visible for inferring weak fields accurately is refuted. Instead, the root of the problem is shown to  be in the signal-to-noise ratio of the observations. If noise is suitably low, then radiative transfer provides the necessary tools for accurate retrievals. Of course, uncertainties will be proportionally larger than when signals are bigger (i.e., stronger fields), but there are no reasons for not trusting the inversion results.

\newpage

\appendix
\section{Appendix. Optimum practical initialization}
\label{sec:appendix}

The center of gravity technique was introduced by \citet[][see also \citeauthor{1979A&A....74....1R} \citeyear{1979A&A....74....1R}]{1967AnAp...30..513S}. The longitudinal component of the magnetic field ($B_{\rm LOS} = B \cos \gamma$) and the line-of-sight (LOS) velocity are provided in this technique by the semi-difference and the semi-sum of the centers of gravity of $S_{+} \equiv I + V$ and $S_{-} \equiv I - V$. Specifically, if $\lambda_{+}$ and $\lambda_{-}$ stand for such centers of gravity, then, assuming a unity magnetic filling factor,
\begin{equation}
\label{eq:blos}
B_{\rm LOS} = \beta_B \frac{\lambda_{+} - \lambda_{-}}{2}
\end{equation}
and
\begin{equation}
\label{eq:bloscofg}
v_{\rm LOS} = \beta_v \frac{\lambda_{+} + \lambda_{-}}{2},
\end{equation}
where $\beta_B = 1/C$, $\beta_v = c/\lambda_0$, and $C = 4.67 \cdot 10^{-13} \lambda_0^2 \, g_{\rm eff}$, with $\lambda_0$ the rest, central wavelength of the line in \AA; $c$ stands for the speed of light. In practice, $\lambda_{+}$ and $\lambda_{-}$ are calculated through
\begin{equation}
\label{eq:lammasmen}
\lambda_{\pm} \equiv \frac{\displaystyle \sum_{i=1}^q S_{\pm} (\lambda_i) \, \lambda_i}{\displaystyle \sum_{i=1}^q S_{\pm} (\lambda_i)}.
\end{equation}

According to \citet{2004ASSL..307.....L}, in the weak field regime, the azimuth of the magnetic field can be approximated by
\begin{equation}
\label{eq:azimuth}
\varphi \simeq \frac{1}{2} \arctan \frac{U}{Q}
\end{equation}
and the circular (Stokes $V$) and linear ($L\equiv \sqrt{Q^2+U^2}$) polarizations can be approximated respectively by Eq.\ (\ref{eq:vpropmag}) for all wavelengths and by 
\begin{equation}
\label{eq:linpol}
L \simeq \frac{3}{4} \Delta\lambda_B^2 \, \bar{G}\, \sin^2 \gamma\, \frac{1}{\Delta\lambda} \left( \frac{\partial I_{\rm nm}}{\partial \lambda} \right)
\end{equation}
for the line wings, where $\Delta\lambda \equiv \lambda - \lambda_0$, that is, the distance to the line center. Both $g_{\rm eff}$ and $\bar{G}$ depend on the quantum numbers of the two atomic levels involved in the transition. If we write Eq.\ (\ref{eq:lande}) as
\begin{equation}
\label{eq:lande2}
g_{\rm eff} = \frac{1}{2} g_{\rm s} + \frac{1}{4} g_{\rm d} d,
\end{equation}
with $g_{\rm s} = g_{\rm u} + g_{l}$, $g_{\rm d} = g_{\rm u} - g_{l}$, then $\bar{G}$ can be written as $\bar{G} = g_{\rm eff}^2 - \delta$, with \begin{equation}
\label{eq:deltadef}
\delta = \frac{1}{80} g_{\rm d}^2\, (16 s -7 d^2 - 4),
\end{equation}
with 
\begin{equation}
\label{eq:ssdef}
s \equiv [j_u (j_u +1) + j_l (j_l +1)]
\end{equation}
and
\begin{equation}
\label{eq:sdef}
d \equiv [j_u (j_u +1) - j_l (j_l +1)].
\end{equation}

Now, from Eqs.\ (\ref{eq:linpol}) and (\ref{eq:vpropmag}), we can write
\begin{equation}
\label{eq:loverv}
\frac{L}{V^2} = \frac{3}{4} \frac{\bar{G}}{g_{\rm eff}^2} \frac{1}{\Delta\lambda} \, \tan^2 \gamma \frac{1}{\partial I_{\rm nm}/\partial \lambda}
\end{equation}
and recast Eq.\ (\ref{eq:vpropmag}) as
\begin{equation}
\label{eq:vpropmag2}
V \simeq -C B_{\rm LOS} \, \frac{\partial I_{\rm nm}}{\partial \lambda}.
\end{equation}

The two last equations finally give
\begin{equation}
\label{eq:tangam}
\tan^2 \gamma = \left| \frac{4\, L\, g_{\rm eff}^2\, \Delta\lambda}{3\, \bar{G}\, C\, B_{\rm LOS}} \right|
\end{equation}
since $B_{\rm LOS}$ has been independently obtained trough Eq.\ (\ref{eq:blos}). Therefore, using only one or the average of some wavelength samples in the line wings, we obtain rough estimates for the magnetic inclination angle through Eq.\ (\ref{eq:tangam}) and for the magnetic azimuth through Equation (\ref{eq:azimuth}). Having $\gamma$ and $B_{\rm LOS}$ provides an estimate for $B$ as well. Note that the strict validity of the equations is unimportant because our only aim is to find approximate guesses for the inversion code to start.

\newpage


\section*{Acknowledgements}
\label{sec:acknowledgements}

We warmly thank the careful review by Andr\'es Asensio Ramos and by Juan Manuel Borrero. Their comments and suggestions are greatly appreciated. We are also thankful to Alberto Sainz Dalda for making us aware of the k-means clustering technique potential and for the bibliographic references. The accurate and very professional astrophysical English editing by Terry Mahoney is thankfully acknowledged. This work has been partially funded by the Spanish Ministerio de Econom\'{\i}a y Competitividad, under projects number ESP2013-47349-C6 and ESP2014-56169-C6, including European FEDER funds.

\newpage



\bibliography{refs}

\end{document}